\documentclass[natbib]{svjour3}                     
\smartqed  
\usepackage{graphicx}
\usepackage{aps-bibstyle}  
\usepackage{rotating}
\usepackage{lscape}
\usepackage{pdfpages}
\journalname{Space Science Reviews}
\begin{document}
\title{Reaction Networks For Interstellar Chemical Modelling: Improvements and Challenges
}


\author{V. Wakelam \and I.W.M. Smith \and E. Herbst \and J. Troe \and W. Geppert \and H. Linnartz \and K. \"Oberg \and E. Roueff \and M. Ag\'undez \and P. Pernot \and H. M.  Cuppen \and J. C. Loison \and D. Talbi 
}


\institute{V. Wakelam \at
              Universit\'e de Bordeaux, Observatoire Aquitain des Sciences de l'Univers, BP 89, F-33271 Floirac Cedex, France \\
              CNRS, UMR 5804, Laboratoire d'Astrophysique de Bordeaux, BP 89, F-33271 Floirac Cedex, France  \\
              \email{wakelam@obs.u-bordeaux1.fr}           
           \and
           I. W. M. Smith \at
              University Chemical Laboratories, Lensfield Road, Cambridge CB2 1EW, UK
   \and
   E. Herbst \at
   Departments of Physics, Astronomy, and Chemistry, The Ohio State University, Columbus, OH 43210 USA
   \and
J. Troe \at
Institut f\"ur Physikalische Chemie, Universitt Gttingen, Tammannstr. 6, D-37077 Gttingen, Germany and Max-Planck-Institut f\"ur \\
Biophysikalische Chemie, Am Fassberg 11, D-37077 Gttingen, Germany
 \and
   W. Geppert \at
   Department of Physics, University of Stockholm,Roslagstullbacken 21, S-10691 Stockholm 
     \and
     H. Linnartz \at
      Sackler Laboratory for Astrophysics, Leiden Observatory, University of Leiden, P.O. Box 9513, NL 2300 RA Leiden, The Netherlands
      \and
      K. \"Oberg \at
Harvard-Smithsonian Center for Astrophysics, 60 Garden Street, MS 42 Cambridge, MA 02138\\
      Sackler Laboratory for Astrophysics, Leiden Observatory, University of Leiden, P.O. Box 9513, NL 2300 RA Leiden, The Netherlands
 \and
 E. Roueff  \at
 LUTH, UMR 8102 CNRS, Universit Paris 7 and Observatoire de Paris, Place J. Janssen, 92195 Meudon, France
 \and
 M. Ag\'undez \at
 LUTH, Observatoire de Paris-Meudon, 5 Place Jules Janssen, 92190 Meudon,
France
 \and
        P. Pernot \at
           Univ Paris-Sud, Laboratoire de Chimie Physique, UMR 8000, Orsay, F-91405 \\
          CNRS, Orsay, F-91405
          \and
      H. M. Cuppen \at
             Sackler Laboratory for Astrophysics, Leiden Observatory, University of Leiden, P.O. Box 9513, NL 2300 RA Leiden, The Netherlands
  \and
  J.-C. Loison\at
   Institut des Sciences Mol\'eculaires, CNRS UMR 5255, Universit\'e Bordeaux I, 33405 Talence cedex, France 
    \and
    D. Talbi \at
   Universit\'e Montpellier II - GRAAL, CNRS - UMR 5024, place Eug\`ene Bataillon, 34095 Montpellier, France
   }

\date{Received: date / Accepted: date}

\maketitle

\begin{abstract}
We survey the current situation regarding chemical modelling of the synthesis of molecules in the interstellar medium. The present state of knowledge concerning the rate coefficients and their uncertainties for the major gas-phase processes -- ion-neutral reactions, neutral-neutral reactions, radiative association, and dissociative recombination -- is reviewed. Emphasis is placed on those 
key 
reactions that have been identified, by sensitivity analyses, as 'crucial' in determining the predicted abundances of the species observed in the interstellar medium. These sensitivity analyses have been carried out for gas-phase models of three representative, molecule-rich, astronomical sources: the cold dense molecular clouds TMC-1 and L134N, and the expanding circumstellar envelope IRC +10216. Our review has led to the proposal of new values and uncertainties for the rate coefficients of many of the key reactions. The impact of these new data on the predicted abundances in TMC-1 and L134N is reported. Interstellar dust particles also influence the observed abundances of molecules in the interstellar medium. Their role is included in gas-grain, as distinct from gas-phase only, models. We review the methods for incorporating both accretion onto, and reactions on, the surfaces of grains in such models, as well as describing some recent experimental efforts to simulate and examine relevant processes in the laboratory. These efforts include experiments on the surface-catalysed recombination of hydrogen atoms, on chemical processing on and in the ices that are known to exist on the surface of interstellar grains, and on desorption processes, which may enable species formed on grains to return to the gas-phase. 

\end{abstract}

\section{Introduction}
\label{intro}

One of the major challenges in the field of astrochemistry is to understand fully the synthetic mechanisms that are responsible for the gaseous molecules that have been observed in the interstellar medium (ISM). Table 1 lists the molecules that have been definitely identified in the ISM, principally by observing spectra of two kinds: (i) transitions between rotational levels observed in emission using ground-based telescopes that are sensitive at millimetre wavelengths, and (ii) transitions between vibrational levels observed in absorption at infrared wavelengths. Because of absorption in the Earth's atmosphere, many of the latter observations have been made with instruments mounted on satellites, such as the Infrared Space Observatory (ISO) and the Spitzer Space Telescope, which are also able to detect ices such as water ice on cold interstellar dust particles.  Recent discoveries from the Herschel satellite have also been included. A large fraction, but by no means all, of the molecules listed in Table~\ref{tab1new} are detected in the cold cores of dense interstellar clouds (ISCs) where, despite their name, the molecular density is quite low by terrestrial standards.  The dominant gas-phase molecule in these sources is molecular hydrogen, with the second-most abundant molecule, CO, down in abundance by a factor of 10$^{4}$. The high relative concentration of molecular hydrogen is due both to an efficient formation mechanism from hydrogen atoms on interstellar dust particles and the fact that hydrogen is the most abundant element in the universe.

As well as being used to identify molecules that are present in the ISM, these spectroscopic observations can be used to estimate the abundances of the observed molecules, usually expressed as abundances relative to that of molecular hydrogen.  The analysis of the rotational emission spectra is non-trivial due to complications such as heterogeneity of physical conditions in the source, the possibility of optical thickness in the spectral lines, and the fact that the analysis leads directly only to a column density along the line of sight in the upper (emitting) rotational level \citep{2009ARA&A..47..427H}.  Converting this column density to a fractional abundance is itself a non-trivial matter since it is difficult to study molecular hydrogen, because it is non-polar.  The resulting fractional abundances are then typically uncertain to at least a factor of $\sim 3$.

Just as in atmospheric chemistry, attempts to further improve our understanding of how interstellar molecules are created in the observed abundances (that are estimated) involve two other interconnected kinds of scientific endeavor: (i) the construction of chemical models, which utilise large networks of elementary chemical and some physical processes, and (ii) experimental and theoretical efforts to obtain the information about these processes which is required as input to the models. Of primary importance are the coefficients which describe the intrinsic rate of all the elementary processes in the model, as well as the products of these processes. To define a rate coefficient, consider the process in which C atoms and H$_2$ molecules radiatively combine to yield CH$_2$.  The loss of carbon atoms can be expressed by the differential equation:
\begin{equation}
\label{ch2}
{\rm -d[C] / d}t  =  k_{\rm RA}{\rm [C] [H_2]}
\end{equation}
Here, $k_{\rm RA}$ is the rate coefficient for radiative association, and [C] and [H$_2$] are the instantaneous concentrations of C atoms and H$_2$ molecules. Concentrations are usually given in cm$^{-3}$. Consequently, the units of $k_{\rm RA}$ and of the rate coefficients for all ``second-order'' processes are cm$^{3}$ s$^{-1}$. The values of rate coefficients generally depend on temperature and in astrochemical compilations it is usual to express the temperature-dependence of each rate coefficient by stating the values of three parameters $\alpha$, $\beta$ and $\gamma$, where
\begin{equation}\label{rate_equa}
k(T)  =  \alpha (T / 300)^\beta \exp(-\gamma / T)
\end{equation}
However, only rarely is there sufficient experimental or theoretical information available to express the temperature-dependence of $k(T)$ with much confidence.
For instance, models often rely on extrapolations of $k(T)$ out of its measurement range, which incurs very large uncertainties\citep{Hebrard:2009}.

The level of success of the models is tested by comparing the predicted molecular abundances with those derived from the spectroscopic observations.
Despite some successes, the discrepancies that are found are frequently very large: greater than an order of magnitude. 
Of course, it must be recognized that both the predicted and observed abundances are uncertain. 
One recent development has been the attempt to quantify the size of the uncertainties in the predicted abundances by Monte Carlo methods, as discussed later in this article. 
Clearly, this procedure, along with estimates of the uncertainty in abundances derived from the spectroscopic observations, allows one to decide whether or not a particular discrepancy between observed and predicted abundances is significant. 
A second development has been the use of models to identify 'key' processes in the network; that is, elementary reactions which strongly influence the predicted abundances of one or more observed species 
and their precision 
\citep{2005A&A...444..883W,2006A&A...451..551W,2008ApJ...672..629V,2009A&A...495..513W}. 
At least, in principle, this information should encourage experimentalists and theoreticians to concentrate their efforts on reactions that are astrochemically crucial.

The present review has its origins in a program supported by the International Space Science Institute in Bern, Switzerland. The authors of the current paper, along with some others whose contributions are acknowledged at the end of this article, held two meetings to discuss the current state-of-knowledge in respect of the synthesis of interstellar molecules. Some emphasis was placed on the importance of not only identifying the key astrochemical reactions, making the best possible estimates of their rate coefficients, but also of trying to quantify the uncertainties associated with these estimates so that some confidence could be placed in the significance of any disagreements between observed and predicted abundances. These efforts took place in parallel with a second related one: the development of a new database for astrochemistry, named KIDA (Kinetic Database for Astrochemistry) which is now opened to the community : http://kida.obs.u-bordeaux1.fr/. 
Although estimates of uncertainties have been given in previous databases (e.g., Umist Database for Astrochemistry, UDfA http://www.udfa.net), KIDA goes further in that, for each key reaction, the entry includes a data sheet that gives reasons for the recommended value of the rate coefficient and its uncertainty, as well as references to any experimental and theoretical work on the reaction.

\begin{table}
\caption{Reported Interstellar and Circumstellar Gas-Phase Molecules {\bf on July 2010}. \label{tab1new}}
\begin{tabular}{llllll}
\hline
\hline
N  = 2 & H$_{2}$ & CH &  CH$^{+}$ & NH & OH \\
& HF & C$_{2}$ & CN & CO & CS \\
& CP & NO & NS & SO & HCl\\ 
 & NaCl & KCl & AlCl  & PN & SiN \\
 & SiO & SiS & CO$^{+}$ & SO$^{+}$  & PO \\
 & SH & AlF & FeO & SiC & CF$^{+}$ \\
& LiH & N$_2$ & SiH & O$_2$ & AlO \\
 & OH$^+$ & CN$^-$ & & & \\
\hline
N = 3 & H$_{3}^{+}$ & CH$_{2}$ & NH$_{2}$ & H$_{2}$O & H$_{2}$S \\ 
& CCH&            HCN & HNC & HCO & HCO$^{+}$ \\ 
& HOC$^{+}$ & HN$_{2}^{+}$&            HNO & HCS$^{+}$ & C$_{3}$ \\ 
& C$_{2}$O & C$_{2}$S & OCS&            MgCN & MgNC \\ 
& NaCN & SO$_{2}$ & N$_{2}$O & SiCN&            CO$_{2}$\\ 
& c-SiC$_{2}$ & SiNC & AlNC & HCP & C$_2$P \\
& AlOH & H$_2$O$^+$ & H$_2$Cl$^+$ & & \\
\hline
N = 4 & CH$_{3}$ &NH$_{3}$ & H$_{3}$O$^{+}$ & H$_{2}$CO & H$_{2}$CS \\ & l-C$_{3}$H 
        & c-C$_{3}$H & HCCH & HCNH$^{+}$ & H$_{2}$CN\\ & HCCN  & HNCO    
     & HOCN     & HCNO & HNCS  \\ & HSCN & C$_{2}$CN        & C$_{3}$O
 & C$_{3}$S   &     SiC$_{3}$ \\  & C$_{3}$N$^{-}$ & PH$_3$ & HCO$_2^+$ &  \\
\hline
N = 5 & CH$_{4}$ & SiH$_{4}$ & CH$_{2}$NH & H$_{2}$C$_{3}$ & c-C$_{3}$H$_{2}$ \\ & CH$_{2}$CN 
 &   H$_{2}$NCN & CH$_{2}$CO & HCOOH & C$_{4}$H \\ & HC$_{2}$CN & HC$_{2}$NC
 &   HNC$_{3}$ & C$_{4}$Si\\ & C$_{5}$ & H$_{2}$COH$^{+}$ & C$_{4}$H$^{-}$ & CNCHO \\
\hline
N = 6 & CH$_{3}$OH & CH$_{3}$SH & C$_{2}$H$_{4}$ & CH$_{3}$CN & CH$_{3}$NC \\ 
& HC$_{2}$CHO &      NH$_{2}$CHO & HC$_{3}$NH$^{+}$ & H$_{2}$C$_{4}$ & C$_{5}$H \\ 
& C$_{5}$N & HC$_{4}$H  &    HC$_{3}$CN  &    c-C$_{3}$H$_{2}$O  & CH$_{2}$CNH \\
& C$_5$N$^-$ & & & & \\
\hline
N = 7 & CH$_{3}$NH$_{2}$ & CH$_{3}$CCH & CH$_{3}$CHO & c-CH$_{2}$OCH$_{2}$ & CH$_{2}$CHCN \\& HC$_{4}$CN 
 & C$_{6}$H & CH$_{2}$CHOH & C$_{6}$H$^{-}$& \\
 \hline
N = 8 & HCOOCH$_{3}$ & CH$_{3}$CCCN & HC$_{6}$H & C$_{7}$H   & HOCH$_{2}$CHO \\&  
CH$_{3}$COOH 
 & H$_{2}$CCCHCN & H$_{2}$C$_{6}$ & CH$_{2}$CHCHO  & CH$_{2}$CCHCN \\
\hline
N = 9 & (CH$_{3}$)$_{2}$O & C$_{2}$H$_{5}$OH & C$_{2}$H$_{5}$CN & CH$_{3}$C$_{4}$H &
C$_{8}$H\\ & HC$_{6}$CN 
& CH$_{3}$CONH$_{2}$ & C$_{8}$H$^{-}$ & CH$_{3}$CHCH$_{2}$ &\\
\hline
N $\ge$ 10 & (CH$_{3}$)$_{2}$CO & CH$_{3}$C$_{4}$CN & CH$_{3}$CH$_{2}$CHO & (CH$_{2}$OH)$_{2}$ & HCOOC$_{2}$H$_{5}$ \\& HC$_{8}$CN  & CH$_{3}$C$_{6}$H &  C$_{6}$H$_{6}$& C$_{3}$H$_{7}$CN & HC$_{10}$CN\\   
& C$_2$H$_5$OCH$_3$ & & & & \\
\hline
\end{tabular}
\end{table}

\section{History of chemical modeling}
\label{sec:1}

Chemical modeling of sources in the ISM began in the 1950's following the detection of CH, CN and CH$^+$ via narrow absorption lines towards bright stars.  In a landmark paper,  \cite{1951ApJ...113..441B} discussed quantitatively the density of molecules in interstellar space. Radiative association reactions, dissociative recombination (see Section~\ref{rateco}), and photodissociation and photoionization processes were emphasized, resulting in a reasonable (if ultimately incorrect) explanation for the presence of CH$^+$, but with questions concerning the formation of CH. About 10 years later, \cite{1960MNRAS.121..238M} investigated the possibility of molecular formation by surface reactions on interstellar dust grains, focusing on H$_2$, which had not yet been detected at that time, and CH.

The detection of the $\Lambda$ doubling transition of OH at 18 cm \citep{1963Natur.200..829W,1964Natur.202..989R} in the Galactic Center, followed by the detection at 6 cm of NH$_3$, the first polyatomic molecule \citep{1968PhRvL..21.1701C}, and observational evidence for the presence of H$_2$ towards $\epsilon$ Per and $\xi$ Per from rocket observations \citep{1970ApJ...161L..81C} stimulated interest in the molecular content of the interstellar medium and in the chemical and physical processes involved in the formation of the observed molecules. The advent of radio-millimeter telescopes and the subsequent discovery of many polyatomic species, including molecular ions, in the 1970's helped to put interstellar chemistry on a firmer basis. Not only astronomers but also chemists and physicists realized how promising and rewarding it could be to search for new, often exotic,  molecules (see Table~\ref{tab1new}) and to build chemical schemes with different formation and destruction processes. 

Initially, two main classes of regions were considered for quantitative chemical modeling: diffuse regions, with a total H density of $\approx 10^{2}$ cm$^{-3}$ and a temperature of 50-100 K, and dense regions, with an H$_2$ density of $\approx 10^{4}$ cm$^{-3}$ and a temperature of 10-20 K.  The modeling was of a steady-state nature, in which the time derivative of all concentrations is set to zero. The detailed chemistry of diffuse regions, including  the prototypical $\zeta$ Oph cloud, was studied by \cite{1977ApJS...34..405B}, who included a detailed discussion of molecular hydrogen formation and excitation, and quantitative treatment of photoprocesses driving the chemistry.  The chemistry of dense clouds, now known as cold cores, was investigated by  \cite{1973ApJ...185..505H}, who showed that the chemistry is driven by cosmic rays, which initiate networks of ion-molecule reactions.   A major success was the contribution of chemical modeling to the identification of the HCO$^+$ molecular ion \citep{1974ApJ...188..255H}.  More recent observations have shown that both diffuse and dense sources are often parts of larger heterogeneous objects.  Moreover, some cold dense cores are in the process of collapsing to form stars and planetary systems.  
The collapse is a complicated one hydrodynamically, and can involve shock waves.

The state of the field in the late 1970's was admirably described by two review papers \citet{1976RPPh...39..573D} and \citet{1976RvMP...48..513W}; these describe the various physical and chemical processes relevant to reactions at the surface of interstellar grains and in the gas-phase and are still relevant today.  A few years later, \cite{1980ApJS...43....1P} reported the first large-scale model for gas-phase chemistry, which was based on experimental studies of ion-molecule reactions performed with a variety of experimental techniques including their  ion cyclotron resonance (ICR) facility.   In addition, they provided an extensive list of rate coefficients for chemical reactions involving compounds of carbon, nitrogen and oxygen. Non-equilibrium effects were specifically addressed in a subsequent paper \citep{1980ApJ...239..151P}. 

As far as interstellar modeling is concerned, one should also point to the decisive influence of computational power, which now makes it possible to solve a large number of coupled chemical equations as a function of time from given initial conditions. In addition, numerical techniques, such as the Gear method \citep{1971nivp.book.....G,1987JCoPh..70....1B}, enable stiff problems, with time-dependent processes with orders of magnitude different time scales, to be handled.   These techniques have been important for pseudo-time dependent chemical studies, which  started with \cite{1984ApJS...56..231L}.  In this approach, the chemical evolution is solved independently of the physical conditions, which are taken as fixed and homogeneous, typically with an H$_2$ density of 10$^4$ cm$^{-3}$ and a temperature of 10 K, conditions that are relevant to the "standard" TMC1 cloud, often taken as a reference for the purpose of comparison between models and observations. Such numerical techniques are also used in the chemical interstellar shock models developed by \cite{1985MNRAS.216..775F}, who followed both the physical and chemical evolution in a magneto-hydrodynamic (MHD) shock.  Complex albeit steady-state models have been constructed for so-called Photon-dominated regions (PDR's), which are heterogeneous sources located near stars  \citep{1997ARA&A..35..179H,2007A&A...467..187R}. 

The number of species involved in such chemical models has increased to  400-500 linked by over 4000 reactions, following the spectacular increase of detected interstellar and circumstellar molecules via their rotational (sub)millimeter-wave spectra.  The need for chemical databases became urgent in the 1980's, and the group at Manchester \citep{1991A&AS...87..585M} was the first to publish such a database, which is now available on a website (UDfA at http://www.udfa.net/), with regular updates published in the literature. Other databases are now also available including the OSU (Ohio State University) database (e.g. http://www.physics.ohio-state.edu/$\sim$eric/research.html) and, as discussed later in this article, the already mentioned KIDA database.

Consideration of both steady state and time-dependent studies has demonstrated the non-linear character of the chemical equations describing interstellar chemistry \citep{1993ApJ...416L..87L}. Comparing steady state solutions with the integration of  time-dependent equations has enabled the discovery of multiple solutions and bi-stable behavior. Two stable solutions have been shown to coexist within a typical range of control parameters, which can be "astrophysical" quantities (density, temperature, cosmic ionization rate) or even the values of the reaction rate coefficients with a range of uncertainties. The solutions correspond to different chemistries adequately described by the level of ionization. The High Ionization Phase  favors atomic ions, unsaturated species and radicals, whereas in the more standard Low Ionization Phase, small saturated molecules and molecular ions play a preponderant role. Such a possibility cannot be excluded in the real ISM and it offers a natural way to explain current problems such as the low abundance of molecular oxygen as well as  the amount of atomic carbon readily detected in dense clouds.

A new stage was reached during the 1990's with the wealth of information obtained using the Infrared Space Observatory (ISO), launched by the European Space Agency in 1995 and operated for 3 years. Among the outstanding results obtained by the various instruments on board this satellite are the determination of the composition of the icy mantles of dust grains,  the measurement of rotational transitions of H$_2$  and other transitions such as the C$^+$ and O fine-structure lines,  and the ubiquity of certain vibrational features thought to be due to  polycyclic aromatic hydrocarbons (PAH's). These and other results have played a determining role in defining new needs for further developments in astrochemical models.  

The need for including surface processes, eventually triggered by photon irradiation, particle bombardment or thermal processing, has been raised, both to understand the composition of icy mantles \citep{2000ApJ...536..347G} and to find an alternative to gas-phase processes, which can fail to reconcile model predictions with observations for some specific molecules. An additional impetus for including gas-grain chemistry was the need to understand the chemical complexity resulting from the detection of increasingly complex organic molecules \citep{1992ApJS...82..167H}. However, gas-grain models rely on many hypotheses concerning the energy barriers associated with both surface diffusion and surface reactions. These barriers are poorly constrained, and the energies that are assumed for them have a considerable impact on the results. An additional difficulty in gas-grain models arises from the stochastic nature of gas-grain chemistry \citep{1982A&A...114..245T}: under certain conditions, the rate of reaction is dominated by the accretion rate of species onto individual grains rather than by the rate of diffusion  
so that one should explicitly introduce the probability of the number of adsorbed species on the surface of one grain. Different ways have been considered to take these effects into account, including Monte Carlo techniques \citep{1997ApJ...482L.203C,2001ApJ...562L..99C} and master equations \citep{2001A&A...375.1111G,2001ApJ...553..595B}. \cite{2004PhRvL..93q0601L} and \citet{2007ApJ...658L..37B} have shown subsequently that moment equations with an appropriate cut-off up to second-order could adequately describe gas-grain reactions.  A simpler, but semi-empirical method has been introduced, in which the rate equations are modified to reproduce some of the effects of stochastic approaches \citep{1998ApJ...499..234C, 2008A&A...491..239G}.

\section{Gas-phase reaction rate coefficients and uncertainties} \label{rateco}



\begin{table}
\caption{Number of reactions of different types in the gas-phase that are included in the OSU kinetic
database for astrochemistry (osu-09-2008). \label{osu_react}}
\begin{center}
\begin{tabular}{llr}
\hline
Type of process & Example & Number in OSU\\
\hline
Gas-grain interactions &  		H  +  H  + grain  $\rightarrow$ H$_2$  + grain	 & 14\\
Direct cosmic ray processes &  H$_2$  +  $\zeta$  $\rightarrow$  H$_2^+$  +  e$^-$	& 11\\
Cation-neutral reactions	& H$_2^+$  +  H$_2$  $\rightarrow$  H$_3^+$  +  H	& 2933\\
Anion-neutral reactions	 & C$^-$  +  NO  $\rightarrow$  CN$^-$  +  O	& 11\\
Radiative associations (ion)& 	C$^+$  +  H$_2$  $\rightarrow$  CH$_2^+$  +  h$\nu$	& 81\\
Associative detachment &	C$^-$  +  H$_2$  $\rightarrow$  CH$_2$  +  e$^-$	& 46\\
Dissociative neutral attachment&	O  +  CH  $\rightarrow$  HCO$^+$  +  e$^-$	& 1\\
Neutral-neutral reactions&	C  +  C$_2$H$_2$  $\rightarrow$  C$_3$H  +  H  	& 382\\
Radiative association (neutral)&	C  +  H$_2$  $\rightarrow$  CH$_2$  +  h$\nu$	& 16\\
Dissociative recombination&	N$_2$H$^+$  +  e$^-$  $\rightarrow$  N$_2$  +  H	& 539\\
Radiative recombination&	 H$_2$CO$^+$  +  e$^-$  $\rightarrow$  H$_2$CO  +  h$\nu$ & 	16\\
Anion-cation recombination	& HCO$^+$  +  H$^-$  $\rightarrow$  H$_2$  +  CO  &	36\\
Electron attachment	& H  +  e$^-$  $\rightarrow$  H$^-$  +  h$\nu$ &	4\\
External Photo-processes &	C$_3$N  +  h$\nu$  $\rightarrow$  C$_2$  +  CN	& 175\\
Internal Photo-processes &       CO  +  h$\nu$  $\rightarrow$  C  +  O	& 192 \\
\hline
\end{tabular}
\end{center}
\end{table}



Chemical models, designed to improve our understanding of how molecules are synthesised in different regions of the interstellar medium, consist of a large number of `rate equations'; that is, ordinary differential equations (ODEs), each of which, like eq.~(\ref{ch2}),  represents how  given chemical reactions or physico-chemical processes build-up or lessen the concentration of a particular chemical species.  Currently, the OSU  database osu-09-2008  includes 455 different species connected by 4457 rate equations. (The figures for the database for astrochemistry held at UDfA  are very similar.) The 455 species include 187 neutral atoms and molecules, 259 atomic and molecular cations, and 6 atomic and molecular anions. In addition, the concentrations of electrons and of neutral and negatively charged grains are included. In time-dependent models,  assumptions must be made concerning the starting conditions: the temperature, the elemental abundances and the state (atomic, molecular, ionic) of the elements, and then the ODEs are integrated forward in time to generate molecular abundances at different ages of the cloud. There are at least two requirements for any such model to be successful: (i) it should include all those processes that are important in determining the molecular abundances, and (ii) the values of the rate coefficients included in the model should be close to the correct values. It is the second of these two requirements that is considered, in general terms, in this section.

Table~\ref{osu_react} lists the different types of processes or reactions that are included in the OSU database and gives the number of each type of process in the model. The first type of reaction refers to the neutralization process between gas phase cations and negatively charged grains plus the H$_2$ formation on grains. This latter process is artificially included in gas phase models assuming that the H$_2$ formation occurs at half the rate of hydrogen atom sticking on a grain \citep{1984ApJS...56..231L}, i.e. the efficiency of the conversion is one as assumed by \citet{1974ApJ...191..375J}. Although the numbers of individual types of processes vary widely, it should not be inferred that these numbers reflect the importance of different types of processes. In this section, an attempt is made to summarize how the rate coefficients for some of the most important classes of reaction are estimated from the available experimental and theoretical evidence and, equally importantly, how the uncertainties in the rate coefficients are estimated. We shall focus on the data for the following classes of reaction: (i) cation (and anion)-neutral reactions, (ii) neutral-neutral reactions, (iii) radiative association reactions (producing both ionic and neutral products), and (iv) dissociative recombination reactions. These categories include most of the reactions in astrochemical models. 
{\bf At the outset, it should be emphasized that the fraction of these processes for which rate coefficients have been measured or calculated at the low temperatures prevalent in cold cores is extremely small.}
In addition, for those reactions that may proceed to different sets of products, the branching ratios to these different channels are frequently unmeasured. Consequently, the values of rate coefficients included in the models are arrived at by a combination of experiment, theory and common sense, although in many important cases neither experimental nor theoretical work has been performed, so that chemical intuition, perhaps guided by information about chemically related reactions, may be the only way to estimate a rate coefficient. 

If the rate coefficient for a reaction is not to be negligibly small at the temperatures prevalent in cold cores, the value of $\gamma$ in eq.~(\ref{rate_equa}) must be negative, zero, or at least very close to zero. In theoretical terms, this means that there should be no significant barrier on the path of minimum potential energy leading from reactants to products. In general terms, the value of the rate coefficient is likely to reflect the ability of the long-range attraction between the reactants to bring them into close contact.
A second fundamental requirement is that there is an electronically allowed (or at least not strongly forbidden) path that leads from the reactants to products. Usually, this can be checked by applying the spin and orbital correlation rules to discover the symmetry of the potential energy surfaces that correlate with the separated reactants and the separated products \citep{Smith:1980}. The importance of these rules is greatest when both reactants have one or more unpaired electrons. Then potential energy surfaces of different spin multiplicity arise. The surfaces with the lowest spin multiplicity are likely to provide the lowest energy pathway from reactants to products. Collisions on other surfaces are unlikely to lead to reaction and this effect can introduce a factor $< 1$ into an estimate of the rate coefficient.

\subsection{Ion-Neutral Reactions}


The experimental database of rate coefficients for reactions between cations and neutral molecules is large. Following the early studies using ion cyclotron resonance \citep{1980ApJS...43....1P,2008ARAC....1..229S} and flowing afterglow methods, most of the recent kinetic data on ion-neutral reactions have been obtained using selected ion flow tubes (SIFTs)  or some variant of this method.  However, many of these measurements are only at room temperature or over a restricted range of temperatures around room temperature. The CRESU and ion-trap techniques have been applied to a limited number of reactions, yielding rate coefficients down to temperatures similar to those in cold cores.  An important advantage of both these methods, as applied to ion-neutral reactions, is that mass spectrometry is used to observe changes in the concentrations of the reactant ions and the same technique can be used to identify the products of the reaction under investigation.

However, for exothermic ion-molecule system, reactions between ions and polar neutral species have not been studied extensively in the laboratory.  A few studies down to temperatures of $\sim$ 20-30~K \citep{1985ApJ...296L..31A,Marquette:1985,1988CPL...143..130R,Rowe:1988}  and some studies at temperatures above 300 K  \citep{VW01} have been undertaken.  Nevertheless, the rate coefficients of most thermal reactions used in model calculations must be estimated.  These estimates can be used in the complete absence of experimental data or, for those cases where room temperature measurements are available, they can be used at higher and lower temperatures if there is reasonable agreement at 300 K. Detailed calculations of reaction rate coefficients require a knowledge of both the long-range and short-range aspects of the intermolecular potentials governing the reactions \citep{1988rcia.conf....1C}.  For exothermic ion-molecule systems especially, it is often possible to neglect short-range considerations and to focus on the long-range asymptotic potential.

The orbital angular momentum associated with the relative motion of the reactants under the influence of the long-range potential creates so-called centrifugal barriers. Rather than solve the detailed quantum mechanics in the absence of short-range barriers, one can use a capture model, in which the sole criterion for reaction is that the reactants must pass over the centrifugal barrier to enter the short-range portion of the potential \citep{Herbst:1996}. Some additional dynamical barriers are created by anisotropies of the potential \citep{1987JChPh..87.2773T}.
 The capture model thus provides an upper limit to the actual rate coefficient.  For non-polar neutrals, the long-range potential depends solely on the distance $R$ between ion and neutral, which in many instances is simply the R$^{-4}$ potential arising from the interaction of charge and dipole polarizability.  With this potential, the rate coefficient is given by the so-called Langevin expression:
\begin{equation}
k_{\rm L} = 2\pi e (\alpha_{\rm pola}/\mu)^{1/2},
\end{equation}
where $e$ is the electronic charge, $\alpha_{\rm pola}$ is the  polarizability,  $\mu$ is the reduced mass of reactants, and cgs-esu units are utilized so that the units for the rate coefficient are cm$^{3}$ s$^{-1}$.

For polar neutrals, the potential is anisotropic and depends on the orientation of the permanent dipole as well as the distance between reactants so that it is no longer central in nature.  A number of detailed treatments of the long-range problem with the capture approximation have been undertaken, starting with the full quantum mechanical treatment.  At a somewhat simpler level are the rotationally adiabatic approach of \citet{clary1987}, the perturbed rotational states approximation of \cite{1980JPSJ...48.2076S}, and the statistical adiabatic channel approach of Troe and co-workers \citep{1987JChPh..87.2773T,2009IJMSp.280...42M}.  These detailed approaches can be used to calculate rate coefficients as functions of the rotational angular momentum of the neutral reactant or, by suitable averaging, thermal rate coefficients.

A simpler approach for the astrochemist is based on classical trajectory calculations in which the long-range potential as a function of distance and orientation is used \citep{1982JChPh..76.5183S,2009IJMSp.280...42M}.  Thermal rate coefficients determined from these calculations can be fit to analytical expressions depending solely on the reduced mass of the reactants, the polarizability, and the permanent dipole moment of the neutral reactant.  This treatment is classical in nature and is only reliable at temperatures above which rotational motion can be considered classical, typically 10 - 20 K.  At lower temperatures, a semi-classical regime is entered, and at still lower temperatures ($\ll$ 1~K), a fully quantum mechanical regime occurs, as studied in detail by Troe and co-workers \citep[see, e.g.,][]{2009IJMSp.280...42M}.  The transition between classical and semi-classical behavior may be of some importance for cold cloud cores, where the temperatures lie in between the two regimes.  The transition can be easily characterized (Troe 1989, 1996, Maergoiz et al. 2009) but, the classical expression is not seriously wrong until under 10 K.

The best-known formula for the ion-dipolar rate coefficient $k_{\rm D}$ in the classical regime and for linear neutrals is the Su-Chesnavich expression \citep{1982JChPh..76.5183S}.  \citet[][and references therein]{2009IJMSp.280...42M} show that it works reasonably well for neutral species that are linear, symmetric tops, or asymmetric tops, although they have also explored the subtle differences among these structures.
The Su-Chesnavich formula \citep[see, e.g.,][]{2009IJMSp.280...42M} is based on the parameter $x$, defined by
\begin{equation}
x = \frac{\mu_{\rm D}}{(2\alpha_{\rm pola} k_BT)^{1/2}},
\end{equation}
where $\mu_{\rm D}$ is an effective dipole moment of the neutral reactant, which is generally very close to the true dipole moment, and $k_B$ is the Boltzmann constant.  The larger the value of $x$, the larger the rate coefficient. The expressions for $k_{\rm D}$ can be written in terms of the Langevin rate coefficient via
\begin{equation}
k_{\rm D}/k_{\rm L} = 0.4767x + 0.6200,
\end{equation}
if $x \ge 2$, and
\begin{equation}
k_{\rm D}/k_{\rm L} = (x + 0.5090)^{2}/10.526  +  0.9754,
\end{equation}
 if $x < 2$.  The latter reduces to the Langevin expression for  $x=0$.
Alternatively, the expressions can be written in powers of temperature $T$.  For example, for $x \geq 2$
\begin{equation}
\label{equD}
k_{\rm D} = c_{1} + c_{2}{\rm T}^{-1/2}
\end{equation}
where
\begin{equation}
c_{1} = 0.62k_{\rm L}
\end{equation}
and
\begin{equation}
c_{2} = 2.1179\frac{\mu_{\rm D}e} {\sqrt{\mu k_B}}.
\end{equation}
For $x<2$, a more complex expression of the type
\begin{equation}
k_{\rm D} = b_{1} + b_{2}T^{-1/2} + b_{3}T^{-1}
\end{equation}
pertains, where $b_{1}$ is simply the Langevin rate coefficient, while the parameters $b_{2}$ and b$_{3}$ are given by the expressions:
\begin{equation}
b_{2} = 0.4296\frac{\mu_{\rm D}e} {\sqrt{\mu k_B}}
\end{equation}
and
\begin{equation}
b_{3} =   \frac {\mu_{\rm D}^{2} \pi e}{10.526k_B (\alpha_{\rm pola} \mu)^{1/2}}.
\end{equation}
For  $x\ge2$, eq.~(\ref{equD}) shows that if the second term is much greater than the first term,  which is more likely at low temperatures, the expression reduces to $c_{2}T^{-1/2}$, which is used in the OSU database
for non-linear neutral reactants.  For linear polar neutrals, the OSU database assumes that the species are rotationally relaxed ($J \approx 0$) and so uses the so-called ``locked dipole'' expression \citep{1986ApJ...310..378H} which is significantly larger than $k_{\rm D}$. The UDfA database, on the other hand, uses the same temperature dependence but scales the result to a measured or estimated result at 300 K:
\begin{equation}
k_{\rm D, UDfA} = k_{\rm D}(300~{\rm K}) (300/T({\rm K}))^{1/2}.
\end{equation}
One should note that $k_{\rm D}$ does not diverge for T$\rightarrow$ 0~K as would be suggested by the classical expressions (5), (7) and (13). Instead, $k_{\rm D}$ levels off at a value $k_{\rm D}$/$k_{\rm L}$=(1+$\mu_{\rm D}/3\alpha_{\rm pola}$B)$^{1/2}$ where B is the rotational constant (in energy units) of the neutral \citep{1987JChPh..87.2773T,1996JChPh.105.6249T}.

It is recommended that  networks incorporate the Su-Chesnavich formulae discussed here in the situation where no experimental information is available, especially if information is required over a broad temperature range, say, from 10 - 1000 K. It must be recognized, however, that differences from the simpler $T^{-1/2}$ dependence may be small except for the prior use of the locked dipole approximation.  This approximation should be removed from the OSU network for consistency, even if highly polar neutrals have their rotational levels relaxed thermally compared with the kinetic temperature.  If a room temperature measurement is available, the capture theory result can be suitably scaled. Use of the capture theory has been heretofore hindered by the lack of polarizability and dipole moment data, but a new compendium of information for interstellar neutral molecules calculated by \cite{2009ApJS..185..273W} makes the needed information readily available for all species in the networks, no matter how exotic.  Instructions for determining the rate coefficients of ion-polar reactions  in the range 10-1000 K can be found on the OSU URL.


Most of the ion-neutral reactions that have been studied in the laboratory are between singly charged cations and neutral molecules that can be synthesised or purchased quite readily.  It is clear from a study of Table~\ref{tab1new}  that these molecules comprise only a fraction of the neutral species that are potential reactants in dense clouds.  The `key' ion-neutral reactions discussed in 
Section~\ref{review} all involve radical atoms as the neutral reactant.  To study these reactions involves an extra level of experimental difficulty since the atoms need to be generated by flow-discharge techniques and their steady-state concentration estimated. The room temperature rate coefficients for a large number of such reactions have been recently tabulated \citep{2008ARAC....1..229S}. 

A factor that might influence the rate coefficients of reactions between ions and open shell species (radicals) is the presence of spin-orbit (or fine structure) splitting in the case of radicals, which have one or more unpaired electrons. For example, the ground states of the C, N and O atoms are respectively $^3$P, $^4$S and $^3$P. If the ion also has an open electronic shell, then potential energy surfaces (PES's) of different spin multiplicity will correlate with the reactants. This means that, in general, only a fraction of the collisions between the reactants will occur on a surface leading adiabatically to products. In these cases, it is recommended that a factor be introduced to allow for spin degeneracy. To take a specific example, the interaction between N($^4$S) atoms and C$_2$H$_4^+$ ions, which have a spin multiplicity of 2, will lead to PES's with both $S = 1$ and $S = 2$. It is reasonable to suppose that reactions will only occur on the triplet surface so any estimate of the rate coefficient should include a factor (3/8) to allow for this. Many radicals, atomic as well as molecular, possess electronic orbital angular momentum (i.e., $L >$ 1 or $\Lambda >$ 1) and consequently the electronic state will be split by spin-orbit interaction into several components. However, as yet the influence of spin-orbit coupling on rate coefficients is not clear, we recommend that such effects be ignored.

\subsection{Neutral-Neutral Reactions}


It should be emphasized that, of the roughly 150 chemical species  that have been identified in the interstellar medium (see Table~\ref{tab1new}), the great majority are neutral and most of these are either unsaturated (hydrogen-poor) molecules or free radicals. The potential importance of neutral-neutral reactions in the chemistry of the interstellar medium, and in particular in cold cores, was underestimated for a number of years on the basis that many such reactions are known to have significant activation energies (that is, large positive values for $\gamma$ in eq.~{\ref{rate_equa}) with the result that their rate coefficients would be extremely small at the low temperatures encountered in the cold cores of dense clouds.
Indeed, reactions between pairs of neutral molecules, e.g. H$_2$ + D$_2$ $\rightarrow$ 2 HD and C$_2$H$_4$ + C$_2$H$_4$ $\rightarrow$ C$_4$H$_8$, are characterised by very high activation energies (corresponding to several thousands Kelvin) and can be ignored in models of most regions of the ISM. Many reactions between radicals and saturated molecules, e.g. D + H$_2$ $\rightarrow$ HD + H and CN + H$_2$ $\rightarrow$ HCN + H, have activation energies corresponding to 500 - 5000 K, and they can also be ignored in modeling the cold cores although there are a few exceptions; e.g., CN + NH$_3$ $\rightarrow$ HCN + NH$_2$.

All of the 382 neutral-neutral reactions included in the OSU database (osu-09-2008) involve, at least one unstable species; that is, one free radical. In only 99 of these reactions, is the other reactant a stable species: for example a saturated molecule, such as H$_2$, CH$_4$, or H$_2$S, or an unsaturated molecule, such as C$_2$H$_2$ or C$_2$H$_4$. The reactants in the other 283 reactions are both unstable. The experimental database for the latter class of bimolecular reactions is much sparser than for the former, particularly at low temperatures, reflecting the experimental difficulties in measuring the rate coefficients for radical-radical reactions. For many reactions of both these types, a starting point for estimating the rate coefficients at low temperature can be the extensive reviews and evaluations of kinetic data for gas-phase reactions that have been compiled for the purposes of modelling the terrestrial atmosphere (\citet{2004ACP.....4.1461A,2006ACP.....6.3625A,2007ACP.....7..981A,2008ACP.....8.4141A}; http://www.iupac-kinetic.ch.cam.ac.uk and \citet{Sander06}; http://jpldataeval.jpl.nasa.gov/download.html).



Essentially all the kinetic studies of neutral-neutral reactions at temperatures below 100 K employ either the continuous flow CRESU technique or the pulsed Laval nozzle variant of the technique. In a recent review \citep{Canosa:2008}, it is stated that 73 reactions have been studied using the continuous CRESU method to reach temperatures as low as 13 K, but more generally ca. 25 K, whereas 20 have been studied employing pulsed Laval nozzles, which only generate temperatures as low as 53 K. Although there is certainly not a perfect overlap between the neutral-neutral reactions which have been studied to low temperatures using the CRESU method and those reactions whose rate coefficients are required for astrochemical modeling, the measurements have done much to identify the factors that control reaction rates at low temperatures, making it possible to predict rate coefficients, or to extrapolate measured rate coefficients to much lower temperatures, with a fair degree of confidence.

Based on a survey of the results  for reactions between atoms or radicals with alkenes and alkynes, two criteria have been proposed \citep{2006FaDi..133..137S}  in order to decide whether reactions of this type will be fast; that is, have rate coefficients of ca. $10^{-10}$ cm$^3$ s$^{-1}$ at 10 K. The first is if the observed rate coefficient at 298 K is $> 10^{-11}$ cm$^3$ s$^{-1}$, the second is if the difference between the ionisation energy of the molecule and the electron affinity of the radical is less than 8.75 eV. These criteria have been supported by recent measurements on the reaction of O($^3$P) atoms with a series of alkenes \citep{Sabbah:2007}.

Despite the CRESU and pulsed Laval studies, the experimental database of rate coefficients for reactions between neutral free radicals is sparse, especially at low temperatures. In general, the interaction between two radicals will give rise to more than one PES, as discussed above. Application of the correlation rules will show whether there are PESs that lead adiabatically to particular sets of products in specified electronic states. In general, the two radical reactants may form a transient bond (an electron from each radical `pairing up') and this interaction will lead to the lowest PES  or possibly PESs. The lowest PES will probably exhibit a deep `well' that can be accessed by a reaction path that exhibits no barrier. However, states of other multiplicities may be created and may not lead to products. One effect is that, if two radicals can undergo an exothermic reaction to yield two products, then the reaction is likely to occur rapidly, even at low temperatures. However, in estimating the rate coefficient, one should allow for the fact that some significant fraction of collisions will occur on PESs that either don't correlate adiabatically with the products or which do so over PESs with significant energy barriers. This feature of radical-radical reactions necessitates the introduction of a factor $<$ 1 which depends on the electronic degeneracies of the reactants and of the surface or surfaces over which reaction is judged to occur.


In order to indicate how the values of rate coefficients for neutral-neutral reactions, and the uncertainties in these rate coefficients, are chosen, we take as examples four ``key'' reactions listed later in the paper: those between (i) CN and NH$_{3}$, (ii) CN and HC$_3$N, (iii) O and C$_2$, and (iv) N + C$_3$.

The reaction between CN radicals and NH$_3$ molecules was one of the first two neutral-neutral reactions examined at low temperatures by the CRESU technique \citep{1994JChPh.100.4229S}. The rate coefficient exhibits an unusually strong negative temperature-dependence; that is, it increases by a factor of more than 10 as the temperature is lowered from 295 to 25 K, the lowest temperature achieved in the experiments. The experimental value of the rate coefficient was assigned a statistical error (at the level of 95\% 
confidence limits) of ca. 5\%. 
As with most direct kinetic experiments on elementary reactions, no information was obtained about the products of this reaction; the progress of reaction was followed by observing the decrease in concentration of the CN radicals. A recent theoretical analysis \citep{2009PCCP...11.8477T}, using a two-transition-state model and quantum chemical calculations of the minimum energy path for the reaction, reproduced the absolute value of the rate coefficient and its temperature-dependence, and concluded that the reaction proceeds exclusively to HCN and NH$_2$, with zero yield of NCNH$_2$ + H. This conclusion was subsequently supported by experimental measurements \citep{blitz09} that demonstrate that the branching ratio to the latter set of products is certainly less than 5\%.
 Almost uniquely, the rate coefficient for the reaction CN + NH$_3$ $\rightarrow$  NH$_2$ + HCN at 10 K, $k = 5.0 \times  10^{-10}$ cm$^{3}$ s$^{-1}$, can be assigned an uncertainty as low as $\Delta {\rm log}_{10} (k) =  0.15$; this uncertainty arises mainly from the extrapolation of the experimental data from 25 to 10 K. The reaction to form NCNH$_2$ + H is taken not to occur.

The reaction between CN radicals and HC$_3$N molecules represents the class of reactions where there are no low temperature measurements but there is an experimental determination of the rate coefficient at 298 K, in this case yielding $k$(298 K) = $1.7 \times  10^{-11}$ cm$^3$ s$^{-1}$ \citep{1989CPL...155..347H}. On the basis of both this rate coefficient and the relative values of the ionization energy of HC$_3$N and the electron affinity of CN, the arguments of \citet{2006FaDi..133..137S} suggest that this reaction has no barrier and should be fast at low temperatures. The absence of any barrier to reaction is supported by ab initio calculations \citep{petrie04}. The reaction is assigned a rate coefficient of $3.5 \times  10^{-10}$ cm$^3$ s$^{-1}$ at 10 K with an uncertainty corresponding to $\Delta {\rm log}_{10} (k) =  0.6$ at this temperature.

The reactions of O atoms with C$_2$ and N atoms with C$_3$ are examples of reactions for which there are no experimental data in the literature. In the first of these reactions, CO and C atoms can be produced by a spin-allowed, highly exothermic reaction: O($^3$P) +  C$_2$($^{1}{\rm \Sigma}_{\rm g}^{+}$)  $\rightarrow$ CO($^{1}{\rm \Sigma}^{+}$) + C($^3$P); $\Delta H^{o}_{298} =  -480.8$ kJ mol$^{-1}$. By the criteria of \citet{2006FaDi..133..137S}, this reaction would be expected to have no barrier. It is assigned values of $k = 3.0 \times  10^{-10}$ cm$^3$  s$^{-1}$ at 10 K and $\Delta {\rm log}_{10} (k) =  0.5$ at this temperature. The reaction between N atoms and C$_3$ molecules in their ($^{1}{\rm \Sigma}_{\rm g}^{+}$) ground state is problematical. Reaction to produce {\rm CN($^{2}{\rm \Sigma}^{+}$) + C$_2$($^{1}{\rm \Sigma}_{\rm g}^{+}$)} is exothermic but spin-forbidden. However, the reaction producing C$_2$ in its low-lying ($^{3}{\rm \Pi}$) excited state is also exothermic but spin-allowed. On the other hand, N atoms are generally unreactive to unsaturated species, such as C$_2$H$_2$ and C$_2$H$_4$, in singlet electronic states, which is consistent with the rules proposed by \citet{2006FaDi..133..137S}, because N atoms actually have a negative electron affinity. Taking these considerations into account, this reaction is assigned a small rate coefficient at 10 K, $k = 1.0 \times  10^{-13}$ cm$^3$ s$^{-1}$, with large uncertainty, $\Delta {\rm log}_{10} (k) =  1.0$.

\subsection{Radiative association reactions}
\label{rareact}

In association reactions, two atomic or molecular species, say A and B, combine to create one new molecular product, AB. In the ISM, such processes are important, since they allow the growth of species from smaller fragments. However, even in dense clouds, the gas density is too low to allow the collisional stabilization of any initially formed adduct, which will have sufficient internal energy to re-dissociate, by collision with some other species. Consequently, the only mechanism by which the adduct can lose energy and generate stable AB is for it to radiate. The importance of radiative association has long been recognized: partly in respect of the reactions of C$^{+}$ and CH$_{3}^+$ with H$_{2}$ \citep{1977ApJ...213..696H, Gerlich92}. In both these cases, the reactions forming two products (i.e., CH$^{+}$ + H and CH$_{4}^+$ + H) are endothermic, so radiative association reactions are the only way to generate larger ions containing carbon and hydrogen. 

Although Table~\ref{osu_react} lists separately the numbers of radiative association processes in the OSU model that involve (i) ion + neutral reactants, and (ii) neutral + neutral reactants, the mechanism for radiative association is the same whether or not one of the reactants is charged. The mechanism involves the initial formation of an energised complex by the association of the two reactants, followed either by the re-dissociation of the complex or the loss of sufficient energy by the spontaneous emission of radiation so that the complex then no longer contains sufficient internal energy to re-dissociate; that is
\begin{equation}
{\rm A^{(+)} + B \rightleftharpoons \lbrace AB^{(+)} \rbrace ^* \rightarrow  AB^{(+)}} + h\nu
\end{equation}
In this scheme, the superscript (+) indicates that the reactant A may be neutral or charged. The superscript * indicates that the initially formed complex will possess sufficient internal energy for $\lbrace {\rm AB}^{(+)} \rbrace ^*$ to re-dissociate to the reactants A$^{(+)}$ + B. A simple steady-state treatment of this mechanism shows that the second-order rate coefficient for radiative association, $k_{\rm RA}$, is given by:
\begin{equation}
\label{kra}
	k_{\rm RA}   =   k_{\rm ass}  \lbrace k_{\rm rad} /(k_{\rm rad} + k_{\rm diss}) \rbrace
\end{equation}
In this expression, $k_{\rm ass}$  is the second-order rate coefficient for the initial formation of $\lbrace {\rm AB}^{(+)} \rbrace ^*$, $k_{\rm rad}$ is the first-order rate coefficient for the radiative stabilisation of $\lbrace {\rm AB}^{(+)} \rbrace ^*$, and $k_{\rm diss}$ is the first-order rate coefficient for re-dissociation of the energised complex $\lbrace {\rm AB}^{(+)} \rbrace ^*$ to reactants. This last rate coefficient depends on the strength of A$^{(+)}$B bond and on the size of the system, that is the number of atoms in the $\lbrace {\rm AB}^{(+)} \rbrace$ complex. The more atoms, the more vibrations in $\lbrace {\rm AB}^{(+)} \rbrace$ in which the energy can be distributed, and the slower re-dissociation will be. $k_{\rm diss}$ also depends strongly on the internal energy in $\lbrace {\rm AB}^{(+)} \rbrace ^*$ or, in thermal systems, the temperature, so that rate coefficients for radiative association are expected to show a strong negative dependence on temperature. The equation~(\ref{kra}) is for a thermal (canonical) system;  more detailed microcanonical formulations using phase space theory are also available  \citep{1988rcia.conf...17B}.  For systems in which there is a competitive two-body exit channel, the denominator of eq.~(\ref{kra}) must also include a term for adduct dissociation into products. This term need not reduce the rate coefficient of radiative association to an immeasurably small value if there is a barrier in the exit channel that lies below the energy of reactants.  It is also possible that association and normal exothermic channels can occur via different, unrelated reaction pathways.

For small species, $k_{\rm diss}$ $>>$ $k_{\rm rad}$, and radiative association is an inefficient process; that is, it occurs in only a small fraction of the collisions in which $\lbrace {\rm AB}^{(+)} \rbrace ^*$ complexes are formed. In these circumstances, eq.~(\ref{kra}) simplifies to:
\begin{equation}
	k_{\rm RA}   =   k_{\rm rad}  \lbrace k_{\rm ass} /k_{\rm diss} \rbrace
\end{equation}
Canonical treatments using this equation can make use of the approximation that $k_{\rm ass} /k_{\rm diss}$ is a thermal equilibrium constant \citep{1988rcia.conf...17B}. Moreover, in interstellar environments, radiative association is only a significant process in reactions where either one reactant is the (relatively) abundant H$_2$ or when $\lbrace {\rm AB}^{(+)} \rbrace ^*$ is `large': that is, it has more than ca. 10 atoms, so that $ k_{\rm diss} \le k_{\rm rad}$ and $k_{\rm RA} \sim k_{\rm ass}$.  

The most difficult part of estimating rate coefficients for radiative association -- and this applies to ion-neutral as well as neutral-neutral systems -- is generally estimating $k_{\rm rad}$. In those cases where only the ground electronic state of the AB$^{(+)}$ species is involved, the radiative process comprises transitions from vibrational levels in $\lbrace {\rm AB}^{(+)} \rbrace ^*$ to vibrational levels in AB$^{(+)}$.The rate of radiative emission in this case can be obtained from the equation \citep{1982CP.....65..185H}
\begin{equation}
\label{rad}
 k_{\rm rad} = \frac{1}{\rho_s(E)} \sum_{i=1}^{\rm modes} A_{i,1,0} \sum_{n_i=1}^{n_{\rm max}} n_i \rho_{s-1} (E-n_i h\nu_i)
\end{equation}
where $E$ is the internal energy in the complex; $\rho_{s} (E)$ and $\rho_{s-1} (E-n_i h\nu_i)$  are the density of states  associated with s oscillators at energy E and with s - 1 oscillators at energy $(E-n_i h\nu_i)$, the energy $n_i h\nu_i$ being `locked' in the ith oscillator. Values of A$_{i,1,0}$, the Einstein coefficient for spontaneous emission from the $n_i = 1$ to $n_i = 0$ level of the ith oscillator, can frequently be obtained from theory or, in some cases, from experimentally determined infrared absorption coefficients. Values of $k_{\rm rad}$ are generally found to lie within the range 10 - 1000 s$^{-1}$. Because the spontaneous emission coefficients depend on the cube of the emission frequency, higher values of $\rm k_{rad}$ are found for molecules containing hydrogen atoms because they will possess some relatively high vibrational frequencies.
One major source of uncertainty in estimating rate coefficients for radiative association arises from the possible role of electronically excited states of AB$^{(+)}$. If such states correlate with A$^{(+)}$ + B, then the value of $k_{\rm rad}$ may be dramatically increased, even if electronic transitions between the electronically excited state and the ground state are not fully allowed.  Such a possibility occurs for the C$^{+}$ + H$_{2}$ system \citep{1977ApJ...213..696H}.

Although similar mechanisms for radiative association operate, whether or not one of the reactants is electrically charged, laboratory measurements -- indeed laboratory measurements at very low temperatures -- have been possible only for ion-neutral systems, but not for neutral-neutral reactions.  Ion-neutral radiative association can be studied down to very low temperature using ion traps \citep{Gerlich09}. The reaction  C$^+$  +  H$_2$ $\rightarrow$  CH$_2^+$  +  h$\nu$ has been studied using both n-H$_2$ and p-H$_2$.  At 10 K, the rate coefficient is $7\times 10^{-16}$ cm$^3$ s$^{-1}$ for n-H$_2$ and $1.7\times 10^{-15}$ cm$^3$ s$^{-1}$ for p-H$_2$, with uncertainties of only ca. 50\%.  These results are in reasonable agreement with theory \citep{Gerlich92}.

Especially in the case of the radiative association between neutral species, where there is little in the way of direct experiments, the rate coefficients have to be estimated in one of two ways, both of which require a theoretical estimate of the rate coefficient, $k_{\rm rad}$, for the radiative emission step.  First, one can make use of experimental data on the related process of collisionally-assisted association. In these reactions, the radiative stabilisation step in the mechanism represented by eq. (15) is replaced by one in which energy is removed from the energised complexes, $\lbrace {\rm AB}^{(+)} \rbrace ^*$, in collisions with a `third-body species (M)'. Then eq. (16) for $k_{\rm RA}$  can be replaced by
\begin{equation}
 k_{\rm 3rd} {\rm [M]} = k_{\rm M}{\rm [M]} \lbrace k_{\rm ass} / k_{\rm diss} \rbrace
\end{equation}
Here $ k_{\rm M}$ is a second-order rate coefficient for the collisional process in which energy is removed from $\lbrace {\rm AB}^{(+)} \rbrace ^*$ and $k_{\rm 3rd}$ is a third-order rate coefficient for collisionally-assisted association under conditions where $k_{\rm diss} >> k_{\rm M}{\rm [M]}$. It should be mentioned that the implicit assumption is made that individual collisions can stabilize the adduct; in reality, master equation techniques, in which many collisions might be needed, are preferable. Secondly, when there is no experimental data on the collisionally-assisted association process, the ratio $\lbrace k_{\rm ass} / k_{\rm diss} \rbrace$, which corresponds to an equilibrium population of internal states in the energized complex, can be estimated using thermal, quasi-thermal, and phase-space approaches \citep{1988rcia.conf...17B} as long as there is structural and vibrational information about the adduct. In general, theoretical treatments are at best accurate to an order-of-magnitude, for both ion-neutral and neutral-neutral systems. If there is a larger disagreement for a system that has been studied experimentally, it is likely that the structure of the theoretical adduct is incorrect or that the adduct is stabilized electronically.  Use of three-body data can also pose an additional problem, since many experiments are performed in a near-saturated limit, in which the third-order rate coefficient obtained is only a lower limit to the true value.

An association process in which a neutral species with a large electron affinity, typically a radical, and an electron stick together to form an anion via emission of a photon is known as radiative attachment.  This reaction is thought to be the primary one for the production of the interstellar anions -- C$_{3}$N$^{-}$, C$_{4}$H$^{-}$, C$_{6}$H$^{-}$, and C$_{8}$H$^{-}$ -- observed to date \citep{2008ApJ...679.1670H}.  At present, there have been no measurements and no detailed calculations of radiative attachment rate coefficients for these systems.  A statistical theory to determine the rate coefficient, akin to that for radiative association, was developed by \citep{1981Natur.289..656H} and can be expressed in a form similar to the radiative association rate coefficient in eq.~(\ref{kra}).    Here the attachment proceeds by way of a temporary adduct, which radiates away energy as in radiative association (see eq.~(\ref{rad})).  If it is assumed that the formation of the adduct occurs with s-wave electrons only, phase-space theory can be used to derive a simple form for $k_{\rm att}$.  Use of this theory shows that the larger detected interstellar anions are formed on nearly every collision between radicals and electrons, with a rate coefficient in excess of 10$^{-7}$ cm$^{3}$ s$^{-1}$ \citep{2008ApJ...679.1670H}.  There is, however, a large (two order-of-magnitude) discrepancy between model and observations when the theory is used for the radiative attachment of C$_{4}$H$^{-}$ \citep{2008ApJ...685..272H}.  More detailed calculations are needed. These may follow the very detailed treatment of collisional and radiative electron attachment of SF$_6$ which forms SF$_6^-$ \citep{2007Troe2,2007Troe1}.

\subsection{Dissociative recombination reactions}

Dissociative recombination (DR) denotes the process whereby a molecular ion recombines with a free electron and the energy released is dispersed through dissociation of the intermediate neutral into two or more fragments, with conversion of any excess energy into kinetic and internal energy of the fragments. The importance of DR is two-fold: firstly, it can serve as the final step in the synthesis of neutral molecules by transforming a protonated species, which was previously produced in a series of ion-neutral reactions, into the unprotonated species. Such reaction sequences have been invoked as synthetic pathways for a multitude of different interstellar molecules; e. g., HCl \citep{1985ApJ...295..501B}:

\begin{equation}
{\rm H_2  +  Cl^{+}  \rightarrow  HCl^{+}  +  H,}
\end{equation}
\begin{equation}
{\rm H_{2}  +  HCl^{+}  \rightarrow  H_{2}Cl^{+}  +  H,}
\end{equation}
\begin{equation}
{\rm H_{2}Cl^{+}  +  e^{-}  \rightarrow  HCl  +  H.}
\end{equation}
In this mechanism, reaction (21) corresponds to the DR.
Secondly, protonation followed by DR can serve as a very efficient destruction mechanism for molecules with high proton affinities, such as nitriles.
Several ions themselves, such as cyclic C$_3$H$_3^+$,
are sufficiently unreactive in their environments that DR constitutes their sole destruction process \citep{1998P&SS...46.1157K}. DR reactions therefore form an integral part in chemical networks of interstellar clouds . Since the abundances of important species predicted by models can be greatly affected by the rates and product branching fractions of DR reactions, knowledge of these parameters is vital for accurate predictions.


Thermal rate coefficients of DR processes vary, but not to an excessively large extent. In fact, most of them lie in the range between $1 \times 10^{-7}$ and     $4 \times 10^{-6}$~cm$^3$ s$^{-1}$ at 300~K, with larger ions tending to show higher rate coefficients. However, many experimental measurements of DR rate coefficients were performed at room temperature and it is not entirely trivial to extrapolate these parameters to the low temperatures encountered in cold dense cores. As a barrier-less electron-ion reaction, DR reactions should display a temperature dependence of the thermal rate coefficients of T$^{-1/2}$ , which is also assumed in models if no contrary evidence exists. Nevertheless, as experiments have shown, deviations from this behavior are common: usually the temperature dependence is stronger, leading to higher values of $k$ at very low temperatures for a variety of reasons including resonances and potential crossings. 

Branching fractions of DR reactions have proven to be surprising and often counterintuitive to chemical common sense. According to an early theory formulated by  \cite{1986ApJ...306L..45B}, DR reactions should preferably follow the product pathway involving the least rearrangement of valence orbitals. In the case of protonated neutrals, this implies formation of the unprotonated parent substance and a hydrogen atom. Although some ions such as HCO$^+$ follow that rule, many other important species such as H$_3$O$^+$  \citep{2000ApJ...543..764J} and protonated methanol (CH$_3$OH$_2^+$) do not \citep{2006FaDi..133..177G}. In the DR of these species, break-up processes of the intermediate neutral into three fragments dominate, which should be highly unlikely according to Bates's theory.  Another early theory, that of \cite{1978ApJ...222..508H}, uses phase-space theory and can only handle three-body channels if they occur via breakup of a primary species formed in a two-body channel. This theory is reasonably successful for the products of H$_{3}$O$^{+}$ + e$^{-}$ but is in general inaccurate. Therefore one has to rely on experimental measurements, which are hitherto unavailable for many ions, so many  branching ratios used in models are still educated guesses. 

Several experimental methods have been employed to determine the rate coefficients and branching ratios of DR reactions. The first studies 
were carried out using a stationary afterglow (where the ions are produced in a discharge followed by subsequent ion-neutral reactions to form the desired ionic species, thus in the $"$afterglow$"$ of the discharge) \citep{1949PhRv...76.1697B}. 
The faults in this approach were partially overcome by the more novel flowing afterglow method. 
Flowing afterglow methods 
have been employed with success for many systems. However, they also have their limitations. Most studies are carried out at room temperature and therefore, as discussed above, have limited validity for the interstellar medium, although recently experiments on the DR of H$_3^+$ have been performed at 77 K \citep{2009PhRvA..79e2707G}. Furthermore, it is difficult to determine the branching ratios of a DR reaction, since rarely can all the product channels be quantified. 
These facts led to the employment of merged beam techniques to investigate DR. Magnetic storage rings, like ASTRID (Aarhus, Denmark), TSR (Heidelberg, Germany) and CRYRING (Stockholm, Sweden), have proven especially successful. 
However, experiments in storage rings do have their shortcomings. Firstly, there is usually no real control of the rovibrational states of the ions, although in the case of small ions (e. g. H$_3^+$) calibrated ion sources have been proven to produce ions with a distribution between different states corresponding to temperatures relevant to the ISM \citep{2003Natur.422..500M}. Secondly, mass selection does not allow separation of isomer-pairs such as HCO$^+$ and HOC$^+$, which can be produced simultaneously in the ion source and could show different behavior upon DR. The fact that both storage ring and afterglow methods have their merits and faults is painfully illustrated by the fact that for the DR of CH$_5^+$ ion storage ring measurements yielded a predominance of the CH$_3$ + 2H
 channel \citep{2009PhRvA..79c0701Z}, whereas flowing afterglow studies found the CH$_4$ + H channel to dominate \citep{mo10}. Imaging measurements showed that the CH$_3$ + 2H production proceeds in a two step mechanism, first leading to H and a highly excited CH$_4$ molecule, which could subsequently dissociate to yield CH$_3$ + H. At the higher pressures in the flowing afterglow experiments, the CH$_4$ intermediate can be deactivated by collisions with buffer gas, leading to a higher branching fraction of the CH$_4$ + H pathway.

Given the problems associated with experimental determinations of the branching ratios and rate coefficients of DR reactions, one could wish to resort to predictions by high-quality ab initio calculations. However, due to the involvement of highly excited states and complicated potential surfaces, such computations are very challenging. Also, a large number of open electronic states exist, which might be close in energy, bringing about break-down of the Born-Oppenheimer approximation. Furthermore, not only bound states, but also free electron states have to be included \citep{2008drmi.book.....L}, making such studies very time-consuming. Although theoretical investigations have been performed for the DR of HCO$^+$ \citep{2006PhRvA..74c2707M}, it is quite reasonable to assume that such efforts will, at least in the foreseeable future, be limited to rather small systems, such as for H$_{3}^{+}$ \citep{2007JChPh.127l4309D}.

Finally, it should be mentioned that, if electrons undergo radiative attachment efficiently to large neutrals such as PAH's, it might be possible that the abundance of negatively-charged PAH's exceeds the electron abundance.  In this case, it would be necessary to include mutual neutralization processes between small cations and large anions in interstellar models.  Unfortunately, the amount of information concerning such processes is miniscule.  More laboratory studies are certainly needed.

\section{Uncertainty analysis, uncertainty propagation and sensitivity analysis}
\label{uncanal}

Because chemical models contain hundreds to thousands of chemical reactions, the model predictions can be strongly affected by uncertainties in the rate coefficients. Dealing with uncertainties in modeling is a three-stage process:

\begin{enumerate}
\item Uncertainty Analysis (UA) consists in identifying the uncertain inputs
of the model and quantifying their uncertainty. The type of information
to be issued at this stage is partly conditioned by the next stages:
it can be a simple standard deviation or a full probability density
function;
\item Uncertainty Propagation (UP) is used to transfer the uncertainty in
the model input(s) to the model output(s);
\item Sensitivity Analysis (SA) enables one to identify input (key) parameter(s) with a large contribution to the uncertainty in the model output(s).
\end{enumerate}
The methodological framework presented in Fig. \ref{fig:Flow-diagram},
according to \citet{DeRocquigny08}, summarizes the operations and
information flow involved in a full analysis.

\begin{figure}

\noindent \begin{centering}
\includegraphics[width=\textwidth]{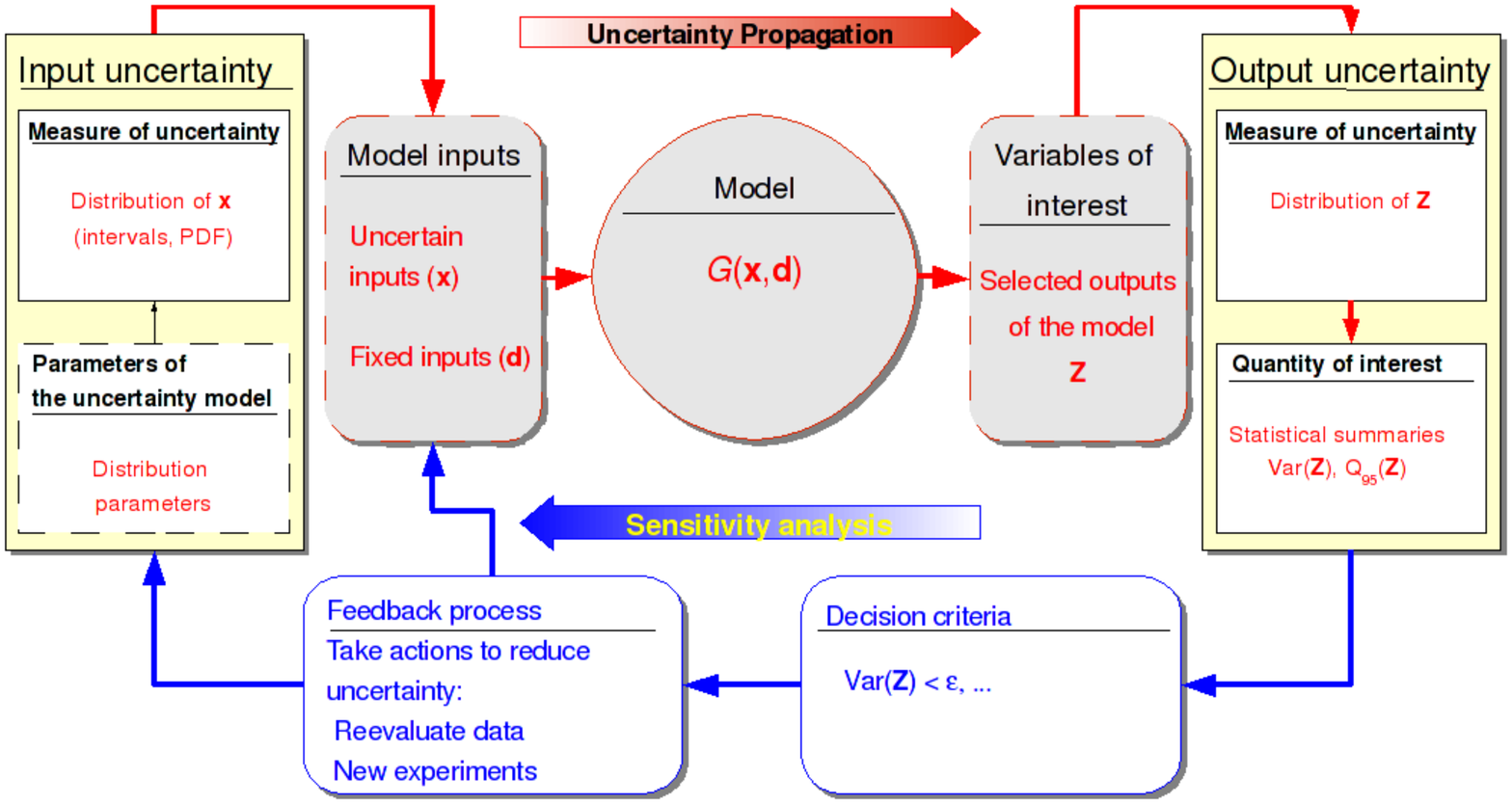}
\par\end{centering}

\caption{\label{fig:Flow-diagram}Flow diagram of Uncertainty Propagation and
Sensitivity Analysis}

\end{figure}

\subsection{Uncertainty Analysis}

UA is  challenging  for the modeller, because it often requires proficiency in the experimental techniques used to evaluate the quantities of interest. 
A general problem is that the reference databases generally lack quantified uncertainty statements about the stored data, or at least properly quantified for the purposes of UP/SA. 
As discussed in this article, KIDA is an attempt to address this shortcoming.

SA results are quite sensitive to this \emph{elicitation} step: too
small uncertainties lead to an over estimation of prediction precision,
whereas too large uncertainties can lead to useless predictions. 
At the present stage, we are still in need of clearly defined guiding rules,
and groups of experts, containing both experimentalists and modellers, to
work on this topic, as has been done in respect of the
Earth's atmospheric chemistry \citep{Sander06,2004ACP.....4.1461A}.

Following the work of \citet{1998P&SS...46..491D} and \citet{2003A&A...398..335D}
on giant planets, exhaustive analysis of these issues was started
for Titan's atmosphere. An evaluation of the uncertainties for neutral-neutral
reactions \citep{Hebrard:2006,2007P&SS...55.1470H} and for ion-neutral reactions
\citep{2007P&SS...55..141C,Carrasco2007,2008P&SS...56.1644C} has been undertaken, in order to prioritize reactions in need of better characterization,
possibly by laboratory measurements. Application to Titan\textquoteright{}s
atmosphere shows that uncertainty in the chemical rate coefficients strongly reduces the predictivity of atmospheric models \citep{2007P&SS...55.1470H}.
A particular issue in this system (which is also relevant to astrochemistry)
is the use of extrapolation for 90\% of the reaction rates to the
low temperatures of Titan's atmosphere (50-200~K). 
The task of providing reasonable uncertainty on extrapolated data is nevertheless an unavoidable step in order to prioritize key reactions, in need of better determination \citep{Hebrard:2009}.

A similar approach has been undertaken by Wakelam and coworkers for astrophysical objects \citep{2005A&A...444..883W,2006A&A...451..551W,2008ApJ...672..629V,2009A&A...495..513W}, as presented in the following sections.

\subsection{Uncertainty propagation}

Uncertainty propagation is a tool to quantify the precision of model
predictions. Because of their central role in all fields of science
and engineering, UP methods have been standardized at the international
level \citep{GUM,GUMSupp1}. They are of two kinds: (i) combination
of variances, as described in the Guide to the Expression of Uncertainty
in Measurements (called {}``the GUM'') \citep{GUM}; and (ii) propagation
of distributions, as described in the first supplement to the GUM (GUM-Supp1)
\citep{GUMSupp1}.

Denoting $X_{i}$ as the $i^{th}$ input of a model $G$, and $Z$ as an output
of this model ($Z=G(X_{1},X_{2},\ldots,X_{n})$), the GUM  stipulates
that $u_{Z}$, the standard uncertainty of $Z$, should be calculated
by the following formula:\begin{equation}
u_{Z}^{2}=\sum_{i,,j=1}^{n}\left(\frac{\partial G}{\partial X_{i}}\right)_{\overline{X}}\left(\frac{\partial G}{\partial X_{j}}\right)_{\overline{X}}{\rm cov}(X_{i},X_{j})\label{eq:prop-var}\end{equation}
where ${\rm cov}(X_{i},X_{j})$ is an element of the covariance matrix of
input parameters, with variances as the diagonal elements ${\rm cov}(X_{i},X_{i})=u_{X_{i}}^{2}$.
The so-called sensitivity coefficients $\left(\partial G/\partial X_{i}\right)_{\overline{X}}$
are evaluated at the mean value of input parameters $\overline{X}$.

This method is exact if the model $G$ is a linear function of the
inputs, and if the error distribution of these inputs is symmetric.
For small relative uncertainties (typically less than 30\%), it provides
generally an excellent approximation of $u_{Z}$, even for non-linear
models.

However, astrochemistry and atmospheric chemistry have to deal with
uncertainties much larger than 30\% \citep{Hebrard:2006,2008ApJ...672..629V},
and because of the non-linearity of photochemical models, eq. (\ref{eq:prop-var})
cannot be relied upon. 
It nevertheless reveals that there are two essential components to uncertainty propagation: the sensitivity of the model to the inputs; and the amplitude of input uncertainty, which should be present in any other UP method.

The method of propagation of distributions requires the definition of probability density functions (pdf) for all the parameters \citep{GUMSupp1}. 
The reference equation for the propagation of distributions is
\begin{equation}
p(Z=a)=\int {\rm d}X_{1}{\rm d}X_{2}...{\rm d}X_{n}\,\delta\left(a-G(X_{1},X_{2},...,X_{n})\right)\, p(X_{1},X_{2},...,X_{n})\label{eq:prop-dist}
\end{equation}
where $a$ is a value in the range of $Z$.
This equation provides the pdf of $Z$ , $p(Z)$ by transforming the pdf of the
inputs, $p(X_{1},X_{2},...,X_{n})$, through the model. 
It can rarely be solved analytically. 
Instead, the GUM-Supp1 recommends the use of Monte Carlo Uncertainty Propagation  (MCUP) methods (see also Section~\ref{ischem}). 
The principle is simple: one generates a number $N_{\rm run}$ of random input sets from their pdf, and for each set, one evaluates the value of the output, thus generating a representative sample of the output pdf $p(Z)$. 
This sample is then used to estimate quantities of interest.

In the hypothesis of independent inputs with Normal/Gaussian type
uncertainties, the same ingredients are necessary to generate the
input random sample as for the method of combination of variances;
i.e., average values and standard uncertainties. Other distributions
might require more complex elicitation procedures, especially in the
case of correlated inputs \citep[\emph{e.g.} branching ratios][]{Carrasco2007}.

\subsection{Sensitivity analysis}
\label{sa}

Sensitivity Analysis (SA) methods are not as standardized as UP methods:
a variety of methods are more or less adapted to specific
scenarios, but globally, SA consists in \emph{the apportionment of output
variance between the uncertain inputs} \citep{Saltelli00,Saltelli04}.

In this work, we are interested in the identification of key parameters
being responsible for a large part of the output variance. 
In the case of a linear model, this could be done by comparing the contributions
of inputs in eq. \ref{eq:prop-var}. 
The ingredients are thus the model sensitivity factors and the input variances/covariances. 
Combination of both terms means that an output might be insensitive to a very
uncertain input, or very sensitive to a fairly well-known parameter.
This remains true for nonlinear models, and both contributions remain
essential. 
Sensitivity analysis has therefore to be based on careful uncertainty analysis and uncertainty propagation steps.

A convenient SA method can be combined with MCUP through the
calculation of input/output correlation coefficients \citep{Helton06}.
Pearson's correlation coefficient
\begin{equation}
\label{CCPearson}
C(X_{j},Z)=\frac{\sum_{l}(X_{j}^{l}-\overline{X_{j}})(Z^{l}-\overline{Z})}{\sqrt{\sum_{l}(X_{j}^{l}-\overline{X_{j}})^{2}\sum_{l}(Z^{l}-\overline{Z})^{2}}}
\end{equation}
is often used \citep{2005A&A...444..883W,2006A&A...451..551W,2008ApJ...672..629V,2009A&A...495..513W},
where the sums run over the sample length ($l=1,N_{run}$). Correlation
coefficients vary between -1 and 1, and influential parameters are
identified through the comparatively large absolute values of their
correlation coefficient with the output $\left|C(X_{i},Z)\right|$.
Spearman's rank correlation coefficient is considered more appropriate
for nonlinear models \citep{Helton06,Dob2009}, albeit our experience
with sensitivity analysis of chemical networks is that the differences
between both methods are generally minor.

Other SA methods, based on specific sampling patterns, can be more
economical to run if one does not require UP results \citep{Saltelli00}.
In the present work, both MCUP and SA stages are of interest, and
the Correlation Coefficients SA method is a computationally inexpensive
byproduct of MCUP.

In the context of sensitivity analysis, a \emph{key reaction} has
a reaction rate uncertain enough to contribute significantly to the
model output uncertainty \citep{2005A&A...444..883W,2006A&A...451..551W,2008ApJ...672..629V,Dob2009,Hebrard:2009,2009A&A...495..513W}.
Identification of such key reactions enables one to point out those in
need of further study, either by an evaluation committee, or by new
experiments or calculations. The improvement of the precision of model
predictions is a slow iterative process between stages of sensitivity
analysis and database updates, as new key reactions are expected to
emerge when the previous ones have been updated \citep{Hebrard:2009}.

In order to select key reactions, a target has to be defined.  
This could be the abundance of a single species of interest. 
In that case, one evaluates and compares the input/output correlation coefficients
for a single output species \citep{2005A&A...444..883W,2008P&SS...56.1630D}. 
If one is interested in a global improvement of model precision, an additional
criterion is necessary, as for instance selecting reactions presenting
large correlation coefficients with a large number of output concentrations
\citep{2005A&A...444..883W,2006A&A...451..551W,2008ApJ...672..629V,Hebrard:2009,2009A&A...495..513W}. 
The key reactions can then be ranked according to the number of species they affect.

\section{Gas-phase molecular cloud chemistry}
\label{ischem}


To improve the chemical modeling of cold cores, we applied sensitivity methods to identify "key reactions".  A detailed review of the existing experimental and theoretical data on these reaction rate coefficients was then conducted to propose new values or confirm the ones already used in the models. We also studied the impact of the proposed new rate coefficients on the cold core chemistry. These three steps are described below.

\subsection{Sensitivity analysis: identification of key reactions}
\label{senanal}

\begin{figure}
\includegraphics[width=\textwidth]{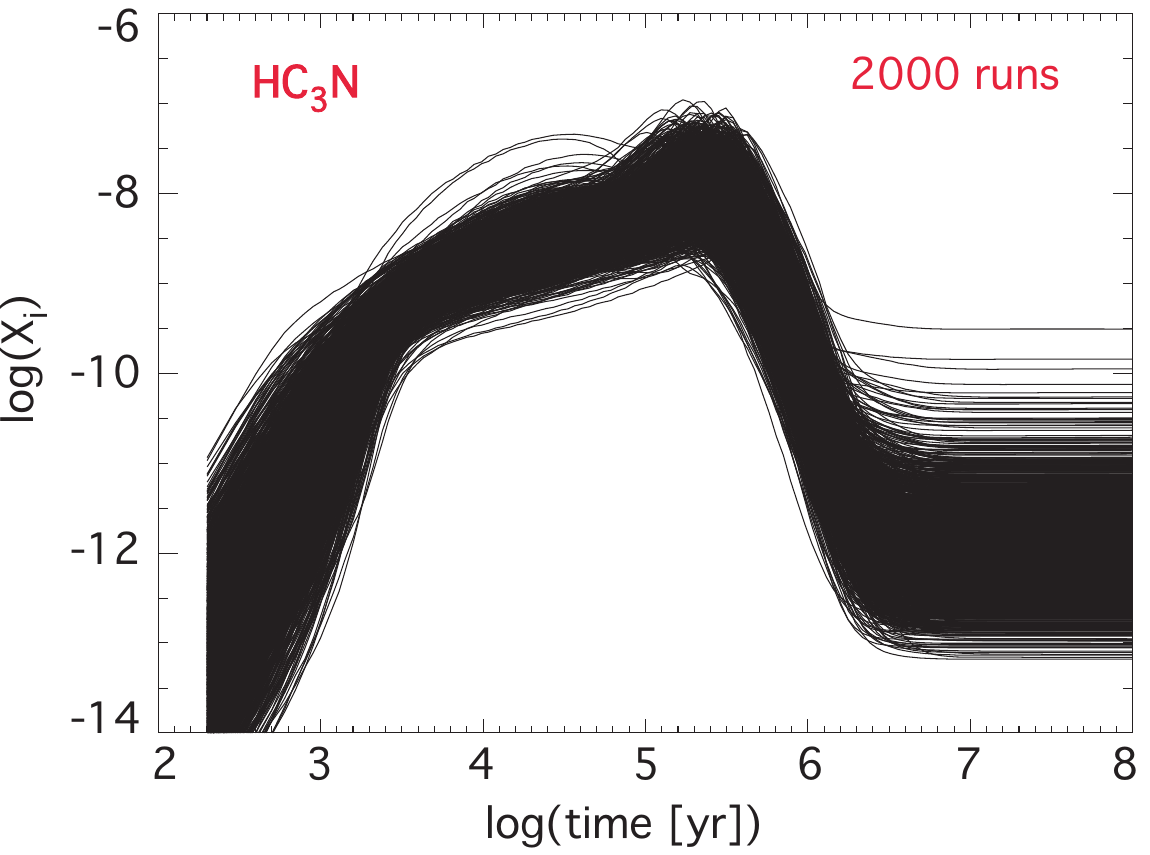}
\caption{Abundance of HC$_3$N as a function of time. Each black line is the result of one run (one set of reaction rate coefficients). The density is $2 \times 10^4$~cm$^{-3}$, species are initially in the atomic form (except for H$_2$) and the rate coefficients are varied within their uncertainty range, as defined in the database. \label{fig1}}
\end{figure}

To identify key reactions, we used Monte Carlo sensitivity methods as described in \citet{2009A&A...495..513W} and this review. Briefly, we randomly varied all the reaction rate coefficients within a certain factor. For each set of rate coefficients, we computed the chemical evolution of the cloud and thus obtained $N$ chemical evolutions. For each species $j$ and each chemical reaction $i$, we calculated Pearson correlation coefficients (see Section \ref{sa}, Eq. \ref{CCPearson}):


\begin{equation}
\label{Pcc}
P^i_j(t)=C(\log(X_j(t)),\log(k_i)),
\end{equation}
where $X_j(t)$ is the abundance for species $j$ and $k_i$ is the rate constant of reaction $i$. 
The larger the value of $P^i_j$, the stronger the correlation and the more important the reaction $i$ for the abundance of the species $j$.

For this work, we used the nahoon chemical model \footnote{$\rm http://www.obs.u-bordeaux1.fr/amor/VWakelam/$} with the osu-09-2008 network. In addition, we allowed gas-phase species to stick onto dust grains and to be evaporated from the surfaces by thermal desorption and cosmic-ray heating. We assumed a fixed homogeneous temperature of 10~K, a visual extinction  $A_{\rm V}$ of 10, and a cosmic-ray ionization rate $\zeta_{\rm H_2}$ of $1.3 \times 10^{-17}$~s$^{-1}$. We considered two densities: $2 \times 10^4$ and $2 \times 10^5$~cm$^{-3}$, which refer to the total nuclear density of hydrogen. The typical low-metal elemental abundances were used \citep[see Table 1 in][]{2008ApJ...680..371W}. 
We considered two types of initial conditions. In the first, all the species are in the atomic form except for hydrogen, which is entirely molecular, while in the second, half of the ionic carbon is in CO, and  the other half is still in C$^+$. 

To sample the reaction rate coefficients, we used two different sets of uncertainty factors. In the first set, the uncertainty factor is defined in the database, where most reactions are assigned a factor of 2, as originally suggested in the UDfA database, but others, studied experimentally, have a smaller factor of 1.25 or 1.5  All radiative association reactions have a larger uncertainty factor of 10.  In the second set, the uncertainty factor is fixed at 10 for all reactions. The goal of this second set is to see the effect of varying the rate coefficients over a larger range of values.
 
Considering the different densities, initial conditions and uncertainty factors, we constructed 8 different models and for each of these models, we made 2500 runs by randomly varying the rate coefficients using a log-normal distribution. As an example, we show the results for HC$_{3}$N in one model (an H density of $2 \times 10^4$~cm$^{-3}$, atomic initial conditions, and the uncertainty factors as defined in the database) in Fig.~\ref{fig1}. 

In Table~\ref{tab1}, we report the reactions that have a significant impact on more than one species. To determine these reactions, we count for each reaction the number of species with a correlation coefficient larger than 0.3 and an abundance larger than $10^{-11}$. The abundance criterion is chosen because these species can be observed in dense cold cores and thus used to put constraints on the model. We used two different times: $10^5$ and $10^6$~yr, covering the probable age of most cold cores. A more thorough approach would be to use a larger number of times, because some reactions important at earlier stages may contribute to the synthesis of the larger molecules at the time chosen. We use a Boolean approach: either a species contributes to the importance of a reaction or not. Another possibility, which is considered in Section 6,  is to sum the correlation coefficients for all species. Then, the larger the sum, the more important the reaction.  In both cases, the result depends on the chosen uncertainty factor in the rate coefficient.  Each method has its pros and cons.  In this current study, we checked that the result of the analysis is not significantly changed if the second approach is used. The list of key reactions is very similar for the different models; only the relative importance of the reactions really changes. 

 Tables~\ref{tab1} and \ref{tab2} give the key reactions in the following case: density of $2 \times 10^4$~cm$^{-3}$, species initially in the atomic form (except for H$_2$) and rate coefficients varied within their uncertainty range, as defined in the database. The 7 other models we ran gave the same reactions but in a different order of importance except for the first reaction in Table  \ref{tab1}.   The association reaction C + H$_2$ $\rightarrow$ CH$_2$ + h$\nu$, which heads the list, influences about 73 abundant species at $10^5$~yr. Although association reactions can have small rate coefficients (see Section~\ref{rareact}), if the reactants are abundant, these reactions can play an important role. In Table~\ref{tab2}, we show the key reactions, divided as to type, for individual observable species. Here again, we considered only species with a correlation coefficient larger than 0.3 and an abundance larger than $10^{-11}$ at $10^5$ or $10^6$~yr.

\begin{table}
\caption{List of key reactions by number of species influenced.  \label{tab1}}
\begin{center}
\begin{tabular}{lc}
\hline
\hline
Reaction & Strongly affected species$^{1}$\\
\hline
C + H$_2$ $\rightarrow$ CH$_2$ + h$\nu$	&73\\
CH$_3^+$ + H$_2$ $\rightarrow$ CH$_5^+$ + h$\nu$	&18\\
C$_2$H$_2^+$ + H$_2$ $\rightarrow$ C$_2$H$_4^+$ + h$\nu$	&C$_2$H$_2$O, C$_2$H$_3$, C$_2$H$_2^+$, C$_2$HO$^+$, C$_2$H$_2$N$^+$, C$_2$H$_4^+$\\
CH$_3^+$ + CO $\rightarrow$ C$_2$H$_3$O$^+$ + h$\nu$	&C$_2$H$_2$O, C$_2$H$_3$O$^+$\\
C$_2$H$_4^+$ + e$^-$ $\rightarrow$ C$_2$H$_3$ + H&	C$_2$H$_2$O, C$_2$H$_3$\\
HSiO$^+$ + e$^-$ $\rightarrow$ SiO + H& Si, SiO\\
HSiO$^+$ + e$^-$ $\rightarrow$ Si + OH 	&Si, SiO\\
C$_3$H$^+$ + H$_2$ $\rightarrow$ C$_3$H$_3^+$ + h$\nu$ &C$_3$H$_2$, H$_2$C$_3$, C$_3$H$^+$, C$_3$H$_2^+$, C$_3$H$_3^+$, H$_3$C$_3^+$\\
C$_3$H$^+$ + H$_2$ $\rightarrow$ H$_3$C$_3^+$ + h$\nu$ 	&C$_3$H$_2$, H$_2$C$_3$, C$_3$H$^+$, C$_3$H$_2^+$, C$_3$H$_3^+$, H$_3$C$_3^+$\\
CH$_3^+$ + HCN $\rightarrow$ C$_2$H$_4$N$^+$ + h$\nu$	&C$_2$H$_2$N, HC$_3$N, C$_2$H$_3$N, C$_2$H$_4$N$^+$\\
C$_4$H$_2^+$ + H $\rightarrow$ C$_4$H$_3^+$ + h$\nu$&	C$_4$H$_2$, C$_5$H, C$_6$H$_6$, C$_4$H$_3^+$\\
CH$_3^+$ + NH$_3$ $\rightarrow$ CH$_6$N$^+$ + h$\nu$	&CH$_3$N, CH$_5$N\\
C$_4$H$_2^+$ + O $\rightarrow$ HC$_4$O$^+$ + H	&C$_3$O, HC$_4$O$^+$\\
\hline
\end{tabular}
\end{center}
$^{1}$Right column indicates the number or names of the species influenced strongly by each reaction.
\end{table}

\begin{table}
\caption{List of key reactions for specific species 
\label{tab2}}
\begin{center}
\begin{tabular}{cc}
\hline
\hline
Neutral-Neutral reactions	&  
Affected species \\
\hline
C + C$_3$O $\rightarrow$ C$_3$ + CO&	C$_3$O \\
C + OCN $\rightarrow$ CO + CN&	OCN \\
H + CH$_2$ $\rightarrow$ CH + H$_2$	 &CH \\
O + CN $\rightarrow$ CO + H	&CN \\
N + CN $\rightarrow$ C + N$_2$	&CN \\
O + NH $\rightarrow$ NO + H	&NH \\
O + C$_2$ $\rightarrow$ CO + C	&C$_2$ \\
O + C$_2$H $\rightarrow$ CO + CH&	C$_2$H \\
O + C$_3$H $\rightarrow$ C$_2$H + CO&	C$_3$H \\
N + C$_3$ $\rightarrow$ CN + C$_2$	&C$_3$  \\
N + NO $\rightarrow$ N$_2$ + O	&NO \\
O + NH$_2$ $\rightarrow$ HNO + H&	NH$_2$\\
O + HNO $\rightarrow$ NO$_2$ + H&N$_2$O\\
O + HNO $\rightarrow$ N$_2$O + H&	N$_2$O\\
CN + NH$_3$ $\rightarrow$ NH$_2$CN + H	& NH$_2$CN\\
O + C$_3$N $\rightarrow$ CO + C$_2$N&	C$_3$N\\
N + C$_4$N $\rightarrow$ CN + C$_3$N&	C$_4$N\\
N + C$_4$H $\rightarrow$ C$_4$N + H	&C$_4$N\\
N + C$_2$N $\rightarrow$ CN + CN&	C$_2$N\\
CN + HC$_5$N $\rightarrow$ NC$_6$N + H&	NC$_6$N\\
CN + HC$_3$N $\rightarrow$ NC$_4$N + H	& NC$_4$N\\
\hline
Association reactions	 & 
Affected species \\
\hline
C$_4$H$_2^+$ + HC$_3$N $\rightarrow$ C$_7$H$_3$N$^+$ + h$\nu$&	HC$_7$N\\
HS$^+$ + H$_2$ $\rightarrow$ H$_3$S$^+$ + h$\nu$	&H$_2$S\\
HCO$^+$ + H$_2$O $\rightarrow$ CH$_3$O$_2^+$ + h$\nu$&	CH$_2$O$_2$\\
C$^+$ + H$_2$ $\rightarrow$ CH$_2^+$ + h$\nu$&	C$_3$\\
S + CO $\rightarrow$ OCS + h$\nu$&	OCS\\
CH$_3^+$ + HC$_3$N $\rightarrow$ C$_4$H$_4$N$^+$ + h$\nu$&	CH$_3$C$_3$N\\
CH$_3^+$ + HC$_5$N $\rightarrow$ C$_6$H$_4$N$^+$ + h$\nu$	&CH$_3$C$_5$N\\
Si$^+$ + H$_2$ $\rightarrow$ SiH$_2^+$ + h$\nu$&	HNSi\\
\hline
	Ion-neutral reactions	&\
Affected species \\
\hline
C$_2$H$_3^+$ + O $\rightarrow$ C$_2$H$_2$O$^+$ + H&	C$_2$H$_2$O$^+$ \\
C$^+$ + S $\rightarrow$ S$^+$ + C	&H$_2$CS \\
C$_5$H$^+$ + N $\rightarrow$ C$_5$N$^+$ + H	&C$_5$H$_2$N$^+$, HC$_5$N \\
C$_2$H$_4^+$ + N $\rightarrow$ C$_2$H$_2$N$^+$ + H$_2$	&C$_2$H$_2$N$^+$ \\
C$_4$H$_2^+$ + S $\rightarrow$ HC$_4$S$^+$ + H &	C$_4$S \\
H$_3^+$ + C $\rightarrow$ CH$^+$ + H$_2$	&C$_3$H$_3$ \\
H$_3^+$ + O $\rightarrow$ OH$^+$ + H$_2$	&H$_3$O$^+$ \\
C$_2$H$_3^+$ + N $\rightarrow$ C$_2$NH$^+$ + H$_2$&	C$_2$NH$^+$ \\
\hline
	Dissociative recombination	& 
Affected species \\
\hline
HC$_4$O$^+$ + e$^-$ $\rightarrow$ C$_3$O + CH	 & C$_3$O\\
H$_2$NC$^+$ + e$^-$ $\rightarrow$ HNC + H	&HNC, H$_2$NC$^+$\\
C$_5$H$_2$N$^+$ + e$^-$ $\rightarrow$ C$_5$N + H$_2$&	C$_5$H$_2$N$^+$\\
HC$_4$S$^+$ + e$^-$ $\rightarrow$ C$_4$S + H	&C$_4$S\\
H$_2$CO$^+$ + e$^-$ $\rightarrow$ CO + H + H	&H$_2$CO$^+$\\
CNC$^+$ + e$^-$ $\rightarrow$ CN + C&	CNC$^+$\\
HC$_4$S$^+$ + e$^-$ $\rightarrow$ C$_4$S + H&	C$_4$S\\
HC$_3$S$^+$ + e$^-$ $\rightarrow$ C$_3$S + H&C$_3$S\\
HC$_3$S$^+$ + e$^-$ $\rightarrow$ C$_2$S + CH	&C$_3$S\\
\hline
\end{tabular}
\end{center}
\end{table}

\subsection{Reaction review}
\label{review}

\begin{figure}
\includegraphics[width=\textwidth]{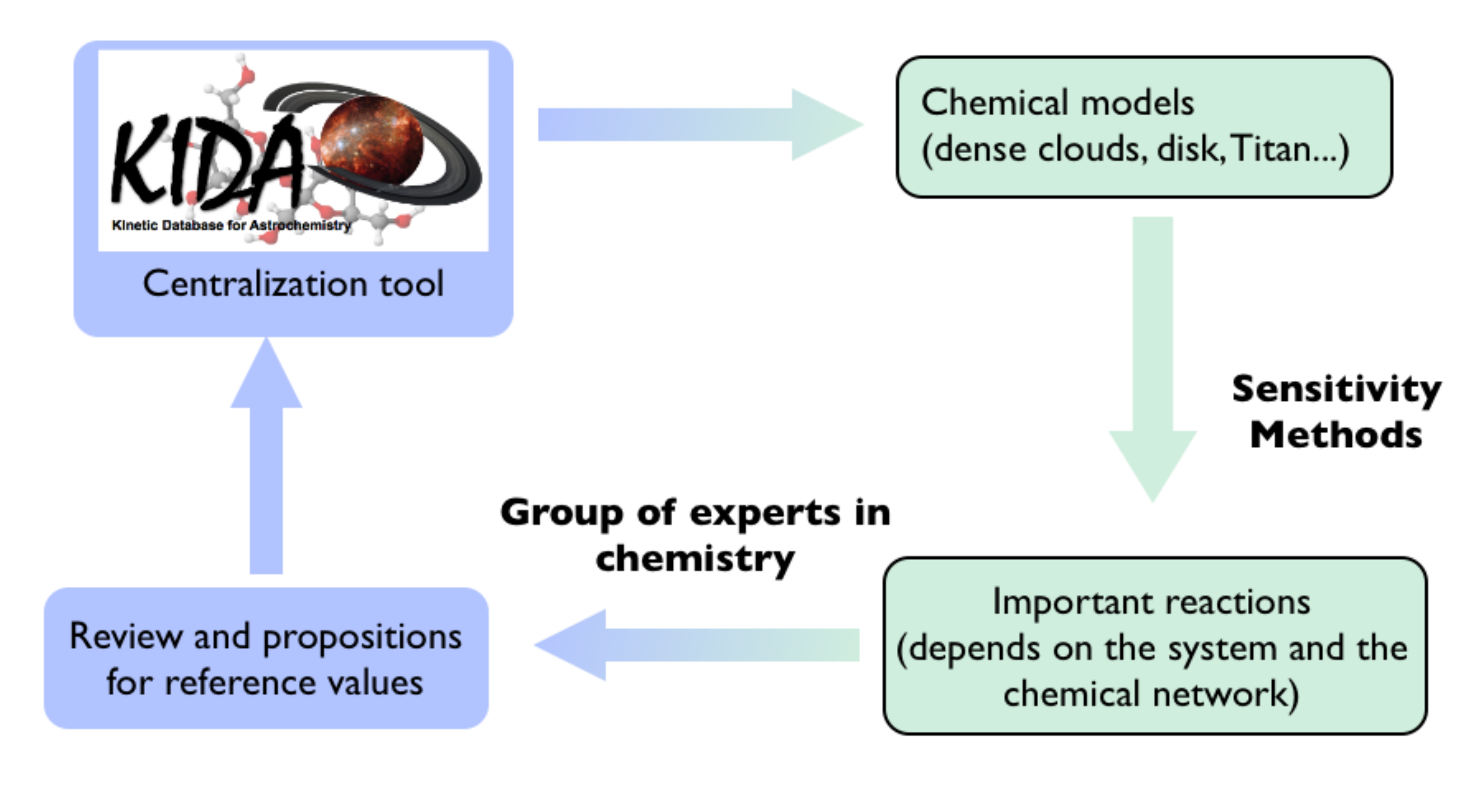}
\caption{The construction of the KIDA database. \label{kida_loop}}
\end{figure}

Of the 59 reactions that have been identified as particularly crucial in the chemistry of cold cores in Tables~\ref{tab1} and \ref{tab2}, 34 have been reviewed and new values of the rate coefficients are proposed for 22 of these reactions. The new rate coefficients that we propose as a result of this review are compared in Tables \ref{tab3}, {\ref{tab4}, \ref{tab5}, and \ref{tab6}, according to type of reaction,  with the values that are currently in the OSU and UDfA databases. We have considered rate coefficients for temperatures between 10 and 298~K. In the tables, a designation of zero means that the rate coefficient is immeasurably small.  

The results of this effort have been included in a new KInetic Data base for Astrochemistry (KIDA), which is available at http://kida.obs.u-bordeaux1.fr. This database differs from the OSU and UDfA data bases in that, at least for the key reactions identified in this article, the reasons for the choice of particular values of the rate coefficients and their uncertainties are explained in DataSheets (accessed by clicking on the entry `Details' in the first page devoted to a particular reaction). {\bf An example of datasheet for the reaction H$_3^+$ + O is given in Appendix~\ref{appendA}.} Figure~\ref{kida_loop} illustrates the construction of this database.

These DataSheets follow the form adopted by the IUPAC Subcommittee for Gas Kinetic Data Evaluation, which reviews and evaluates kinetic, photochemical and heterogeneous data for atmospheric chemistry (http://www.iupac-kinetic.ch.cam.ac.uk). 
For each reaction the DataSheet includes:
\begin{enumerate}
\item 	 the possible product channels, where appropriate,  and the electronic states of the reactants and products;
\item 	the enthalpies of reaction for the possible reaction channels;
\item 	a Table giving experimental (and theoretical) values of the rate coefficients, and the values given in previous reviews, including the OSU and UDfA data bases;
\item 	comments on the experimental and theoretical studies that have been performed;
\item 	recommended values of both the rate coefficients and their estimated uncertainties for temperatures between 10 and 298~K except as mentioned in the text,  and comments (and justification) for the choices that have been made.
\end{enumerate}

When we have an estimation for the uncertainty factor F$_{\rm unc}$, we have indicated it in the Tables \ref{tab3}, {\ref{tab4}, \ref{tab5}, and \ref{tab6} otherwise, the symbol {\bf -} is used.
There are three reactions that have been examined -- C + H$_2$ $\rightarrow$ CH$_2$ + h$\nu$ and C$^+$ + S $\rightarrow$ S$^+$ + C -- 
but for which we do not give recommendations at the present time, since no experiment or detailed calculation has been done at low temperature. The impact of their rate coefficients on the abundances of the species appears to be especially strong. They will be discussed separately in the next section. Finally that the association reaction S + CO $\rightarrow$ OCS + h$\nu$ is currently being studied theoretically by the group of Gunnar Nyman (G\"oteborg University, Sweden). 

\begin{landscape} 
\begin{table}
\caption{Summary of the reaction review: Neutral-Neutral reactions. \label{tab3}}
\begin{center}
\begin{tabular}{l|lllll|lll|lllll}
\hline
\hline
Reaction & \multicolumn{5}{c|}{UDfA value} & \multicolumn{3}{c|}{OSU value$^a$} & \multicolumn{4}{c}{New value} \\
 & $\alpha^b$ & $\beta$  & $\gamma$ & F$_{\rm unc}$$^c$ & T range & $\alpha^b$ & $\beta$ & F$_{\rm unc}$$^c$ & $\alpha^b$ & $\beta$ & $\gamma$ & 
 F$_{\rm unc}$$^c$ &\\
\hline
C + C$_3$O $\rightarrow$ C$_3$ + CO &	1.0(-10) & 0&0 &  2 & 10-300 & 1.0(-10)  & 0 & 2&	1.0(-10) & 0 & 0 & 4 \\
\hline
C + OCN $\rightarrow$ CO + CN	& 1.0(-10) & 0 &0 & 2 & 10-300 & 1.0(-10)&0	& 2 & 1.0(-10)& 0  &  0 & 5 \\
\hline
{\bf H + CH$_2$ $\rightarrow$ CH + H$_2$ }	& 6.64(-11) & 0 & 0 &1.25 & 300-2500 & 	2.7(-10) &0& 2	& 2.2(-10) & 0 &  0 & 4\\
\hline
{\bf O + CN $\rightarrow$ CO + N} &	4.36(-11) & 0.46 &  -364 & 1.5 & 300-5000 &	4.0(-11) & 0&2 &	2.6(-11) & -0.12 & 0 & 4 \\
\hline
{\bf N + CN $\rightarrow$ C + N$_2$}	&3.0(-10) &0 & 0  & 2 & 298-2500  & 3.0(-10) &0&	2 & 1.0(-10) & 0.18 & 0 & 4\\
\hline
{\bf O + NH $\rightarrow$ NO + H} &	1.16(-10) & 0&0 & 1.25 & 250-300 & 1.16(-10) &0 & 2	& 6.6(-11) & 0 & 0 & 4 \\
{\bf O + NH $\rightarrow$ OH + N}	& 1.16(-10) & 0 & 0 & 1.25 & 250-300 &	1.16(-11) &0 & 2 &	0 & 0 & 0 & - \\
\hline
{\bf O + C$_2$ $\rightarrow$ CO + C}	& 1.0(-10) & 0 &0 & 2 & 10-300 & 1.0(-10) &0&2 &	2.0(-10) & -0.12 & 0 & 3  \\
\hline
{\bf  O + C$_2$H $\rightarrow$ CO + CH}	& 1.7(-11) & 0 & 0 & 1.25 & 290-2500 & 1.7(-11) &0& 2 & 1.0(-10) & 0&0 & 4  \\
\hline
{\bf O + C$_3$H $\rightarrow$ C$_2$H + CO}	& 1.7(-11) & 0 &0  & 2 & 10-300 & 1.7(-11)&0	& 2 & 1.0(-10) & 0 &  0 & 4\\
\hline
N + C$_3$ $\rightarrow$ CN + C$_2$	& 1.0(-13) & 0 & 0 & 2 & 10-300 & 1.0(-13)&0	& 2 & 1.0(-13) & 0 & 0 & 10\\
\hline
N + NO $\rightarrow$ N$_2$ + O&	3.75(-11) & 0& 26  & 1.25 & 100-4000 &	3.0(-11) & -0.6  & 2 & 4.76(-11)$^d$ & -0.35 & 58 & - \\
\hline
{\bf O + NH$_2$ $\rightarrow$ HNO + H}	 &4.56(-11) & 0 & -10 & 1.25 & 200-3000 & 8(-11) &0	& 2 & 6.3(-11)& -0.1 &0 & 3  \\
{\bf O + NH$_2$ $\rightarrow$ NO + H$_2$} & 8.3(-12) & 0 & 0 & 1.25 & 200-3000 & 1.0(-11) & 0 & 2 & 0 & 0 & 0 & - \\
{\bf O + NH$_2$ $\rightarrow$ OH + NH} & 1.39(-11) &  0& 40 & 1.25 & 200-3000 & 2.0(-11) & 0 & 2 & 7(-12) & -0.1  &0 & 3 \\
\hline
{\bf O + HNO $\rightarrow$ NO$_2$ + H } & 1.0(-12) & 0& 0  & 2 & 10-300 & 1.0(-12) & 0 & 2 & 0 &0 & 0 & - \\
{\bf O + HNO $\rightarrow$ NO + OH} & 6.0(-11) & 0 &0  & 1.25 & 10-2500 & 3.8(-11) & 0 & 1.25 & 3.77(-11) & -0.08 & 0 & 4\\
\hline
{\bf CN + NH$_3$ $\rightarrow$ HCN + NH$_2$}	& 3.41(-11) &  -0.9 & -9.9	& 1.25 & 25-761 & 1.38(-11) & -1.14 & 2 & 2.8(-11) & -0.85 & 0 & 1.4 \\
{\bf CN + NH$_3$ $\rightarrow$ NH$_2$CN + H}	& 1.38(-11) & -1.14 & 0 & 2 & 10-300& 1.3(-11) & -1.14 & 2 & 0 &0 & 0& - \\
\hline
{\bf O + C$_3$N $\rightarrow$ CO + C$_2$N} &	4(-11) & 0& 0  &2 & 10-300 &	4(-11) & 0 & 2	& 1.0(-10) & 0 & 0 & 4 \\
\hline
N + C$_4$N $\rightarrow$ CN + C$_3$N& 	1.0(-10) & 0 & 0 & 2 & 10-300	& 1.0(-10) & 0 & 2	& 1.0(-10) & 0 & 0 & 4 \\
\hline
N + C$_2$N $\rightarrow$ CN + CN& 	1.0(-10) & 0 & 0 & 2 & 10-300	& 1.0(-10) & 0 & 2	& 1.0(-10) & 0 & 0 & 4 \\
\hline
{\bf N + C$_2$H $\rightarrow$ C$_2$N + H}& 	1.7(-11) & 0 & 0 & 2 & 10-300	& 1.7(-11) & 0 & 2& 1.0(-10) & 0.18  & 0 & 4 \\
\hline
\end{tabular}
\end{center}
Note: units of rate coefficients are cm$^{3}$ s$^{-1}$. Temperature range of validity is 10 to 298~K for the new and the osu rate coefficients.\\
Rate coefficients for boldfaced reactions have been changed.\\
$^a$ $\gamma$ is 0 for all reactions. \\
$^b$ d(e) means $\rm d \times 10^e$\\
$^c$ Uncertainty factor means that $k$ is between $k$/F$_{\rm unc}$ and $k$*F$_{\rm unc}$ \\
$^d$ This rate coefficient is has been computed by \citet{Honvault:2009} for temperatures between 10 and 500~K. More studies are currently going on and will be reported by Bergeat et al. (in preparation).
\end{table}
\end{landscape} 

\begin{landscape} 
\begin{table}
\caption{Summary of the reaction review: Radiative association reactions. \label{tab4}}
\begin{center}
\begin{tabular}{l|llll|lll|llll}
\hline
\hline
Reaction & \multicolumn{4}{c|}{UDfA value} & \multicolumn{3}{c|}{OSU value} & \multicolumn{4}{c}{New value} \\
 & $\alpha^a$ & $\beta$   & F$_{\rm unc}$$^b$ & T range (K) & $\alpha^a$ & $\beta$ & F$_{\rm unc}$$^b$ & $\alpha^a$ & $\beta$  & $\gamma$ & 
 F$_{\rm unc}$$^b$ \\
\hline
\hline
C$^+$ + H$_2$ $\rightarrow$ CH$_2^+$ + h$\nu$	& 4(-16) &  -0.2 & 2 & 10-300& 4(-16) & -0.2 & 10 & 2(-16) &  -1.3 & 23 & 3 \\
\hline
C$_2$H$_2^+$ + H$_2$ $\rightarrow$ C$_2$H$_4^+$ + h$\nu$& 	2.34(-14) & -1.5 &	1.25 & 10-300& 1.5(-14) &  -1.0 & 10 & 2.9(-14) & -1.5 & 0 & 3 \\
\hline
CH$_3^+$ + H$_2$ $\rightarrow$ CH$_5^+$ + h$\nu$	& 1.3(-14) & -1.0 & 2 & 10-300 & 1.3(-14)& -1.0 & 10 &	4.1(-16) & -2.3 & 30 & 3\\
\hline
\end{tabular}
\end{center}
Note: units of rate coefficients are cm$^{3}$ s$^{-1}$. Temperature range of validity is 10 to 298~K for the new and the osu rate coefficients.\\
Rate coefficients for boldfaced reactions have been changed.\\
$^a$ d(e) means $\rm d \times 10^e$\\
$^b$ Uncertainty factor means that $k$ is between $k$/F$_{\rm unc}$ and $k$*F$_{\rm unc}$ \\

\end{table}

\begin{table}
\caption{Summary of the reaction review: Ion-neutral reactions. \label{tab5}}
\begin{center}
\begin{tabular}{l|lll|ll|llll}
\hline
\hline
Reaction & \multicolumn{3}{c|}{UDfA value} & \multicolumn{2}{c|}{OSU value} & \multicolumn{4}{c}{New value} \\
 & $\alpha^a$   & F$_{\rm unc}$$^b$ & T range (K) & $\alpha^a$ & F$_{\rm unc}$$^b$ & $\alpha^a$ & $\beta$  & $\gamma$ &  F$_{\rm unc}$$^b$ \\
\hline
{\bf H$_3^+$ + O $\rightarrow$ OH$^+$ + H$_2$}	 & 8.4(-10) & 1.5 & 10-41000 & 8.0(-10) & 1.5	& 7.98(-10) & -0.156 & 1.41 & 1.4 \\
{\bf H$_3^+$ + O $\rightarrow$ H$_2$O$^+$ + H }& 3.6(-10) & 1.5 & 10-41000 & 0.  & - & 3.42(-10) & -0.156 & 1.41 & 1.4 \\
\hline
\end{tabular}
\end{center}
Note: units of rate coefficients are cm$^{3}$ s$^{-1}$. Temperature range of validity is 10 to 298~K for the new and the osu rate coefficients.\\
Rate coefficients for boldfaced reactions have been changed.\\
$^a$ d(e) means $\rm d \times 10^e$\\
$^b$ Uncertainty factor means that $k$ is between $k$/F$_{\rm unc}$ and $k$*F$_{\rm unc}$ \\
\end{table}
\end{landscape} 

\begin{landscape} 
\begin{table}
\caption{Summary of the reaction review: Dissociative recombination.  \label{tab6}}
\begin{center}
\begin{tabular}{l|lll|ll|lll}
\hline
\hline
Reaction & \multicolumn{3}{c|}{UDfA value} & \multicolumn{2}{c|}{OSU value} & \multicolumn{2}{c}{New value} \\
 & k(T) & F$_{\rm unc}$$^a$ & T range & k(T) & F$_{\rm unc}$$^a$ & k(T) & F$_{\rm unc}$$^a$ &\\
\hline
& $2.83 \times 10^{-7} (\frac{T}{300})^{-0.65}$ & 1.25 & 10-300 & $6.2 \times 10^{-7} (\frac{T}{300})^{-0.65}$ & 1.25 & $2.83 \times 10^{-7} (\frac{T}{300})^{-0.65}$ & - \\
{\bf HCNH$^+$ + e$^-$ $\rightarrow$ HCN + H}	& rb=0.34	& & & rb=0.4 &&  rb=0.34 &  - \\
{\bf HCNH$^+$ + e$^-$ $\rightarrow$ HNC + H}	& rb=0.34 & & & rb=0.4	& & rb=0.34 & - \\
{\bf HCNH$^+$ + e$^-$ $\rightarrow$ CN + H	 + H} & rb=0.32 & & & rb=0.2	& & rb=0.32 & - \\
\hline
 & $3.0 \times 10^{-7} (\frac{T}{300})^{-0.5}$ & 2& 10-300&   $3.0 \times 10^{-7} (\frac{T}{300})^{-0.5}$ &  2&$2.0 \times 10^{-6} (\frac{T}{300})^{-0.7}$ & - \\
{\bf HC$_5$NH$^+$ + e$^-$ $\rightarrow$ C$_5$N + H$_2$}  & rb=0.5 & & & rb=0.5	& & rb=0.04& -  \\
{\bf HC$_5$NH$^+$ + e$^-$ $\rightarrow$ HC$_5$N + H}  & rb=0.5 & & & rb=0.5	& & rb=0.46 & - \\
{\bf HC$_5$NH$^+$ + e$^-$ $\rightarrow$ HCN + HC$_4$ } & rb=0 & & & rb=0& 	& rb=0.22&  - \\
{\bf HC$_5$NH$^+$ + e$^-$ $\rightarrow$ HNC + HC$_4$ } & rb=0 & & & rb=0	& & rb=0.22& -  \\
{\bf HC$_5$NH$^+$ + e$^-$ $\rightarrow$ HC$_3$N + HC$_2$ } & rb=0 & & & rb=0&	& rb=0.06 & - \\
\hline
 & $1.5 \times 10^{-7} (\frac{T}{300})^{-0.5}$ & 2 & 10-300 & $1.5 \times 10^{-7} (\frac{T}{300})^{-0.5}$  & 2 & $5.0 \times 10^{-7} (\frac{T}{300})^{-0.7}$ & - \\
{\bf H$_2$CO$^+$ + e$^-$ $\rightarrow$ HCO + H}  & rb=0.33 & & & rb=0.33	& & rb=0.30 & - \\
{\bf H$_2$CO$^+$ + e$^-$ $\rightarrow$ CO + H + H  }& rb=0.67 & & & rb=0.67 	& & rb=0.5 & - \\
{\bf H$_2$CO$^+$ + e$^-$ $\rightarrow$ CO + H$_2$ } & rb=0 & & & rb=0	& & rb=0.15 & - \\
{\bf H$_2$CO$^+$ + e$^-$ $\rightarrow$ CH$_2$ + O } & rb=0 & & & rb=0	& & rb=0.05&  - \\
\hline
 & $3.0 \times 10^{-7} (\frac{T}{300})^{-0.5}$ & 2 & 10-300 & $3.0 \times 10^{-7} (\frac{T}{300})^{-0.5}$ & 2 & $4.0 \times 10^{-7} (\frac{T}{300})^{-0.6}$ & - \\
{\bf CNC$^+$ + e$^-$ $\rightarrow$ CN + C } & rb=1 & & & rb=1	& & rb=0.95 & - \\
{\bf CNC$^+$ + e$^-$ $\rightarrow$ C$_2$ + N}  & rb=0 & & & rb=0	& & rb=0.05 & - \\
\hline
 & $5.6 \times 10^{-7} (\frac{T}{300})^{-0.76}$ & 2 & 10-300 & $5.6 \times 10^{-7} (\frac{T}{300})^{-0.76}$ & 2 & $5.6 \times 10^{-7} (\frac{T}{300})^{-0.76}$ & - \\
C$_2$H$_4^+$ + e$^-$ $\rightarrow$ C$_2$H$_3$ + H  & rb=0.11& &  & rb=0.11 & & rb=0.11 & - \\
C$_2$H$_4^+$ + e$^-$ $\rightarrow$ C$_2$H$_2$ + H$_2$  & rb=0.06& &  & rb=0.06 & & rb=0.06&  - \\
C$_2$H$_4^+$ + e$^-$ $\rightarrow$ C$_2$H$_2$ + H + H  & rb=0.66 & & & rb=0.66 & & rb=0.66 & - \\
C$_2$H$_4^+$ + e$^-$ $\rightarrow$ C$_2$H$_2$ + H + H$_2$ & rb=0.1 & & & rb=0.1 & & rb=0.1& -  \\
C$_2$H$_4^+$ + e$^-$ $\rightarrow$ CH$_4$ + C  & rb=0.01& &  & rb=0.01 & & rb=0.01 & - \\
C$_2$H$_4^+$ + e$^-$ $\rightarrow$ CH$_3$ + CH  & rb=0.02 & & & rb=0.02 & & rb=0.02 & - \\
C$_2$H$_4^+$ + e$^-$ $\rightarrow$ CH$_2$ + CH$_2$  & rb=0.04 & & & rb=0.04 & & rb=0.04 & - \\
\hline
 & $3.0 \times 10^{-7} (\frac{T}{300})^{-0.5}$ & 2 & 10-300 & $3.0 \times 10^{-7} (\frac{T}{300})^{-0.5}$ & 2 & $3.0 \times 10^{-7} (\frac{T}{300})^{-0.5}$ & - \\
HOSi$^+$ + e$^-$ $\rightarrow$ SiO + H  & rb=0.5 & & & rb=0.5 & & rb=0.5& - \\
HOSi$^+$ + e$^-$ $\rightarrow$ Si + OH  &  rb=0.5& & & rb=0.5&&  rb=0.5 & - \\
\hline
\end{tabular}
\end{center}
Note: units of rate coefficients are cm$^{3}$ s$^{-1}$; rb stands for branching fraction. Temperature range of validity is 10 to 298~K for the new and the osu rate coefficients.\\
Rate coefficients for boldfaced reactions have been changed.\\
$^a$ Uncertainty factor means that $k$ is between $k$/F$_{\rm unc}$ and $k$*F$_{\rm unc}$ \\

\end{table}
\end{landscape} 

\subsection{Impact of new rate coefficients}

We separate this discussion into four parts.  In the first part, we show the impact of the updated reactions listed in Tables~\ref{tab3}, \ref{tab4}, \ref{tab5} and \ref{tab6}. We compare the computed abundances with observations in dense clouds in the next part. Then two reactions, C + H$_2$ $\rightarrow$ CH$_2$ + h$\nu$ and C$^+$ + S $\rightarrow$ C + S$^+$, will be considered separately because a correct evaluation of their rate coefficients at low temperature requires detailed calculations which have not been carried out as part of the present study. We will however demonstrate how sensitive the abundances of the species  are to the rate coefficients of these reactions assuming reasonable minimum and maximum values.  For this section, we will use the following physical conditions: a temperature of 10 K, an H density of $2\times 10^4$~cm$^{-3}$, and an $A_{\rm V}$ of 10. 

\subsubsection{Impact of updated reaction coefficients}

\begin{figure}
\includegraphics[width=\textwidth]{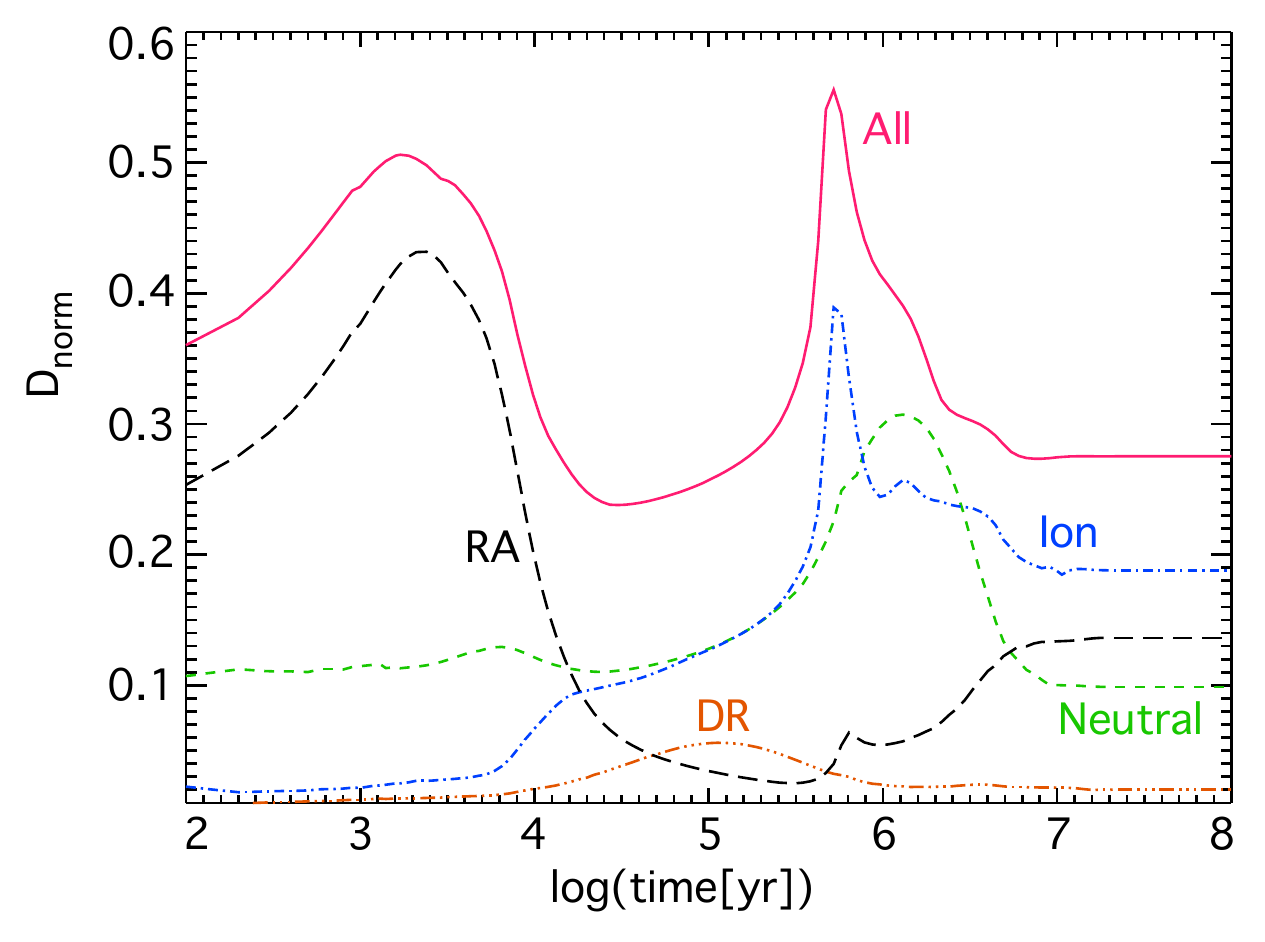}
\caption{The $D_{norm}$ parameter (see text) as a function of time when the different types of reactions are updated: DR stands for dissociative recombination; RA, for radiative association; Neutral, for neutral-neutral reactions; Ion, for ion-neutral reactions; and All, for the  case in which all rate coefficients are updated. } \label{D_tout}
\end{figure}

\begin{figure}
\includegraphics[width=\textwidth]{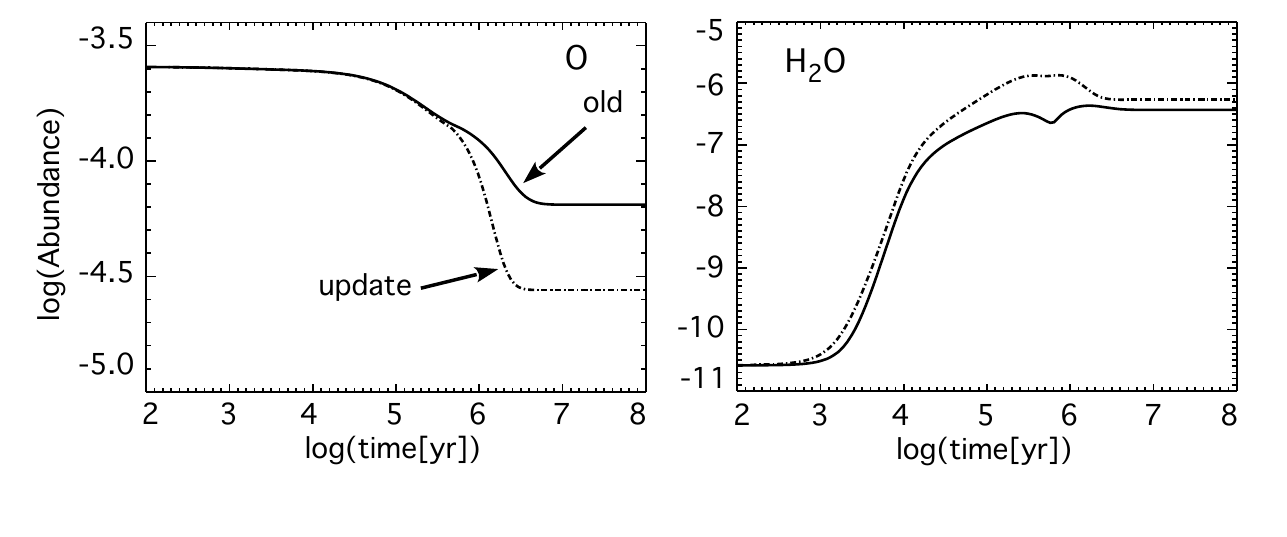}
\caption{O and H$_2$O abundances as a function of time predicted by the model in the case of the osu-09-2008 network (solid line) and when the ion-neutral reaction H$_3^+$ + O rate coefficient has been updated (dashed line). } \label{ion_mol}
\end{figure}

In order to study the impact of the new rate coefficients, we have calculated at each time a function $D_{norm}(t)$, defined by the equation  
\begin{equation}
 D_{norm}(t) = \frac{1}{N} \sum_j |{\rm log}(X^j_{\rm old}(t))-{\rm log}(X^j_{\rm update}(t))|
\end{equation}
with $X^j_{\rm old}$ the abundance of species $j$ computed with osu-09-2008 and $X^j_{\rm update}(t)$ the abundance of species $j$ computed with the updated networks. The sum $N$, which is 102 for all networks, is over all the species whose abundance is larger than $10^{-11}$ at $10^5$~yr. In addition to the osu-09-2008 network, we have used 5 other networks. In four of the five networks, we have updated separately each class of reaction: neutral-neutral reactions from Table~\ref{tab3}, radiative associations from Table~\ref{tab4}, ion-neutral reactions from Table~\ref{tab5} and dissociative recombinations from Table~\ref{tab6}. In the fifth network, we have updated all the types of reactions at the same time. 
 
The parameter $D_{norm}$ is shown in Fig.~\ref{D_tout} for the different networks. 
The changes in the rate coefficients for the  radiative association reactions are more important for times earlier than $10^4$~yr whereas ion-neutral and neutral-neutral reactions impact the abundances of the species especially between $10^5$ and $10^7$~yr. In particular, there is a peak of $D_{norm}$ at $5\times 10^5$~yr due to the change in the rate coefficient for the H$_3^+$ + O reaction (since it is the only updated ion-neutral reaction) that affects strongly the abundance of many species including C, C$^+$, O, O$_2$, SO, CH$_2$ and H$_2$O at that time (see Fig.~\ref{ion_mol}). This reaction is indeed the main destruction path of the atomic oxygen. The abundance of O is then decreased whereas the ones of H$_2$O and O$_2$ are increased. The update of the branching ratios does not affect strongly the results at steady-state since OH$^+$ is rapidly converted into H$_2$O$^+$ by reacting with H$_2$. 

When neutral-neutral reactions are updated, the species that are most affected are C$_2$, NO, C$_2$H, C$_3$H and C$_5$ (at $10^6$~yr). These species have been affected by the changes in the reaction rate coefficients of O + C$_2$, N + NO, O + C$_2$H and O + C$_3$H respectively as they are the most important reaction for each of these species.  Among these species, the sensitivity of the C$_5$ abundance is an example of the combined effect of different rate coefficients since we did not find a unique "key" reaction as done in Table~\ref{tab2}. 
The impact of the update in the rate coefficients for dissociative recombination is the smallest.

\subsubsection{Agreement with observations}

\begin{figure}
\includegraphics[width=\textwidth]{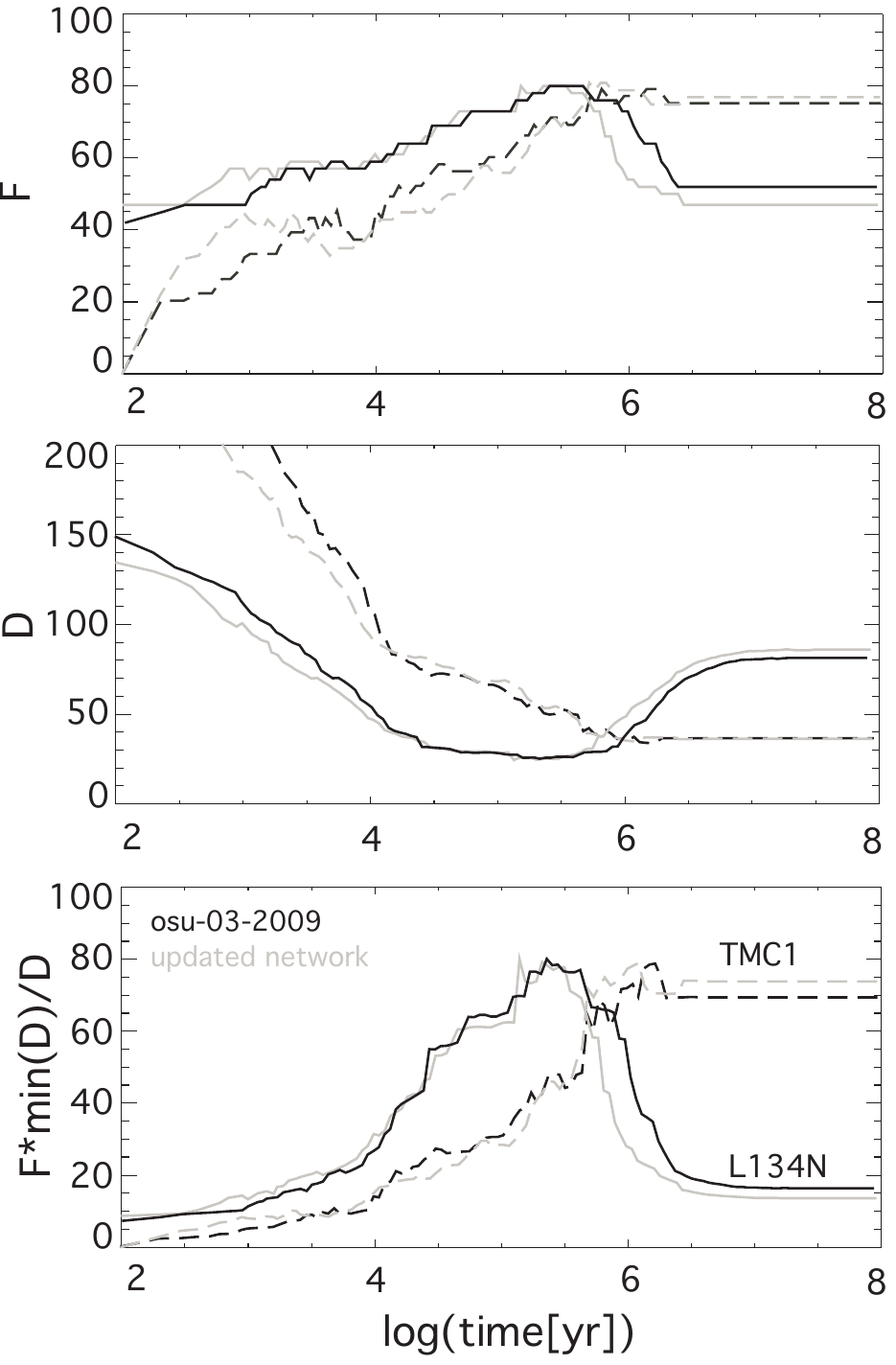}
\caption{Comparison between model and observations (see text for details). Solid lines are for L134N whereas dashed lines are for TMC1. Black lines show the results using the osu-09-2008 network whereas gray lines are used for the updated network.} \label{const_cloud}
\end{figure}

\begin{table}
\caption{Molecules  observed in L134N and TMC1 not reproduced by the models at the best ages indicated in the table. }
\begin{center}
\begin{tabular}{cc|cc}
\hline
\hline
\multicolumn{2}{c}{L134N} & \multicolumn{2}{c}{TMC1}  \\
 osu-09-2008 & update & osu-09-2008 & update \\
 $3\times 10^5$~yr & $2\times 10^5$~yr & $1\times 10^6$~yr & $5\times 10^5$~yr \\
 \hline
 NO & CN & CN & CN \\
 H$_2$S  &  H$_2$S  & OH & OH \\
 CH$_3$OH &  CH$_3$OH & CH$_3$OH & C$_2$H \\
 CH$_3$CHO & CH$_3$CHO & CH$_3$CHO & CH$_3$OH \\
 HC$_7$N & HC$_7$N & C$_3$H$_3$N & CH$_3$CHO \\
 HCO$^+$ & HCO$^+$ & C$_3$H$_4$ & C$_3$H$_3$N \\
HCS$^+$ &HCS$^+$ & HCS$^+$ & C$_3$H$_4$ \\
SiO & SiO & HCNH$^+$ & HCS$^+$ \\
 & & SiO & SiO  \\
 & & CH$_3$C$_3$N &  HC$_9$N \\
& & HC$_9$N & \\
 \hline
\end{tabular}
\end{center}
\label{comp_clouds}
\end{table}%

We compared observed and modeled abundances for the dense clouds L134N and TMC1, which are normally used as references for these studies \citep{2007A&A...467.1103G}. The observed abundances -- composed of 42 species in L134N and 53 species in TMC1 -- are given in Tables 3 and 4 of \citet{2007A&A...467.1103G}. Upper limits on SiO abundance are considered in addition \citep[$2.6\times 10^{-12}$ for L134N and $2.4\times 10^{-12}$ for TMC1,][]{1989ApJ...343..201Z}.   Since the abundances have been published without estimation of the observational error, we assume an error of a factor of three for all molecules. To obtain the error bars on the theoretical abundances, we used Monte Carlo simulations to study the propagation of uncertainties in reaction rate coefficients \citep{2005A&A...444..883W}. The error in the theoretical abundance (2$\sigma$) is then defined as the envelope containing 95.4\% of the abundance distribution. The agreement between observed and modeled abundances is then achieved when an overlap exists between these two quantities taking into account both types of error bars. Using this definition of the agreement, we computed three parameters as a function of time. The first parameter, labeled $F$,  is the percentage of molecules that is in agreement. The second parameter is the distance of disagreement $D$ (in log units, similar to $D_{norm}$ eq. 26 but not normalized) for the species where we do not have agreement. $D$ is the sum, over all such species, of the difference between modeled and observed closest error bars. For example, assuming that the observed abundance of a species is defined by $\rm X_{obs, min}$ and $\rm X_{obs, max}$ and the modeled one by $\rm X_{mod, min}$ and $\rm X_{mod, max}$ at time $t,$ if the abundance of a species is over-estimated by the model, the distance of disagreement for this species is $\rm |log(X_{mod, min}) - log(X_{obs, max})|$. Finally, the last parameter allows us to take into account both the agreement and the disagreement with the model by multiplying F(t) with min($D$)/$D$(t), with min($D$) the minimum of $D$ over all times.  This last parameter tends to sharpen the times at which agreement is best.

We computed the three parameters for four different models. For both clouds, we used the same physical parameters: a temperature of 10~K, an H$_2$ density of $10^4$~cm$^{-3}$, an $A_{\rm V}$ of 10 and a $\zeta_{H_2}$ of $1.3\times 10^{-17}$~s$^{-1}$.   All species are initially in  atomic form, except for hydrogen, which is assumed to be already completely molecular. Different elemental abundances were chosen for the two clouds. For L134N, we used the typical low-metal elemental abundances \citep[see Table 1 in][]{2008ApJ...680..371W}. For TMC1, the observation of large quantities of C-rich species seems to indicate that the C/O elemental ratio is larger than unity. As in previous studies, we lowered the elemental abundance of oxygen to obtain C/O equal to 1.4. For both clouds, two networks were tested: osu-09-2008 and the updated network described in the previous section. Fig.~\ref{const_cloud} shows the three parameters (F, $D$ and F*min($D$)/$D$) as a function of time for the two clouds and each network. The best agreement for L134N is not changed significantly. The best time is still around $(2-3)\times 10^5$~yr using both networks. For TMC1, we have an agreement only slightly better: 79\% of the observed species are reproduced with the updated network against 77\% with osu-09-2008. Specifically, more CH$_3$C$_3$N and HCNH$^+$ are now produced, moving their predicted abundances closer to the observed values, whereas C$_2$H is now underproduced by the model. The best times are also slightly shifted to earlier times: $4.7\times 10^5 - 6.3\times 10^5$~yr using the updated network and $5.7\times 10^5 - 1.5\times 10^6$~yr with osu-09-2008 although the agreement stays high until steady-state. Species not reproduced by the different models are listed in Table~\ref{comp_clouds} for both clouds. 


\subsubsection{The C + H$_2$ association reaction}

\begin{figure}
\includegraphics[width=\textwidth]{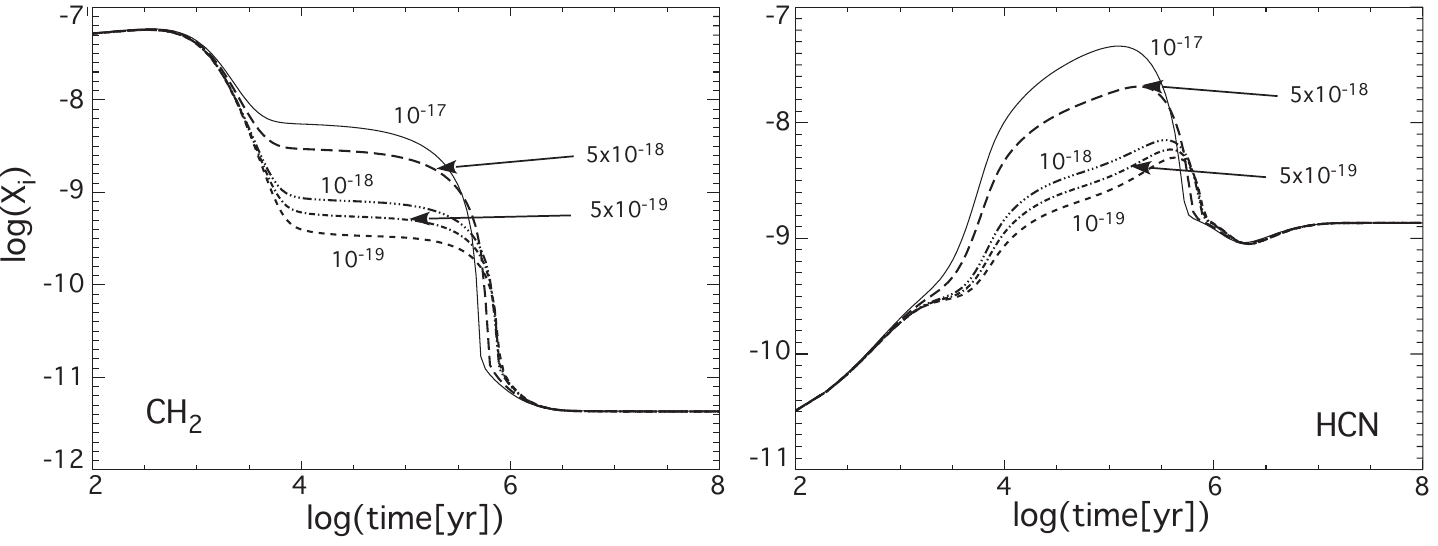}
\caption{CH$_2$ and HCN abundances as a function of time for the dense cloud conditions. Different curves were obtained with different values of the C + H$_2$ $\rightarrow$ CH$_2$ + h$\nu$ reaction rate coefficient. The units of the rate coefficient are cm$^3$ s$^{-1}$.} \label{C_H2}
\end{figure}

The radiative association reaction C + H$_2$ $\rightarrow$ CH$_2$ + h$\nu$ has been found to be the most important one for dense cloud chemistry in the 10$^{5-6}$ yr era, since it contributes significantly to the production of CH$_2$, one of the first carbon-containing molecules. The rate coefficient for this reaction recommended in the OSU and UDfA databases is $10^{-17}$ cm$^3$ s$^{-1}$ and comes from \citet{1980ApJS...43....1P}. An estimate based on the supposition that radiative stabilisation involves only infrared transitions between the vibrational levels of CH$_2$ (see the datasheet in KIDA for more details) yields a value of $6\times 10^{-19}$ cm$^3$ s$^{-1}$ at 10~K. However, it is possible that the internal levels close to the dissociation limit of CH$_2$ are of mixed electronic character allowing an increase in the spontaneous emission rate from these levels. Therefore, we have carried out calculations with the value of this rate coefficient varied between $10^{-19}$ and $10^{-17}$ cm$^3$ s$^{-1}$ in order to examine the sensitivity of the results to this reaction. Note that more detailed calculations are needed for this reaction in order to have a better estimate.

In Fig.~\ref{C_H2}, we show the CH$_2$ and HCN abundances as a function of time calculated by the updated model when different values of the rate coefficient for formation of CH$_2$ are employed. HCN is an example of molecules observed in dense clouds and its second main production reaction at $10^5$~yr is N + CH$_2$. The results are shown for five values of the C + H$_2$ $\rightarrow$ CH$_2$ + h$\nu$ rate coefficient: $10^{-19}$, $5\times 10^{-19}$, $10^{-18}$, $5\times 10^{-18}$ and $10^{-17}$ cm$^3$ s$^{-1}$. The abundances of the species are much more sensitive to the rate coefficient when it is larger than $10^{-18}$. For most molecules,  the abundances calculated by the model are smaller between $10^4$ and $10^6$~yr for smaller rate coefficients. To reproduce the observed abundance of HCN  in TMC-1, which is $2\times 10^{-8}$ \citep{1992IAUS..150..171O}, a rate coefficient larger than $10^{-18}$ cm$^3$ s$^{-1}$ is better. It is however risky to constrain rate coefficients by comparing the modeled abundances with the ones observed in the interstellar medium since the differences that are found when a single rate coefficient is changed will depend on the values of the other rate coefficients and of the reactions currently present in the network. The change in other rate coefficients or the addition or removal of reactions from the network may change the "best" rate coefficient for C + H$_2$ $\rightarrow$ CH$_2$ + h$\nu$. Detailed quantum calculations would be required to know at least if the rate coefficient of this reaction is larger than $10^{-18}$ cm$^3$s$^{-1}$ and such detailed calculations are beyond the scope of this article.

\subsubsection{The C$^+$ + S charge exchange reaction}

\begin{figure}
\includegraphics[width=\textwidth]{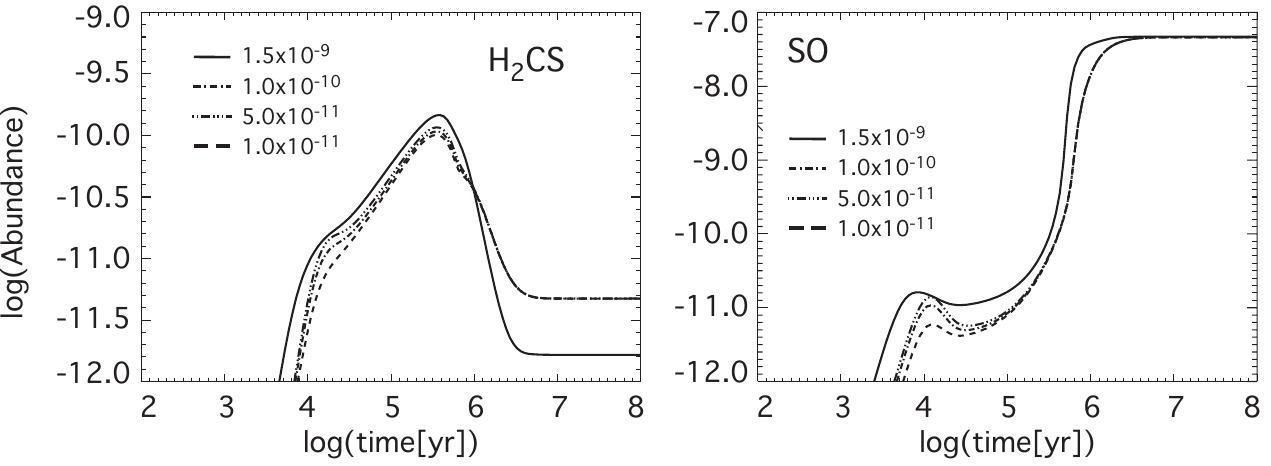}
\caption{H$_2$CS and SO abundances as a function of time for cold dense clouds.  Different curves were obtained with different values of the C$^+$ + S $\rightarrow$ S$^+$ + C reaction rate coefficient: $1.5\times 10^{-9}$ (osu-09-2008), $1\times 10^{-10}$, $5\times 10^{-11}$ and $1\times 10^{-11}$ cm$^3$ s$^{-1}$. } \label{C+_S}
\end{figure}

\begin{figure}
\includegraphics[width=\textwidth]{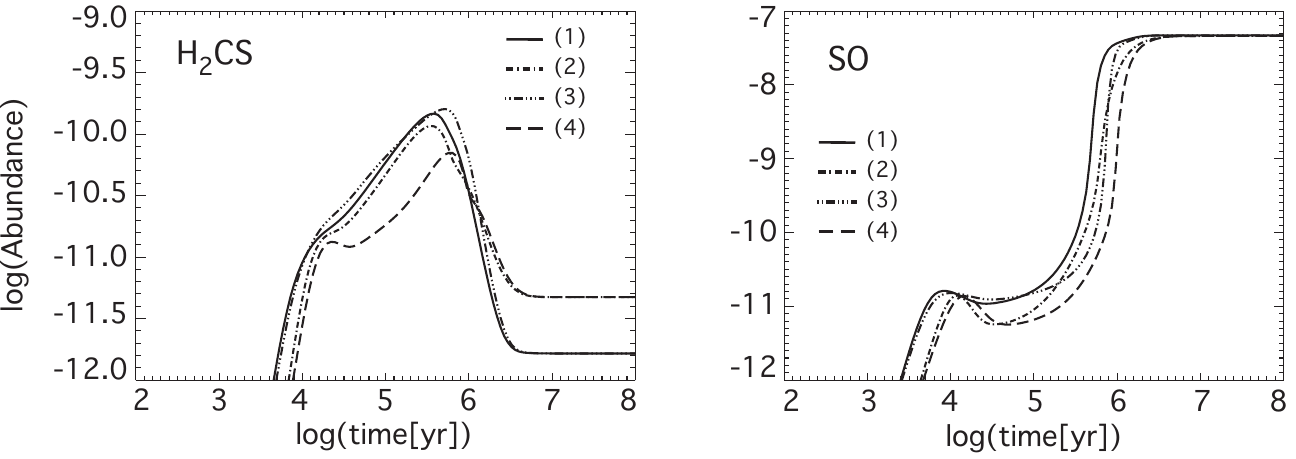}
\caption{H$_2$CS and SO abundances as a function of time for cold  dense clouds. Different curves were obtained with different values of two rate coefficients:  (1) the rate coefficients of C + H$_2$ and C$^+$ + S are $10^{-17}$ and $1.5\times 10^{-9}$ respectively, (2) C + H$_2$ is $10^{-17}$ and C$^+$ + S is $10^{-10}$, (3) C + H$_2$ is $10^{-18}$ and C$^+$ + S is $1.5\times 10^{-9}$ and (4) C + H$_2$ is $10^{-18}$ and C$^+$ + S is $10^{-10}$. The units of the rate coefficient are cm$^3$ s$^{-1}$.  \label{C+_S_C_H2}}
\end{figure}

With an accurate quantum chemical treatment for the potential surface, followed by semi-classical dynamics, the rate coefficient for the C$^+$ + S charge transfer reaction has been calculated to be $5.7\times 10^{-12}$ cm$^3$s$^{-1}$ at 1000~K and $2\times 10^{-12}$ at 500~K \citep{2008CPL...467...28B}.  These calculations at high temperature tend to show that the $1.5\times 10^{-9}$ cm$^3$s$^{-1}$ value used in the astrochemical models for the 10-100 K temperature range is overestimated. A definitive conclusion needs a treatment that takes into account the quantum effects at very low temperature, which could increase, probably not drastically, the C$^+$ + S rate coefficient. In the meantime,  we suggest, for temperatures between 10 and 100 K,  upper and lower limits for the rate coefficient of respectively $10^{-10}$ and $10^{-11}$ cm$^3$ s$^{-1}$.

In Fig.~\ref{C+_S}, we show the abundance of  H$_2$CS and SO as  functions of time calculated for four values of the rate coefficient: $1.5\times 10^{-9}$ (osu-09-2008), $1\times 10^{-10}$, $5\times 10^{-11}$ and $1\times 10^{-11}$ cm$^3$s$^{-1}$. The sensitivity does not appear to be very strong. If we, however, also vary the rate coefficient of C + H$_2$ , we then have a significant combined effect, warned about in the previous section.   This is shown in Fig.~\ref{C+_S_C_H2} where the abundances of the species have been computed with four sets of rate coefficients (in units of cm$^{3}$ s$^{-1}$) for C + H$_2$ and C$^+$ + S, respectively: (1) those in osu-09-2008, i.e. $10^{-17}$ and $1.5\times 10^{-9}$, 2) $10^{-17}$ and  $10^{-10}$, 3) $10^{-18}$ and  $1.5\times 10^{-9}$ and 4) $10^{-18}$ and $10^{-10}$.

\section{IRC +10216 chemistry}

In addition to the chemistry of cold cores, the chemistry of certain 
circumstellar regions is also of interest. IRC +10216  is a relatively nearby  carbon-rich (elemental [C]/[O] $>$ 1) asymptotic giant branch (AGB) star
which is undergoing a mass loss process that produces an extended
circumstellar envelope. 
It is also one of the richest molecular sources in the sky with more
than 70 molecules discovered to date in its circumstellar gas. 
The hot and dense regions near the stellar surface
contain stable carbon-bearing molecules such as C$_2$H$_2$, HCN,
CS, ... with abundances determined by thermodynamic rather than kinetic considerations. As the
gas expands it starts to be exposed to the interstellar
ultraviolet radiation field so that molecules are
photodissociated, and the resulting radicals participate in rapid
gas phase chemical reactions forming new molecules. {\bf The fact that
the chemistry of this source is relatively simple in terms of
geometry (it may be modelled as an uniform
spherical outflow) makes IRC +10216 an ideal target on which to
apply and test gas phase chemical models \citep{1993ApJ...410..188C,1994A&A...288..561M,1998ApJ...502..898D,2000MNRAS.316..195M,2006ApJ...650..374A,2009ApJ...697...68C}. The reader may notice however that uncertainties in the initial chemical composition (especially the initial abundance of C$_2$H$_2$) and UV field may also affect the computed species abundances as already noticed by \citep{1994A&A...288..561M}. }

\subsection{Chemical model and sensitivity analysis}

Here we report a sensitivity analysis in parallel to the one for cold cores.
We computed the chemical composition of the gas as
it expands from a radius of 10$^{15}$ cm to one of
10$^{18}$ cm, while the  gas density decreases as 
the inverse square of radius and the temperature as a power law of it
with the exponent depending on the distance from the central star \citep{2006ApJ...650..374A}.

We utilized the udfa06 chemical network \citep{2007A&A...466.1197W}
with the rate coefficient uncertainties
compiled in this database.  Because IRC +10216 harbors the largest
variety of molecular anions yet observed and because of the impact of
reactions with anions for the chemistry of this source \citep{2009ApJ...697...68C}, the chemical network was enhanced by a set of
reactions involving anions taken from the osu\underline{
}01\underline{ }2009\footnote{See
\texttt{http://www.physics.ohio-state.edu/$\sim$eric/research.html}}
database, for which we assumed an uncertainty in the rate coefficients
of a factor of 2. {\bf The photodissociation rate of CO was computed including self-shielding according to the treatment of \citet{1988ApJ...328..797M}, and the error was assumed to be low (a factor 1.25).} The sensitivity analysis was carried out by
running 1000 different chemical models in which the reaction rate
constants are randomly varied within their uncertainties using a lognormal distribution \citep{2005A&A...444..883W}.



\begin{figure}
\includegraphics[angle=-90,width=\textwidth]{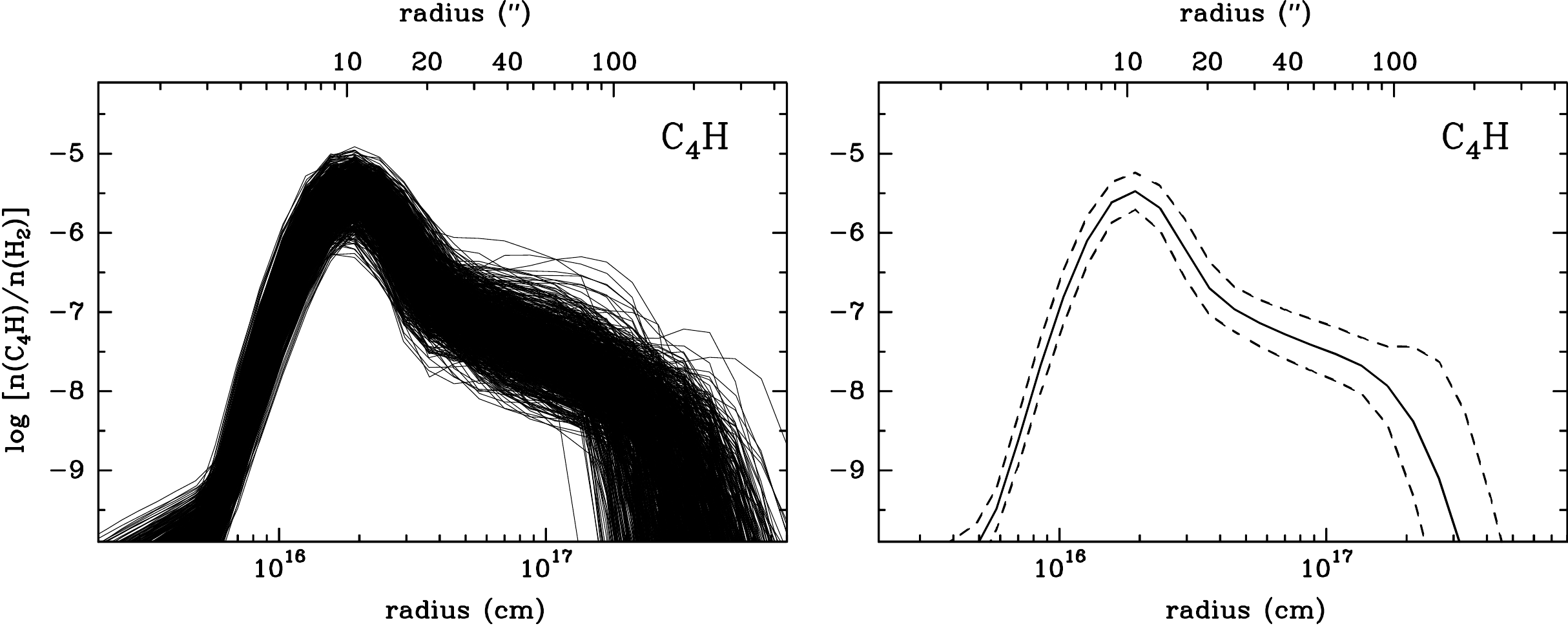}
\caption{Abundance of C$_4$H in IRC +10216 as a function of
radius. \emph{Left}: as calculated with the 1000 chemical models.
\emph{Right}: mean abundance (solid line) and 1$\sigma$ standard
deviation (dashed lines).} \label{fig-c4h-abun}
\end{figure}

As an example, in Fig.~\ref{fig-c4h-abun} we
show as a function of radius the abundance distribution as well as the mean abundance and
 1$\sigma$ deviation of C$_4$H, a species
which is formed in the outer envelope.  The
uncertainty in the abundance increases with radius, or with time
(in an expanding envelope a given radial position may be
interpreted as a certain temporal instant), which occurs because
 the dispersion in the abundances due to
the rate coefficient uncertainties gets larger.

\begin{figure}
\includegraphics[width=\textwidth]{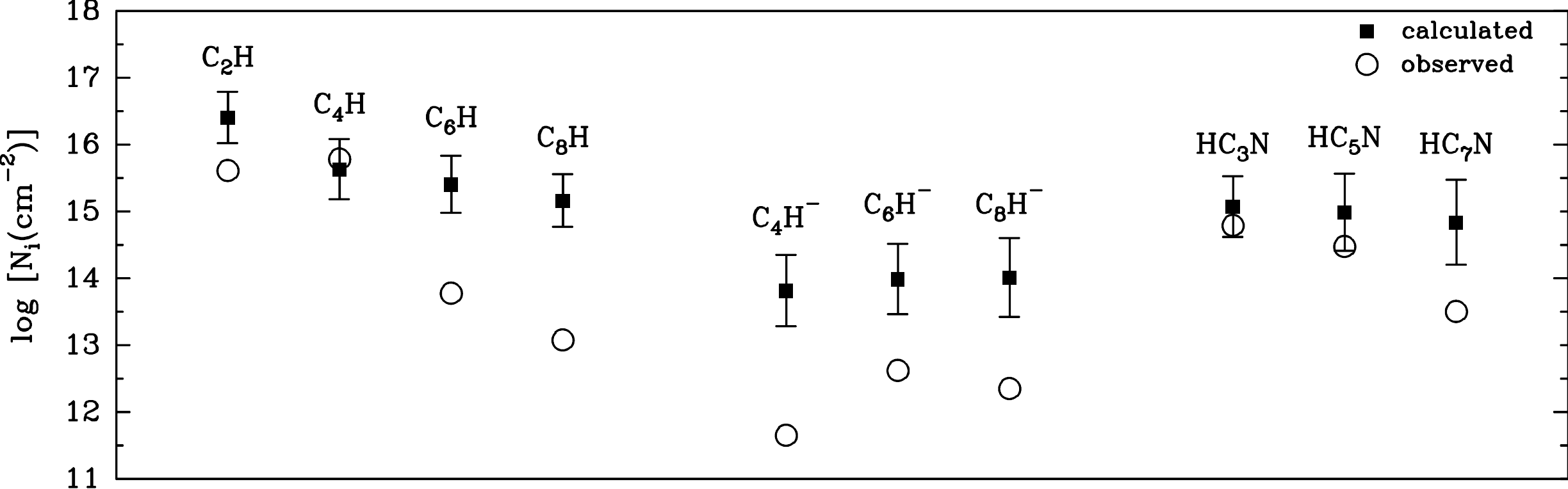}
\caption{Calculated (squares) and observed (circles) column densities of various
carbon-chain molecules in IRC +10216.} \label{fig-cd-irc10216}
\end{figure}

In the case of a circumstellar envelope such as IRC +10216, it is
more appropriate to discuss the uncertainty propagation of the
reaction rate coefficients in terms of the total column densities
rather than  the abundance at a given radius.  In
Fig.~\ref{fig-cd-irc10216} we plot the total column densities
(twice the radial column densities from 10$^{15}$ to 10$^{18}$ cm)
calculated with the chemical model for various carbon-chain
molecules, with their 1$\sigma$ error due to the uncertainties in
the reaction rate coefficients, and compare them with the observed
column densities. We note that the observed column densities
for long carbon-chains are noticeably lower than the calculated
ones within their respective errors, something which was also found
to occur in previous chemical models of IRC +10216 \citep{2000MNRAS.316..195M} and which is probably due to incompleteness in the
chemical network or to wrong reaction rates of key reactions
involving these species.
The observed column densities obviously have
 errors but are less likely to be the cause of large
disagreements unless some fundamental molecular parameters such as
the permanent dipole moment are severely wrong.

\subsection{Most critical reactions}

\begin{table}
\begin{center}
\footnotesize \caption{List of key reactions for the
chemistry of IRC +10216.  
\label{table-crit-reac}}
\begin{tabular}{lrrl}
\hline\hline
Reaction $i$                                   & F$_{\rm unc}$ & $\sum_j P_j^i$ & Affected species $j$ \\
\hline
N$_2$ + h$\nu$ $\rightarrow$ N + N                     & 10~~ & 12.9         & N, N$_2$, HC$_2$N, C$_3$N, C$_3$N$^-$, C$_2$N, SiNH, ... \\
C$_2$H$_2$ + h$\nu$ $\rightarrow$ C$_2$H + H           & 2~~  & 12.7         & C$_2$H$_2$, C$_4$H$_2$, C$_3$H, C$_8$H$_2$, HC$_3$N, c-C$_3$H$_2$, ... \\
C$_2$ + h$\nu$ $\rightarrow$ C$_2^+$ + e$^-$           & 2~~  & 12.4         & C$_2$H$_2^+$, C$_2$H$^+$, C$_2^+$, C$_2$H$_3^+$, H, C$_2$H$_4^+$, ... \\
C$_2$H + h$\nu$ $\rightarrow$ C$_2$ + H                & 10~~ & 10.5         & C$_2$H, C$_3$H$_3$, C$_2$, C$_3^+$, C$_4$H$_3$, C$_2$S, ... \\
C + h$\nu$ $\rightarrow$ C$^+$ + e$^-$                 & 2~~  & 9.3          & C$^+$, e$^-$, C, HCP, SiS, H$_3$O$^+$, HCCO$^+$, P, ... \\
H$_2$ + cosmic ray $\rightarrow$ H$_2^+$ + e$^-$       & 2~~  & 8.0          & H$_3^+$, N$_2$H$^+$, HNC, NH$_4^+$, HCO$^+$, HCNH$^+$, ... \\
HCN + h$\nu$ $\rightarrow$ CN + H                      & 1.5  & 7.8          & HCN, HC$_3$N, CNC$^+$, C$_2$H$_3$CN, SiH$_3$, HC$_5$N, ... \\
C$_2$H + C$_4$H$_2$ $\rightarrow$ C$_6$H$_2$ + H       & 2~~  & 7.1          & C$_6$H$_2$, C$_7$H$^+$, C$_7$H, C$_4$H$_2$, C$_6$H, ... \\
CS + h$\nu$ $\rightarrow$ C + S                        & 2~~  & 4.3          & CS, S, HC$_4$S$^+$, C$_4$S, H$_2$CS \\
Si + h$\nu$ $\rightarrow$ Si$^+$ + e$^-$               & 2~~  & 3.7          & Si$^+$, Si, SiNH$_2^+$, SiN$^+$, SiC$_3$, ... \\
C$_2$H + C$_6$H$_2$ $\rightarrow$ C$_8$H$_2$ + H       & 2~~  & 3.4          & C$_8$H$_2$, HC$_9$N, C$_6$H$_2$, C$_7$H, C$_6$H, HC$_7$N, ... \\
O + C$_2$ $\rightarrow$ CO + C       & 2~~  & 3.0          & C$_2$H$_2^+$, C$_2$H$^+$, C$_2^+$, H, C$_2$, C$_4$H$^+$, ... \\
C$^+$ + SiS $\rightarrow$ SiS$^+$ + C       & 2~~  & 3.0          & SH, SiS$^+$, SO, S$_2$, SO$^+$, SiS, ... \\
NH$_3$ + h$\nu$ $\rightarrow$ NH$_2$ + H               & 1.5  & 2.7          & NH$_3$, NH$_2$, H$_2$CN$^+$, HNO, NH$_2$CN \\
C$_4$H$_2$ + h$\nu$ $\rightarrow$ C$_4$H$_2^+$ + e$^-$ & 2~~  & 2.6          & C$_4$H$_2^+$, C$_6$H$_4^+$, CH$_3$C$_4$H$^+$, C$_8$H$_3^+$, H$_3$C$_9$N$^+$, C$_4$S \\
CN + SiH$_4$ $\rightarrow$ HCN + SiH$_3$              & 1.5  & 2.5          & SiH$_4$, SiH$_3$, H$_2$SiO, SiCH$_2$, H$_2$SiO$^+$ \\
S + h$\nu$ $\rightarrow$ S$^+$ + e$^-$                 & 2~~  & 2.4          & C$_4$S$^+$, HC$_2$S$^+$, S$^+$, C$_3$S$^+$, NS$^+$ \\
CN + C$_4$H$_2$ $\rightarrow$ HC$_5$N + H              & 2~~  & 2.4          & HC$_5$N, C$_5$N, H$_2$C$_5$N$^+$, CH$_3$C$_5$NH$^+$, CH$_3$C$_5$N \\
H$_2$ + SiC$_2$H$^+$ $\rightarrow$ SiC$_2$H$_3^+$ + h$\nu$            & 2~~  & 2.3          & SiC$_2$H$_3^+$, SiC$_2$H$_2$, SiC$_2$H, SiC$_3$H$_2^+$, SiC$_3$H \\
\hline
\end{tabular}
\end{center}
\end{table}

We have used the
Pearson correlation coefficients $P_j^i$ (defined in an analogous manner
to
eq.~(\ref{Pcc})) with column
densities instead of abundances. Table~\ref{table-crit-reac}
collects the most critical reactions for the chemistry of IRC
+10216, as defined by  the largest $\sum_j P_j^i$ where the sum
is restricted to species $j$ with a column density in excess of
10$^{10}$ cm$^{-2}$ and with $P_j^i > 0.3$. Here we have used the second method discussed in section 5.1, i.e. the sum of $P_j^i$ for each reaction.
The list of reactions
is dominated to a large extent by photoprocesses, since the
chemistry of the outer envelope of IRC +10216 is very similar to that found in
photon dominated regions, except that it is a
carbon-rich chemistry.   Also critical are some
neutral-neutral reactions as well as the cosmic ray ionization of
H$_2$, and to a lesser extent ion-neutral reactions. Reactions
involving anions are not very critical compared with the types of
reactions mentioned above, according to the adopted criteria.
Among them the electron radiative attachment to large radicals to
form the corresponding anions are the most critical ones, affecting
mainly the column densities of both the radical and the
resulting anion. {\bf The CO photo-dissociation rate is not very important in our case because of shelf-shielding effects and its low uncertainty (F=1.25). }

The types of reactions critical in the circumstellar gas
of IRC +10216 are markedly different from those for dark clouds such
as TMC-1, where ion-neutral reactions dominate the chemistry and
photoprocesses are inhibited by the large visual extinction of the
cloud. From Table~\ref{table-crit-reac} it seems clear that a
critical review of the photoprocesses, currently
evaluated as a function of the visual extinction through
parametric expressions, is the best initial step in order to
reduce the global uncertainty in the chemical composition of IRC
+10216 as calculated by chemical modelling.

\section{Surface chemistry in models}

Interstellar chemical models that contain both gas-phase and grain-surface chemistry are known as gas-grain models.  In the standard approach, reactions on grain surfaces are treated by rate equations, although the rate coefficients are different from their gas-phase counterparts \citep{1992ApJS...82..167H}.  More advanced methods utilize stochastic approaches, since these are more accurate when the average number of reactive species on an individual grain approaches unity, and handle irregular surfaces more appropriately, as discussed in Section~\ref{sec:1}.  In this discussion, we emphasize the basic rate equation method and its uncertainties.   In dealing with small particles, it is useful to  define concentration in terms of numbers of a given species per grain.  Other possibilities are to use the bulk concentration of adsorbates and fractional abundances relative to gas-phase hydrogen density.  The former is obtained by multiplying the number of species per grain by the  grain concentration.  More common units in surface science include numbers of monolayers, and areal concentration (cm$^{-2}$). 

Consider a dust particle with $N$ surface sites,  defined as minima in a potential surface where adsorption takes place preferentially.  Here, we limit consideration  to the diffusive or Langmuir-Hinselwood mechanism of reaction in which two molecules undergo diffusion until finding one another and reacting.      Whether the reactants are accreted from the gas, or form on the surface from processes such as photodissociation, they are assumed to be thermalized before they react.    If  there are $N(A)$ molecules of species A and $N(B)$ molecules of species B on a grain, and if these two species react to form molecule C when both species occupy the same potential minimum, the rate of formation of C is given by the equation \citep{1992ApJS...82..167H,1998ApJ...499..234C}.
\begin{equation}
\label{diff}
dN(C)/dt = ( k_{\rm hop,A} + k_{\rm hop,B})N(A)N(B)/N,
\end{equation}
where $k_{\rm hop,i}$, the rate of hopping of species $i$ over the potential barrier between two adjacent sites, is itself given by the equation
\begin{equation}
k_{\rm hop,i} = \nu _{0} \exp({-E_{\rm b,i}/T}).
\end{equation}
Here $\nu _{0}$ is the so-called attempt frequency (s$^{-1}$) and $E_{\rm b}$ is the energy barrier in  units of K.   
Equation~(\ref{diff}) can be simplified to
\begin{equation}
\label{rx}
dN(C)/dt = ( k_{\rm diff,A} + k_{\rm diff,B})N(A)N(B) =  k_{\rm AB}N(A)N(B),
\end{equation}
where $k_{\rm diff,i} = k_{\rm hop,i}/N$  is the hopping rate over a number of sites equivalent to the whole grain.  If there is a chemical activation energy in addition to the diffusion barrier, then a factor $\kappa < 1$ must be included. The simplest expression for $\kappa$ is a simple Boltzmann factor containing the chemical activation energy, although more complex expressions involving the competition between diffusion and chemical reaction can also be utilized \citep{Herbst:2008}.

Equations for the rate of appearance/disappearance of all surface and gas-phase species in a gas-grain network must include terms for adsorption of gas-phase species onto dust particles, and desorption of surface species into the gas phase, in addition to chemical reactions of formation and destruction.  The rate of adsorption, in units of the number of species A sticking per single grain per unit time is given by the expression
\begin{equation}
dN(A)/dt = S_{\rm A}(T) v_{\rm A} \sigma _{\rm gr} {\rm [A]},
\end{equation}
where $S_{\rm A}$ is the sticking coefficient, which is a function of temperature, $v_{\rm A}$ is the velocity of species A in the gas-phase, $\sigma_{\rm gr}$ is the cross section of the grain, and [A] is the concentration of species A in the gas-phase.  The equation for desorption depends upon the mechanism; for thermal evaporation, which is normally unimportant in the cold interstellar medium but increases in importance at higher temperatures \citep{2004MNRAS.354.1133C}, it is
\begin{equation}
dN(A)/dt = -k_{\rm evap,A} N(A),
\end{equation}
where the rate coefficient is the same as that for hopping except that the barrier is replaced by the desorption energy $E_{\rm D}$, which is generally considerably larger than $E_{\rm b}$: $k_{\rm evap,A} = \nu_0 \exp{(\rm -E_D/T)}$.

The important parameters in these physical processes are the attempt frequencies, the energy barriers, and the desorption energies.   These parameters are also used in stochastic approaches.  They are more or less obtainable by a variety of methods in surface science, some of which are mentioned below, although estimates are often utilized, especially for the diffusion barriers.  The result is that calculated abundances of surface species are themselves often estimates to an order of magnitude. It would appear that the uncertainty and sensitivity analyses discussed for gas-phase reactions here can be used for surface reactions as well, but there are many additional complications including other mechanisms for reaction than the diffusive one, and rough surfaces instead of the smooth ones with single values for energy barriers and desorption energies.  As more is learned about the chemistry on the surfaces of small particles, it may well be that uncertainty and sensitivity analyses will be useful.

Stochastic methods used in place of the rate equation technique can be subdivided into macroscopic and microscopic (kinetic Monte Carlo) methods.  These are considered below in the discussion of H$_{2}$ formation.


\section{Surface Chemistry:  Experiment}

As an introduction, it must be mentioned that, unlike the situation in the gas phase, the results of laboratory experiments on surface reactions are not as easily converted to rates under interstellar conditions for a variety of reasons.  To study surface chemistry, especially at low temperatures, various research groups have performed a variety of experiments on cold surfaces such as olivine, amorphous carbon, and assorted ices.  In general, the experiments are not carried out on small particles.  The best characterized reaction is the formation of hydrogen molecules from hydrogen atoms \citep{1999ApJ...522..305K}.  The complexities of this process are discussed below.


In addition to H$_{2}$ formation, surface reactions in cold cores are thought to produce simple ices.  
Such ices, detected by comparison of astronomical infrared observations with laboratory data, are comprised primarily of water, but also contain CO, CO$_2$
and several other species, such as CH$_3$OH, NH$_3$, HCOOH, and CH$_4$ in different mixing ratios \citep{2000IAUS..197..135E}.  The formation of more complex species on surfaces, via a variety of mechanisms, is also likely.  In cold sources, reactions between atoms that accrete onto the surface and ices may lead eventually to complex species \citep{2000IAUS..197..135E,2005IAUS..231..237C}.   It is now also thought that terrestrially common organic species, such as dimethyl ether and methyl formate, found in the gas phase of  hot cores, are formed by surface processes involving the combination of radicals, which are produced photochemically from the simpler ices, especially on ices of methanol \citep{2009A&A...494L..13O}.   This process is  followed by desorption as star formation converts a cold core into a hot one \citep{2008ApJ...682..283G}.     The experimental
study of analogs of  these icy grains in space is the topic of a rather new research
field in laboratory solid-state astrochemistry.  

\subsection{H$_{2}$ Formation on Surfaces}

Molecular hydrogen is the most abundant molecule in the ISM. It is a precursor for the formation of more complex molecules through protonated molecular hydrogen, H$_3^+$, serves as an important coolant, and shields itself and other molecules from the harsh interstellar radiation field. This shielding arises from the unusual two-step photodissociation mechanism of H$_{2}$ in the Lyman and Werner  transitions from 912-1100 \AA, which is discrete rather than continuous, so that specific small wavelength regions are depleted strongly.  Under most circumstances, the most efficient formation route of H$_2$ in the gas phase is through three-body reactions where the third body can take some of the 4.5~eV of excess energy that is released in the reaction. In the dilute conditions of the ISM, three body collisions do not occur and these reactions are not possible. Already in 1949 it had been suggested that surface processes on dust particles could play a role in H$_2$ formation~\citep{1949QB4.U8l11p2....}. The grain surface would in this case act as the third body. 

The surface formation of molecular hydrogen can roughly be divided into two regimes. At low temperatures, the hydrogen atoms are weakly bound, or physisorbed, to the grain surfaces and can still easily move on the surface. Once two atoms meet, they will form molecular hydrogen through a barrierless reaction. If the surface temperature becomes too high, the residence time of atoms on the surface becomes so short that, at the low fluxes in the ISM, the chance of two atoms meeting becomes negligible. At these higher temperatures, H$_2$ formation has therefore to proceed through chemisorbed hydrogen atoms. Observations show
indeed that molecular hydrogen is also formed in warmer areas ($>$ 20 K) such as PDRs and
post-shock regions. Here we will make a clear distinction between the two mechanisms and we will discuss the main uncertainties related to both cases.

\subsubsection{Low temperature}
The H$_2$ formation process from physisorbed atoms has been intensely studied over the past years. 
These studies have focused on three different surfaces: silicates, carbonaceous material and water ice. Experimental studies are geared towards obtaining binding energies \citep{1997ApJ...475L..69P,1997ApJ...483L.131P, 1999A&A...344..681P, 2003Sci...302.1943H, 2005JChPh.122l4701H, 2005ApJ...627..850P, 2005CPL...404..187D, 2008A&A...492L..17M,2009PCCP...11.4396F} or excitation energies of the desorbing H$_2$ molecules \citep{2003Ap&SS.285..769P, 2009MNRAS.397L..96C}. Theoretical studies have used the binding energies as input parameter and have focused on the influence of surface roughness and amorphicity \citep{2005A&A...434..599C, 2006MNRAS.367.1757C, 2005MNRAS.361..565C} and the so-called accretion limit \citep{1998ApJ...509L.121C,2001ApJ...562L..99C, 2002A&A...391.1069S,2009ApJ...691.1459V,2001ApJ...553..595B,2001A&A...375.1111G} on the formation efficiency. 

Experimentally, molecular hydrogen formation is studied using mass spectrometry coupled with temperature-programmed desorption (TPD). In an ultra-high-vacuum set-up,  hydrogen atoms are deposited on a cold interstellar grain analogue of one of the above-mentioned materials and the desorption of new molecules as a function of temperature is measured. Usually  beams of partially dissociated H$_{2}$ and D$_{2}$ are used and HD is observed, thereby avoiding the problems associated with the presence of undissociated molecules in the incident beams.  These beams are either produced in a micro-wave or a thermal cracking source. The atoms are generally collisionally cooled before they reach the surface where they can react.
Binding energies can be obtained by fitting a rate equation model to the experimental data where the desorption and diffusion energies are used as fitting parameters. The procedure results in different types of uncertainties. In their work, \citet{1999ApJ...522..305K} stated that they were only able to extract lower limits for the desorption energy of H atoms from the experimental data of \citet{1997ApJ...475L..69P,1997ApJ...483L.131P,1999A&A...344..681P}. Higher values for this energy were found to give degenerate results.
In general, one could state that the desorption energies of stable species like H$_2$ and D$_2$ are relatively straightforward to determine whereas desorption energies of unstable species like H and D are much harder to determine. Unfortunately, the latter are especially important in the determination of the temperature window in which efficient formation of H$_2$ can occur.

Another problem with the fitting procedure is that the formation mechanism is implicitly assumed in the model used for fitting. To use the results from  \cite{1999ApJ...522..305K} again as an example, \cite{2004ApJ...604..222C} and \cite{2005MNRAS.361..565C} were able to fit the same experimental results using different models. \cite{2004ApJ...604..222C} applied a binary model with both physisorption and  chemisorption binding sites, as well as quantum mechanical tunneling and thermal hopping between sites. \cite{2005MNRAS.361..565C} introduced surface roughness in their model. Another crucial ingredient to H$_2$ formation models is the diffusion barrier, which was recently determined to be 22~$\pm$~2~meV for D atoms on porous amorphous water \citep{2008A&A...492L..17M} in agreement with earlier experimental \citep{2003Sci...302.1943H} and theoretical results \citep{1991JChPh..95.6026B}. The determination of diffusion barriers remains experimentally challenging, however, and is one of the largest causes of uncertainty in the modeling of surface chemistry in general. 

The experimentally determined desorption rates show a rather large spread. This spread is not only due to experimental uncertainties but is an intrinsic property of the interstellar grains. \cite{2003Sci...302.1943H,2005JChPh.122l4701H} and \cite{2009PCCP...11.4396F} have shown the importance of surface roughness and porosity for H$_2$ desorption from amorphous solid water ice. More porosity or roughness leads to a wider distribution of desorption energies with a higher mean value. The exact preparation of the water ice substrate is key to the amount of porosity. \cite{2006MNRAS.365..801P} showed a similar effect occurs for silicates. Computer simulations of molecular hydrogen formation under interstellar conditions, \emph{i.e.}, using much lower fluxes than under laboratory fluxes, had already shown that an increase in surface roughness allows H$_2$ formation at higher temperatures \citep{2005A&A...434..599C, 2006MNRAS.367.1757C, 2005MNRAS.361..565C}. Surface irregularities in the form of protrusions, pores or amorphicity can act as sinks which increase the residence time of atoms on the surface whereas atoms on the smooth terraces can easily move around and reach these sinks. A second large uncertainty in surface chemistry is therefore the uncertainty in the exact surface grain structure. It appears that the difference in binding energies between the different materials is smaller than the difference due to the differences in  surface structure. 
  

Unlike the surfaces used in the laboratory, interstellar dust grains are small and generally have a small number of H atoms attached to them. The average number of H atoms per grain can under certain interstellar conditions be unity or less 
\citep{1982A&A...114..245T, 2001ApJ...553..595B, 2004MNRAS.348.1055L}.  Since clearly a minimum of two atoms needs to be present for H$_2$ to form, deterministic models like the 
standard rate equations can break down and one has to resort to stochastic methods. Generally two types of models are applied to replace the rate equation methods: macroscopic Monte Carlo methods \citep{1998ApJ...509L.121C,2001ApJ...562L..99C, 2002A&A...391.1069S,2009ApJ...691.1459V} and master equation treatments \citep{2001ApJ...553..595B,2001A&A...375.1111G,2002A&A...391.1069S}. The latter can be easily linked to rate equations which handle the gas phase chemistry in the most straightforward way. The main disadvantage of this technique is that, as the number of species expands, the equations blow up rapidly. Algorithms have been suggested to make the effect less dramatic \citep{2004PhRvL..93q0601L,2007ApJ...658L..37B}. 
Apart from the stochastic methods, some modifications to the rate equations in a semi-empirical manner have been put forward to mimic some of the results of these methods \citep{1998ApJ...499..234C, 2002P&SS...50.1257C,2008A&A...491..239G}. 
The above-mentioned roughness can be included via microscopic or kinetic Monte Carlo methods \citep{2006PNAS..10312257H}, and appears to extend the regime in which deterministic methods apply, since it increases the surface abundances. Whether simple rate equations will be able to capture this roughness is not yet clear (Cuppen et al. 2010a, submitted).  

\subsubsection{High temperature}
Molecular hydrogen formation at higher temperatures is almost uniquely studied on graphite. This system acts as a model system for both carbonaceous grains and polycyclic aromatic hydrocarbons (PAHs). Hydrogen atoms can form a chemical bond with graphite after they overcome a chemical activation barrier of 0.15-0.2~eV \citep{2006PhRvL..97r6102H,1999CPL...300..157J}. Chemisorbed atoms therefore mostly occur at higher temperature. 

Experimental \citep{2006PhRvL..96o6104H, 2006CPL...425...99A} and theoretical \citep{2003CPL...368..609F,2003JAP....93.3395M,2006PhRvL..96o6104H,2006PhRvL..97r6102H,2006CPL...431..135R} studies show that hydrogen atoms preferentially adsorb in clusters on graphite. Density Functional Theoy (DFT) calculations show that not only the sticking into dimer configurations is more favorable, but also that the binding energy of these dimers is higher than for two individual monomers. More complex structures consisting of three or more hydrogen atoms are found to be favorable as well \citep{2006PhRvL..97r6102H,2009JChPh.130e4704C}. The complex desorption behavior of H$_2$ from graphite as observed by \cite{2002JChPh.117.8486Z} was attributed to the presence of dimers and at high coverage to tetramers \citep{2009CPL...477..285G}.

Hydrogen abstraction from graphite has been studied experimentally \citep{2002CPL...366..188Z,2002JChPh.117.8486Z} and by DFT and Quantum Wave Packet calculations \citep{2000CPL...319..303F, Meijer:2001, 2002SurSc.496..318S, 2002JChPh.116.7158S, Morisset:2004, 2006JChPh.124l4702M}. Since the barrier for diffusion is higher than the desorption barrier for chemisorbed atoms, the formation mechanism is often referred to as an Eley-Rideal reaction where an incoming hydrogen atom reacts with another H atom that is chemically bound to the surface. Computational limitations do not, however, allow a simulation of the full complexity of the H-graphite system. Simplifications could be problematic. Generally either none or only a single C atom on the simulated graphite surface is allowed to relax during the reaction, which could artificially close low barrier reaction channels, such as sticking of hydrogen atoms into specific configurations \citep{2006PhRvL..97r6102H,2006CPL...431..135R}. 
\cite{2008JChPh.128q4707C} showed, in a kinetic Monte Carlo study which did allow for different mechanisms, that H$_2$ is more likely formed through a combination of fast diffusing physisorbed atoms \citep{2007JPC.111} and chemisorbed atoms. The fast moving physisorbed atoms can find a chemisorbed atom to form a dimer with. Part of the excess energy released in the binding process can be used to overcome the barrier for H$_2$ formation and desorption from the formed dimer. 

Under interstellar conditions, we expect some H$_2$ to form upon H atom sticking through this dimer-mediated mechanism. If the cloud is too cold  for the remaining complex structures to desorb, stochastic heating of the carbonaceous grains by UV photons can facilitate this process. Detailed simulations under these conditions are in progress and indeed the dimer-mediated mechanism proves to be dominant for high (gas) temperature (Cuppen et al. 2010b, submitted).

\subsection{Ice chemistry}
The formation of H$_2$ occurs in the submonolayer regime. Molecular hydrogen does not build up in layers but easily desorbs. When less volatile species land or form on the surface, layers of ice form, which can be rather inhomogeneous in composition. This layering gives an extra dimension to the surface chemistry which makes the laboratory results harder to interpret and surface modeling less intuitive. 
  
The new research field of laboratory solid state astrochemistry involves  surface processes such as desorption, reactions and energetic processing. Again models are used to interpret these data. Information on one-step processes like thermal desorption and photodesorption is relatively straightforward to extract. However often a sequence of processes as well as competing processes make it hard to disentangle the individual steps that are used as input for grain surface models. Here we will discuss thermal desorption, surface reactions by means of atom bombardment, and UV photoprocessing of ices. 

\subsubsection{Thermal desorption}
As in the case of hydrogen, the desorption energies determine the temperature regime in which species will be present on the surface of a grain and  available for reaction. The desorption energies of a wide collection of stable species have been determined using the TPD technique. Examples are N$_2$ \citep{2005ApJ...621L..33O}, CO \citep{2005ApJ...621L..33O}, O$_2$ \citep{2007A&A...466.1005A}, H$_2$O \citep{2005JChPh.122d4713B,2001MNRAS.327.1165F}, and CH$_3$OH \citep{2009MNRAS.398..357G}. The desorption energies have been mostly determined for the desorption of pure ices
from different substrates. The difference between the different substrates are rather small and become negligible in the multilayer regime. In the multilayer regime, the molecules desorb with a (near) zeroth order rate whereas they desorb with a (near) first order rate in the monolayer regime.

Since interstellar ices are not homogeneous the desorption of mixed layers is more relevant for astrochemical modeling. However, the introduction of more species in the ice makes the desorption process immediately much more complex. First, the desorption energy can change depending on its surrounding material. Second, the dominant mantle species can prevent other species from desorbing. 
\cite{2004MNRAS.354.1133C} showed, for instance, that molecules like CO and CO$_2$ can become trapped in an ice mantle which consists predominantly of water ice since the desorption of water occurs at much higher temperatures than for CO and CO$_2$. However, at the long timescales available in the ISM, some of these trapped species might be able to escape because of segregation, in which the two main fractions of the mantle slowly separate \citep{2009A&A...505..183O}. This process depends on a large number of parameters including surface temperature, ice composition and mixing ratio. Fortunately, it appears that some simple changes in gas-grain codes could treat the main features of this behavior \citep[][Fayolle et al. in preparation]{2009PhDT.........3O}. These changes would include splitting the grain mantle into two phases: a bulk and surface phase \citep{1993MNRAS.261...83H}. Desorption is only allowed from the latter phase.

\subsubsection{Surface reactions by atomic bombardment}
The experimental studies of ice reactions can be roughly divided into two groups depending on the analysis technique and the thickness regime (submonolayer versus multilayer). As in H$_2$ formation, the reactants and products can be probed mass spectrometically. The surface is initially exposed to a small quantity of the reactants, after which the surface is heated until the products and reactants desorb and are detected (thermally programmed desorption or TPD). Different initial exposures and temperatures can be probed to obtain information on reaction order, etc. The main advantage of this technique is the sensitivity, which allows for submonolayer exposures and which is able to detect all species by their masses. The method has however four major disadvantages: the products cannot be probed \emph{in-situ}, \emph{i.e.}, during the atom bombardment, additional reactions during the heat-up to desorption cannot be excluded, quantifying the desorbing species is not straightforward, and some of the interesting species have equal masses.

A second method is to initially grow an ice of several monolayers and expose this ice to the atom beam \emph{while} Recording Reflection Absorption Infrared Spectra (RAIRS). In this way the reactants and products are probed \emph{in-situ} at the time and temperature that one is interested in, which is the main advantage of this technique. Quantifying the formed product is relatively simple, provided that the RAIRS is calibrated with an independent method. The main disadvantages are that not all species can be detected unambiguously, e.g. because of spectral overlap, and that the sensitivity is less than with the previous technique. So far this technique has only been applied to smaller species, with ethanol the largest product formed \citep{2007A&A...474.1061B}. 
In the recent years, most laboratory reaction studies have been focused on the formation of two molecules: methanol and water. We discuss both briefly below.

Let us start with the formation of surface methanol, which has long been thought by astrochemists to proceed by the hydrogenation of CO ice via successive reactions with atomic hydrogen \citep{1997ApJ...482L.203C}:
\begin{equation}
{\rm CO \rightarrow HCO \rightarrow H_{2}CO \rightarrow H_{2}COH \rightarrow CH_{3}OH.}
\end{equation}
Laboratory data have become available confirming this sequence of reactions in several laboratories using RAIRS as the main analytical technique  \citep{2002ApJ...571L.173W,2009A&A...505..629F}.
Generally, pure CO ice is grown under ultra high vacuum conditions ($<10^{-10}$ mbar). The ice is
subsequently bombarded by H-atoms, which are produced
by  dissociating molecular hydrogen, H$_{2}$, using a microwave or thermal cracker. 
The formation of  H$_{2}$CO and CH$_{3}$OH was detected by both RAIRS and TPD. The first and third reaction step possess barriers to reaction. Since the rate of formation of molecules depends on a sequence of events: deposition of H atoms, diffusion and then reaction, these barriers can only be determined using a detailed model. Moreover, the formation of H$_{2}$CO and CH$_{3}$OH is in competition with H atom desorption and H$_2$ formation. 
A product yield of a few monolayers was obtained, which increases with temperature, below the desorption temperature. This indicates that the H atoms are able to penetrate a few monolayers of the CO ice and that this penetration depth increases with temperature. A microscopic kinetic Monte Carlo algorithm was fitted to a large dataset of H$_{2}$CO and CH$_{3}$OH production and CO loss data as functions of temperature and H-atom fluence  \citep{2009A&A...505..629F}. The reaction barriers were used as fitting parameters. The uncertainties in the obtained chemical barriers are mostly due to uncertainties in diffusion barriers and sticking fractions and experimental conditions such as temperature and fluxes.
Using these barriers, the methanol production can be simulated for
interstellar conditions \citep{2009A&A...508..275C}. This study clearly showed that under interstellar conditions different processes become dominant as compared to the laboratory. Under laboratory conditions H$_2$ formation is dominant whereas under interstellar conditions CO hydrogenation is dominant.  The study further showed that it is crucial to include layering and consider accessibility in surface models in order to correctly simulate surface chemistry. Again a  model that distinguishes between the surface layer and internal layers could help to include this behavior \citep{1993MNRAS.261...83H}.

Water ice can be formed on grain surfaces through three reaction channels: hydrogenation of atomic oxygen, molecular oxygen and ozone \citep{1982A&A...114..245T}. The laboratory of the Universit\'e Cergy-Pontoise has studied all three channels using a Quadrupole Mass Spectrometer as their main analysis technique \citep{2007msl..confE..79D,2008A&A...492L..17M,2009ApJ...705L.195M}. They confirmed that the hydrogenation of these three species indeed results in the formation of water, but quantitative data are not available. 
The hydrogenation of solid O$_2$ is the best studied channel \citep{2008CPL...456...27M,2008ApJ...686.1474I,2009ApJ...701..464O}. \cite{2008ApJ...686.1474I} investigated this reaction for a large range of temperatures. They found very surprising results. Unlike the CO hydrogenation study, where the product yield was only a few monolayers; O$_2$ hydrogenation appeared to result in a much larger yield which showed a strong temperature dependence. This larger yield occurs because of the ability of H atoms to diffuse easily through an O$_2$ ice (Ioppolo et al. in preparation). These results complicate the layering problem as discussed above, since the shielding of molecules is found to be strongly dependent on the species. Furthermore, the experiments showed that the reaction network of three channels that had been initially proposed appears to be more complex and that several of the channels are linked through additional reactions (Cuppen et al. 2010c, submitted).


\subsection{UV-photoprocessing of ices}
Other experiments are directed at the chemical and physical processes that occur during and after UV processing of ice samples.  The flux of UV photons in the laboratory is much greater than the flux in space, so that great care must be taken in the interpretation of these experiments.
 In the Leiden experiment \citep{2005ApJ...621L..33O,2007ApJ...662L..23O},
the ice is irradiated by the output of a broadband hydrogen-microwave
discharge lamp that simulates the interstellar radiation field. It peaks around
121 nm (Ly $\alpha$) and covers 115-170 nm ($\sim$7.5-10.5 eV) with a variable
photon flux covering roughly $(1-10)\times 10^{13}$ photons cm$^{-2}$s$^{-1}$ at
the substrate surface. The resulting ice behavior is monitored using  RAIRS
to quantify changes in ice constituents and TPD to investigate 
 ice desorption and identify the products of ice chemistry.

One of the major outcomes of these stuides is that photodesorption, a non-thermal desorption mechanism, is quite efficient and thus important in the cold interstellar medium \citep[see][for CO, N$_2$, CO$_2$ and H$_2$O ice photodesorption efficiencies, respectively]{2007ApJ...662L..23O,2009A&A...496..281O,2009ApJ...693.1209O}. The investigated molecules photodesorb through  different mechanisms. N$_2$ ice only co-desorbs with other molecules, which is expected since it has no dipole-allowed transition within the wavelength range of the lamp,  while CO ice photodesorbs non-dissociatively. H$_2$O and CO$_2$ are dissociated by the lamp. Following dissociation, theoretical simulations predict that a fraction of the fragments will desorb directly or recombine in the ice and then desorb \citep{2008A&A...491..907A}. Both desorption pathways are observed experimentally.

The photodesorption yields are $\sim$10$^{-3}$ molecules per incident photon for all molecules except for N$_2$, which has a yield that is an order of magnitude lower. CO ice photodesorption is a surface process because of its peculiar mechanism. CO$_2$ and H$_2$O are observed to photodesorb from deeper within the ice, but desorption is still limited to the top 10 layers.  It is also found that for methanol the photodesorption
yield amounts to around 2x10$^{-3}$ desorbed molecules per incident UV photon
\citep{2009A&A...504..891O} .  If one looks at the photodesorption in more detail, the yield $Y$ does depend to some extent on  $T$,  the ice temperature, $x$, the ice thickness and $l(T)$,  the temperature dependent mean-free-path of the excited molecule in the ice.  \cite{2009A&A...496..281O,2009ApJ...693.1209O} empirically determined photodesorption yields $Y$ for CO, CO$_2$ and H$_2$O to be 
\begin{eqnarray}
Y_{\rm CO} =  10^{-3}\left(2.7-(T-15)\times 0.17\right), \\
Y_{\rm CO_2}(T<40\:K) = 10^{-3}\left(1.2(1-e^{-x/2.9})+1.1(1-e^{-x/4.6})\right),  \\
Y_{\rm H_2O} = 10^{-3}\left((1.3+0.032\times T)(1-e^{-x/l(T)}) \right).
\end{eqnarray}
These yields are for multi-layered ices and have absolute uncertainties of $\sim$40--60\%; the current experimental set-up does not allow for (sub-)monolayer desorption determinations.   
Despite their uncertainties, these relatively high photodesorption rates may explain, for example, why CO can be found
in the gas phase at temperatures well below its desorption temperature \citep{2007ApJ...662L..23O,2009A&A...496..281O}, in protoplanerary disks for instance \citep{2009A&A...493L..49H}.

As photodesorption proceeds, photodissociation also occurs and  
new spectral features appear that correspond with reaction products \citep{2009A&A...494L..13O}.
Photochemistry in icy grain mantles was suggested as a path to chemical complexity more than three decades ago \citep{1979Ap&SS..65..215H} and studied since then \citep{1986A&A...158..119D,2005A&A...432..895D,2007AdSpR..40.1628N}.
New experiments in Leiden show that many of the complex organics  observed in
hot cores indeed can be formed through UV irradiation of solid CH$_3$OH and ice mixtures containing methanol \citep{2009A&A...504..891O}.  The UV radiation dissociates methanol into CH$_{2}$OH, CH$_{3}$O, CH$_{3}$, OH, and H.  Depending on ice temperature, some or all radicals react to form more complex molecules such as C$_2$H$_{5}$OH and (CH$_{2}$OH)$_{2}$.  During star formation, ices are always exposed to low levels of UV irradiation because of cosmic ray interactions with H$_{2}$.  After the protostar turns on,  the temperature of the cold material rises and the radicals produced by the photoprocessing diffuse rapidly enough to react to form more complex species, which 
are thermally released into the gas phase \citep{2006A&A...457..927G,2008ApJ...682..283G}.   The laboratory results are currently being fit to a theoretical model that uses rate equations; once the basic rate parameters are understood, they can be used to refine the model of photo-production of complex organics in hot cores (\"{O}berg, private communication).

\section{Summary and conclusions}

Some of the major projects undertaken in connection with this article are: 

(i)	the identification of the 'key' reactions in gas-phase models of the cold cores in dense interstellar clouds and of circumstellar envelopes; that is, those reactions that influence most strongly the predicted  abundances of the major species  observed in these molecule-rich astronomical objects, and their precision;

(ii)	 the careful evaluation of the available information on both the rate coefficients for these processes and the uncertainties in these coefficients;

(iii)	 an examination of the effects of including the re-evaluated rate coefficients on the predicted abundances of  the gas-phase model of cold cores. 

In parallel with the preparation of this article, a new database for astrochemistry (http://kida.obs.u-bordeaux1.fr/) has been prepared. Reasons for choosing the new values of rate coefficients can be found in that database. In running the models with the newly evaluated input data, uncertainties in the predicted abundances have been estimated by running a large number of model calculations with the rate coefficients for each of the reactions in the model chosen randomly within the evaluated uncertainties. The impact of changes in the rate coefficients has been explored for two types of astronomical sources, in which the physical conditions are quite different: cold dense clouds and evolved stars. In the latter case, photodissociation processes, which have not been evaluated here, are shown to be important. The comparison between predicted and observed abundances is slightly improved for the TMC-1 dense cloud by  including the new rate coefficients.  Whatever the current degree of improvement, it is important to include the 'best' values of the rate coefficients in this, and future, model calculations. Outstanding amongst the key reactions is the radiative association of C atoms with H$_2$ to form CH$_2$. This reaction presents an especially difficult problem to both experimentalists and theoreticians, but one that we hope will be tackled in the near future.

In the latter part of our article, we discuss how chemical changes on the surface of interstellar grains can be incorporated into so-called gas-grain models. The advantages and disadvantages of both rate equation and stochastic methods are considered. In addition, the results of recent experiments on surfaces designed to mimic the chemistry on both 'bare' surfaces and those 'coated'  with ices of different composition are reviewed, with some emphasis placed on the mechanisms for the formation of molecular hydrogen.

\begin{acknowledgements}
We acknowledge the "International Space Science Institute" for hosting the meetings of the international team "A new generation of databases for interstellar chemical modeling in preparation for HSO and ALMA". We would also like to thank the following people who took part in one or both of the ISSI meetings but did not contribute a section to this report:  Nathalie Carrasco, Prof. Ewine van Dishoeck, Prof. Dieter Gerlich, Eric H\'ebrard, Liv Hornekaer, Andrew Markwick, Prof. Tom Millar, Anton Vasyunin. V.W. thanks Astrid Bergeat for discussions about N + NO reaction. V.W., JC.L. and D.T. thank the French program PCMI for partial funding of this work.  E. H. wishes to thank the Center for the Chemistry of the Universe (NSF, US) for support of his program in chemical kinetics. VW and IWMS thank the Royal Society for a grant to facilitate their co-operation.
\end{acknowledgements}

\bibliographystyle{aps-nameyear}      
\bibliography{biblio_issi}   

\appendix
\section{Appendix A}\label{appendA}


\section*{Supplementary material}

Author: Ian Smith (University of Cambridge, UK)

$$H_3^+ + O(^3P) \rightarrow OH^+ + H_2 \qquad (1)$$
$$\rightarrow H_2O^+ + H \qquad (2)$$

*Thermodynamic Data*

$\Delta H°_{298}(1) = -57$ kJ mol$^{-1}$ $\qquad\qquad \Delta H°_{298}(2) = -158$ kJ mol$^{-1}$

$\Delta S°_{298}(1) =$ $\qquad\qquad \Delta S°_{298}(2) =$

$K_c(1) =$ $\qquad\qquad K_c(1) =$

Thermochemical data are taken from ref. 3

**Rate Coefficient Data ($k = k_1 + k_2$)**

| $k$ / cm$^3$ molecule$^{-1}$ s$^{-1}$ | $T$ / K | Reference | Comments |
|---|---|---|---|
| *Rate Coefficient Measurements* | | | |
| $(8 \pm 4) \times 10^{-10}$ | 300 | Fehsenfeld, 1976[1] | (a) |
| $(12 \pm 5) \times 10^{-10}$ | 295 ± 5 | Milligan and McEwan, 2000[2] | (b) |
| *Theoretical Evaluations* | | | |
| $11.9 \times 10^{-10} (T/K)/300)^{-0.144}$ | 0 – 70 | Bettens, Hansen and Collins, 1999[4] | (c) |
| $9.4 \times 10^{-10} (T/K)/300)^{-0.320}$ | 70 – 500 | Bettens, Hansen and Collins, 1999[4] | (c) |
| $1.14 \times 10^{-9} (T/K)/300)^{-0.156}\exp(-1.41/T)$ | 5 - 400 | Klippenstein et al., 2010[5] | (d) |
| *Reviews and Evaluations* | | | |
| $(8 \pm 4) \times 10^{-10}$ | all temperatures | UMIST database | |
| $(8 \pm 4) \times 10^{-10}$ | all temperatures | OSU website | |

**Comments**

(a) A very short paper with few experimental details. Flowing afterglow technique: production of $H_3^+$ described in ref. 3, no details of how of O-atom concentration were estimated. No measurements of branching ratio.

(b) Experiment employs a SIFT technique to generate and select $H_3^+$ ions. O atoms from microwave discharge of $O_2$/He. Reaction of $CH_3^+ + O$ (which it is claimed has a well-established rate coefficient) was used to calibrate the O-atom concentrations. Error takes into account uncertainties in estimating O-atom concentrations. Branching ratios are found to be $k_1/(k_1 + k_2) = 0.7$ and $k_2/(k_1 + k_2) = 0.3$.

(c) Results of a quasiclassical trajectory study on an *ab initio* potential energy surface,[4] provides a temperature-dependence for $k$. The branching ratios were calculated to be $k_1/(k_1 + k_2) = 0.94$ and $k_2/(k_1 + k_2) = 0.06$, independent of collision energy.

(d) Kinetics predicted from a combination of transition state theory, trajectory calculations and master equation analysis employing

potential energy surface from high level quantum chemical calculations.

**Preferred Values**

$k(298\ K) = (1.2 \pm 0.3) \times 10^{-9}$ cm$^3$ molecule$^{-1}$ s$^{-1}$

$k(10\ K) = (1.7 \pm 0.4) \times 10^{-9}$ cm$^3$ molecule$^{-1}$ s$^{-1}$

$k(T)\ 1.14 \times 10^{-9}\ (T/K)/300)^{-0.156} \exp(-1.41/T)$ cm$^3$ molecule$^{-1}$ s$^{-1}$

$k_1/(k_1 + k_2) = (0.7 \pm 0.2)$ and $k_2/(k_1 + k_2) = (0.3 \pm 0.2)$, independent of temperature.

*Reliability*
 $\Delta \log k = \pm\ 0.15$

*Comments on Preferred Values*
Both the OSU and UMIST databases accept Fehsenfeld's room temperature value from 1976,[1] and (apparently) assume (a) that reaction (1) to OH$^+$ + H$_2$ is dominant, and (b) that the rate coefficient is independent of temperature. Unfortunately, Fehsenfeld gives no details of how the O-atoms were generated and their concentration estimated. A follow-up paper that was supposed to give these details could not be found.

Milligan and McEwan[2] do give details. In effect, they measure the rate coefficient relative to that for CH$_3$O$^+$ + O. Their assessment of errors seems well founded, their value of $k(298\ K)$ and the branching ratio at 298 K (and the uncertainties in both quantities) are accepted.

Both Fehsenfeld[1] and Milligan and McEwan[2] only give room temperature values of the rate coefficient. The latter authors point out $k(298\ K)$ is close to the value expected from the Langevin model (17 × 10$^{-10}$ cm$^3$ molecule$^{-1}$ s$^{-1}$), on which basis one might expect no temperature-dependence.

Bettens *et al.*[4] provide a thorough theoretical analysis of this reaction. Their findings include: (a) that the capture cross-section is greater than that given by the Langevin model due to the long-range ion-quadrupole forces between H$_3^+$ and O($^3$P); and (b) that only *ca.* 80% of the 'captured' trajectories lead to reaction. They assume that reaction occurs only on the lowest triplet surface correlating with reactants and reduce their calculated value by a factor taking account of the spin-orbit states in O($^3$P$_J$). Together with (a) and (b), the value of $k(298\ K)$ obtained is close to that obtained by Milligan and McEwan[2] and to that estimated by a Langevin analysis.

The calculations of Klippenstein et al.[5] are in fair agreement with those of Bettens et al.[4], both in respect of the absolute values of the rate coefficients and their *T*-dependence, and again with the limited experimental data. They do not provide information on the branching ratio. We adopt their recommended *T*-dependence (in recognition of their more sophisticated treatment of spin-orbit effects) and the branching ratio from ref. 2.

We adopt the values of $k(298\ K)$ and the expression for $k(T)$ from Klippenstein et al. In view of the good agreement between the two sets of calculations and their agreement with the experimental values of k(298 K) the uncertainties in the recommendations are quite small.

(17.03.2010)



\end{document}